\documentclass[12pt]{article}
\usepackage{latexsym}
\usepackage{amsmath}
\usepackage{epsfig,graphics}

\newcommand{\be}{\begin{equation}}
\newcommand{\ee}{\end{equation}}

\newlength{\figsize}
\begin{document}

\begin{titlepage}

\vspace*{0.7in}
 
\begin{center}
{\large\bf Glueballs and k-strings in SU(N) gauge theories : \\
calculations with improved operators\\ }
\vspace*{1.0in}
{Biagio Lucini$^{a}$, Michael Teper$^{b}$ and Urs Wenger$^{c}$\\
\vspace*{.2in}
$^{a}$Institute for Theoretical Physics, ETH Z\"urich,\\
CH-8093 Z\"urich, Switzerland\\
\vspace*{.1in}
$^{b}$Theoretical Physics, University of Oxford,\\
1 Keble Road, Oxford OX1 3NP, U.K.\\
\vspace*{.1in}
$^{c}$NIC/DESY Zeuthen, Platanenallee 6, 15738 Zeuthen, Germany
}
\end{center}

\vspace*{0.55in}

\begin{center}
{\bf Abstract}
\end{center}

We test a variety of blocking and smearing algorithms for constructing
glueball and string wave-functionals, and find some with much improved
overlaps onto the lightest states. We use these algorithms to obtain
improved results on the tensions of $k$-strings in SU(4), SU(6), and
SU(8) gauge theories. We emphasise the major systematic errors that
still need to be controlled in calculations of heavier $k$-strings,
and perform calculations in SU(4) on an anisotropic lattice in a bid
to minimise one of these. All these results point to the $k$-string
tensions lying part-way between the `MQCD' and `Casimir Scaling'
conjectures, with the power in $1/N$ of the leading correction lying
$\in [1,2]$.  We also obtain some evidence for the presence of
quasi-stable strings in calculations that do not use sources, and
observe some near-degeneracies between (excited) strings in different
representations. We also calculate the lightest glueball masses for
$N=2, ...,8$, and extrapolate to $N=\infty$, obtaining results
compatible with earlier work. We show that the $N=\infty$
factorisation of the Euclidean correlators that are used in such mass
calculations does not make the masses any less calculable at large
$N$.

\end{titlepage}

\setcounter{page}{1}
\newpage
\pagestyle{plain}

\section{Introduction}
\label{section_intro}

Consider a lattice with spatial and temporal lattice spacings $a_s$
and $a_t$ (generically referred to as $a$ when equal).  Monte Carlo
calculations of the eigenvalues of the Hamiltonian, $H$, usually
proceed via the calculation of Euclidean correlation functions
\be
C(t)
=
\langle \Phi^{\dagger}(t)\Phi(0)\rangle
=
\sum_n |\langle\Omega|\Phi^{\dagger}|n\rangle|^2 e^{-E_n t}
=
\sum_n |c_n|^2 e^{-a_tE_n n_t}
\label{eqn_corrM}
\ee
where 
\be
H|n\rangle = E_n |n\rangle,
\label{eqn_eigenH}
\ee
and $t \equiv a_t n_t$.  If we were interested in a glueball mass, we
would construct $\Phi$ from products of link matrices around closed
contractible loops, and we would take linear combinations such that
$\Phi$ has the appropriate $J^{PC}$ quantum numbers and ${\vec p}=0$.
To calculate the tension of a $k$-string we would use a
non-contractible loop that winds $k$ times around a spatial torus.
(See Section~\ref{subsection_stringerr} and e.g.
\cite{blmt-kstring,blmt-glue}
for more details.) 

Suppose we are interested in the mass of the state $|l\rangle$.  Since
the statistical errors in these calculations are approximately
independent of $t\equiv a_tn_t$ while, as we see from
eqn(\ref{eqn_corrM}), the desired `signal' drops exponentially in
$n_t$, it is clearly important to construct operators that have
$|c_l|^2 \simeq 1$ (normalising to $\sum |c_n|^2 = 1$) if the
contribution we are interested in, $|c_l|^2 e^{-a_tE_l n_t}$, is to
dominate at the small values of $n_t$ where it is still much larger
than the errors. That is to say, we need an operator $\Phi$ that is a
good approximation to the wave-functional of the state $|l\rangle$.
This is more important the heavier the state.

Simple loops of bare fields are highly local and will project
more-or-less equally on all states. Operators that project mainly on
the lightest physical states should be smooth on the ultraviolet
lattice scale. Fast iterative techniques for constructing such
operators go under the names of blocking
\cite{block}
and smearing
\cite{smear}.
Although the existing techniques have allowed for the accurate
calculation of the lightest masses, even at small $a$, they are not
good enough in all cases. In particular, the overlaps of $k$-strings
are found
\cite{blmt-kstring}
to satisfy
\be
|c_{k}|^2 \simeq \biggl\{ |c_{k=1}|^2 \biggr\}^k
\label{eqn_overk}
\ee
and the mass of the string increases rapidly with $k$.  Thus an
acceptable overlap for the fundamental string, e.g.  $|c_{k=1}|^2 \sim
0.85$, transforms into a uselessly poor overlap for, say, a $k=4$
string.

In this paper we suggest and test some improvements to the common
blocking and smearing techniques. Using these improved methods we
present some illustrative calculations of $k$-string tensions and
glueball masses for groups as large as SU(8).

Using anisotropic lattices with $a_t \ll a_s$ provides a finer
resolution of the $t$-dependence of $C(t)$ and helps to establish the
minimum $t$ at which the lightest mass already dominates $C(t)$
\cite{anisotropic}.
This is of particular importance for heavy masses, such as those of
$k$-strings with $k>1$, where it is undoubtedly the source of the
major systematic error in current calculations.  We perform
calculations of $k=2$ string tensions in SU(4) using anisotropic
lattices with the emphasis on minimising this particular systematic
error.

All these calculations obtain masses from correlations of field
fluctuations. However we know that in the $N = \infty$ limit there are
no fluctuations and correlators factorise.  One might wonder if this
means that mass calculations become impossible as $N \to \infty$
\footnote{We are grateful to Simon Dalley for making
this point to one of us a number of years ago.}.
We analyse this question towards the end of this paper.

In the next section we summarise the lattice `setup'. We then move on
to discuss systematic errors, with the focus on those that are
potentially important in calculations of $k$-string tensions. The
following section embellishes the standard blocking and smearing
algorithms, with detailed numerical tests to establish the extent of
the improvement.  We then summarise the results of our calculations
with these improved operators, as well as some with an anistropic
lattice spacing, first for glueballs and then for $k$-strings, where
we demonstrate how improving the operators has improved upon previous
work. In addition to the lightest, stable $k$-strings we observe
evidence for heavier, quasi-stable strings. Finally, after a general
analysis of mass calculations in the $N \to \infty$ limit, we
summarise, in the concluding section, what we have learned about the
physics, the methods, and what are some of the outstanding problems.

\section{Lattice setup}
\label{section_lattice}

We use hypercubic, periodic lattices of size $L_s^3 L_t$ (using $L$
when there is no ambiguity). We assign SU($N$) matrices, $U_l$, to the
links $l$. (Or equivalently $U_\mu(n)$ for a link emanating in the
$\mu$ direction from site $n$.) For most of our calculations we use
the standard isotropic ($a_s=a_t=a$) plaquette action
\begin{equation}
S = \beta \sum_{p}\{1-{1\over N}{\mathrm {ReTr}} U_p\},
\label{B1}
\end{equation}
where $U_p$ is the ordered product of the SU($N$) matrices around the
boundary of the plaquette $p$. $S$ appears in the Euclidean Path
Integral as $e^{-S}$ and becomes, in the continuum limit, the usual
Yang-Mills action with
\begin{equation}
\beta = {{2N} \over {g^2}}.
\label{B2}
\end{equation}
As we vary $N$ we expect 
\cite{largeN,lattice-thooft}
that we will need to keep constant the 't Hooft coupling, $\lambda$,
\begin{equation}
\lambda \equiv g^2 N  
\label{B3}
\end{equation}
for a smooth large-$N$ limit. Non-perturbative calculations
\cite{blmt-kstring,blmt-glue}
support this expectation.

The anisotropic lattice action we use is 
\begin{equation}
S = 
\beta_s \sum_{p_s}\{1-{1\over N}{\mathrm {ReTr}} U_{p_s}\}
+
\beta_t \sum_{p_t}\{1-{1\over N}{\mathrm {ReTr}} U_{p_t}\}
\label{B4}
\end{equation}
where $p_s$ and $p_t$ are spatial and temporal plaquettes, and the
couplings are related to the anisotropy
\begin{equation}
\xi={{a_t}\over{a_s}}
\label{B5}
\end{equation}
by
\begin{equation}
\beta_s = \xi\beta \ \ \ ; \ \ \
\beta_t = {1\over{\xi}}\beta \ \ \ ; \ \ \
\beta = {{2N}\over{g^2}}
\label{B6}
\end{equation}
at tree level. The actual renormalised anisotropy, $\xi_r$, will
differ from the bare anisotropy, $\xi = \sqrt{\beta_s/\beta_t}$, that
we use in eqn(\ref{B6}) in setting the couplings. If we were only
interested in ratios of masses this would not matter: we can safely
extrapolate $a_tm_a/a_tm_b \equiv m_a/m_b$ to the continuum limit with
an $(a_t m_c)^2$ correction term. However if we wish to calculate a
string tension this is no longer so since the mass, $m_k$, of the
winding $k$-string is given by
\begin{equation}
a_tm_k = a_ta_sL_s \sigma_k 
- 
{{a_t}\over{a_s}}{{\pi}\over{3 L_s}}.
\label{B7}
\end{equation}
Here we include the leading long distance string correction
\cite{string}
to the linear term. So to calculate $a_t\surd\sigma_k$ we need to
calculate the renormalised anisotropy $\xi_r = a_t/a_s$. We shall do
so by calculating string energies for the lowest few non-zero momenta
and assuming the continuum dispersion relation, as described in detail
in Appendix D of ref
\cite{mtd3}.
Different ways of calculating $\xi_r$ will differ by lattice spacing
corrections, but this will not matter if we perform a continuum
extrapolation.

Our simulations are performed with a combination of heat bath and
over-relaxation updates, as described in
\cite{blmt-kstring,blmt-glue}.
We update all the $N(N-1)/2$ SU(2) subgroups of the SU($N$) link
matrices.

\section{Systematic errors}
\label{section_errors}

\subsection{k-strings}
\label{subsection_stringerr}

What is a $k$-string? Consider sources that transform as $\psi(x) \to
z^k \psi(x)$ under a global gauge transformation in the centre of the
group, $z\in Z_N$. Gluon screening cannot change $k$ but can change
one source to another of the same $k$.  The $k$-string is the flux
tube with the smallest string tension in the $k$ class, and so it is a
stable string. All other strings in the $k$ class will decay to it
(for a long enough string).  One finds
\cite{blmt-kstring}
that it corresponds to the totally antisymmetric representation.
Obvious constraints arise from $z^N=1$ and from the fact that $k$ and
$-k$ are charge conjugates. So for $N\leq 3$ we only have $k=1$
strings, for $N\geq 4$ we also have $k=2$ strings etc.

We expect
\begin{equation}
\sigma_k \stackrel{N\to\infty}{\longrightarrow} k\sigma
 \ \ \ ; \ \ \ \sigma\equiv\sigma_{k=1}
\label{eqn_D1}
\end{equation}
and the interesting physical question is how tightly bound are the
$k$-strings at finite $N$. Earlier calculations
\cite{blmt-kstring,PisaK}
have established that the values of $\sigma_k/\sigma$ lie in the range
spanned by the `Casimir Scaling'
\cite{CS}
and `MQCD'
\cite{MQCD}
conjectures
\begin{equation}
{{\sigma_k}\over{\sigma}}
= \begin{cases}
 k(N-k)/(N-1) & : \quad \text{`Casimir scaling'} \\
\sin{{k\pi}\over{N}}/\sin{{\pi}\over{N}} & : \quad \text{`MQCD'}
\end{cases}
\label{eqn_D2}
\end{equation}
(See
\cite{blmt-kstring,PisaK}
for a more detailed discussion.) To be able to firmly exclude one or
both of these possibilities requires an accuracy at the $\leq \pm 2\%$
level on $\sigma_k$. There are at least two sources of systematic
error that can be significant at this level of precision. The first
tends to lead to the value of $\sigma_k/\sigma$ being overestimated,
and increasingly so for increasing $k$ and $N$. The second works
(potentially) in the reverse direction.

\subsubsection{extracting the mass}
\label{subsubsection_massbias}

When we estimate a mass from the correlator in eqn(\ref{eqn_corrM}) we
do so by calculating the effective masses
\be
am_{eff}(t) = -\ln {{C(t)}\over{C(t-a)}}
\label{eqn_D3}
\ee
and identifying an effective mass plateau $t\geq t_{min}$ where
$am_{eff}(t_{min}) = am_{eff}(t), \ t\geq t_{min}$, within errors.
Thus for $t\geq t_{min}$ we can regard $C(t)$ as being given by a
single exponential and we can then use $am_{eff}(t_{min})$ as our
estimate of the actual mass.  (The actual procedure will typically
involve fits over ranges of $t$, but this does not alter the
argument.) Now, since $\sigma_k$ grows with $k$ (and indeed with $N$)
the mass of the corresponding string also grows. For such heavy states
the correlator in eqn(\ref{eqn_corrM}) drops rapidly into the
statistical noise as we increase $t$ and the significance of an
apparent effective mass plateau decreases.  Thus for heavier states we
will typically end up extracting the mass at a value of $t$ that is
too low. Given the positivity of the correlator, this means that the
mass estimate will be somewhat too high. This shift will grow with $k$
because the mass grows with $k$. It is thus a systematic bias that
will lead to an overestimate of $\sigma_k/\sigma$, and the
overestimate will increase with $k$ and $N$.  Since an increase in
statistical errors has the same type of effect, if the higher $N$
calculations have lower statistics (which can easily happen because
they are more expensive) this systematic bias will be further
enhanced. This is further compounded by the fact that the overlap of
the best operator typically decreases with $k$, as in
eqn(\ref{eqn_overk}), so that the real value of $t_{min}$ is higher
for these more massive states.

The solution to this problem is firstly to find improved operators for
which $|c_{k=1}|^2 \simeq 1$ so that $|c_k|^2 \sim 1$ for the values
of $k$ of interest; secondly to find methods that can improve the
statistical accuracy of calculations of heavier masses at larger
$n_t$. Once this is done, calculations with anisotropic lattices may
be very useful in giving us a finer-grain in $t$ so that we can
establish the existence of a plateau with greater certainty.

In this paper we go some way to providing the first part of this
solution, showing how one can significantly improve operator overlaps
in a simple way. We do not address the second part, on how to improve
the statistical accuracy of mass calculations of heavy states, but
note that important progress has very recently been made on this
problem
\cite{meyer}
in a generalisation of the recent multi-level algorithm for Wilson
loops
\cite{LW}.
Finally, we perform a limited anisotropic lattice calculation to
illustrate its utility.

\subsubsection{finite volume correction}
\label{subsubsection_volumebias}

Denoting the lightest mass of a periodic $k$-string of length $l=aL$
by $m_k(L)$, we expect
\cite{string,blmt-kstring}
that at sufficiently large $l$ we will have 
\begin{equation}
am_k(L) \equiv a^2\sigma_k(L)L 
= 
a^2\sigma_k(\infty)L - {{\pi}\over{3L}}
\label{eqn_D4}
\end{equation}
up to higher order corrections in $1/L^2$.  There is good evidence
that this has become a very accurate approximation to $k=1$ strings
once $aL\sqrt{\sigma}\geq 3$
\cite{blmt-kstring},
but the evidence that this is also so for $k>1$ strings
\cite{blmt-kstring}
is very much less precise. Current calculations of $k$-string tensions
are with $aL\sqrt{\sigma}\sim 3$ and assume the validity of
eqn(\ref{eqn_D4}). This raises an important issue because the
magnitude of the shift in $\sigma_k/\sigma$ produced by, for example,
doubling the $O(1/L^2)$ correction to $\sigma_k(L)$ is typically as
large (or larger than) the difference between the two possibilities in
eqn(\ref{eqn_D2}).

Theoretically, there is something one can say about these
corrections. As $N\uparrow$ we expect $\sigma_k \to k\sigma$ so that
the $k$-string is less and less strongly bound. Now, the lightest
energy of the unbound state composed of $k$ independent fundamental
$k=1$ strings is
\begin{equation}
a\tilde{m}_k(L) = k am_{k=1}(L) = 
k a^2\sigma(\infty)L - k {{\pi}\over{3L}}
\label{eqn_D5}
\end{equation}
for large enough $L$. Because of the larger string correction, this
may be lighter than the $k$-string bound state for smaller $L$ and may
therefore be the $k$-string ground state whose mass we are calculating
numerically. One can crudely estimate the critical length $L_c$ below
which this should be so, as the length at which $a\tilde{m}_k(L)$ in
eqn(\ref{eqn_D5}) and $am_k(L)$ in eqn(\ref{eqn_D4}) are equal, i.e.
\begin{equation}
a\sqrt{\sigma}L_c(k;N) 
= \begin{cases}
\sqrt{\frac{\pi}{3}\frac{N-1}{k}} & : \quad \text{`Casimir scaling'} \\
\frac{N}{k}\sqrt{\frac{2}{\pi(1+\frac{1}{k})}} & : \quad \text{`MQCD'}
\end{cases}
\label{eqn_D6}
\end{equation}
for the two possibilities in eqn(\ref{eqn_D2}), and, in the second
case, taking into account only the leading large-$N$ correction. Thus
if we calculate $\sigma_k$ from strings of length $L<L_c(k,N)$ then we
can expect to obtain an underestimate of $\sigma_k/\sigma$.  We note
that for $k=2$ and $N=8$, we have $a\sqrt{\sigma}L_c \sim 2$. This is
getting uncomfortably close to the kind of volume we work on, given
the crude and asymptotic nature of the estimate and the fact that
there may well be an extended transition region between the unbound
and bound states.

One should also consider scenarios where the $k$-string is composed of
a mixture of $k^\prime < k$ strings. And given that all these strings
of length $L$ are on a transverse two torus of dimension $L\times L$
there will be some energy shift for smaller $L$. Nonetheless it is
clear that if eqn(\ref{eqn_D4}) is to be valid for $k\geq 2$ strings,
we will, at large $N$, have to go to larger $L$ than is needed for
$k=1$ strings, and the systematic bias in not doing so appears to
underestimate the value of $\sigma_k/\sigma$. Thus it is important to
do a much more accurate finite volume study than any currently
available.  Such a study will involve longer and thus heavier
$k$-strings and will undoubtedly require something like the multilevel
algorithm
\cite{meyer}
to be possible. Until such studies are performed any attempt to
calculate the $k$-string tensions at the few percent level, as we are
trying to do when we wish to differentiate between Casimir Scaling and
MQCD, must be considered provisional.

\subsubsection{topology and ergodicity}
\label{subsubsection_qerr}

As $N$ increases (and also as $a$ decreases) the lattice fields are
increasingly trapped in a given topological charge sector. This is
because of the well-known suppression of small instantons at large $N$
\cite{blmt-glue,ox-Q,pisa-Q,GW}.
Thus in SU(8) our sequence of 45000 $16^4$ lattice fields at
$\beta=45.7$ never changes from $Q=0$. And our 50000 $12^4$ SU(8)
fields at $\beta=44.85$ only have a very few changes of $Q$. Is our
volume large enough that the effect of this loss of ergodicity in the
global topology on the calculated masses is negligible, or not?

To answer this question we separate our sequence of SU(8) lattice
fields at $\beta=44.85$ into subsequences with $Q\simeq 0$, $|Q|=1$
and $|Q|\simeq 2$. (The `$\simeq$' indicates that the subsequences
contain a very few fields with a different $Q$.)  Within these
subsequences we perform calculations of the lightest $k=1$ and $k=2$
string masses, and we list these values in Table~\ref{table_meffQ}. We
see that the masses appear to be unaffected by the constraint of being
in a fixed topological charge sector. This reassures us that, at the
level of our precision in this paper, this is not a significant source
of error. However this is something that will need checking again if
one is doing a much more accurate calculation, and for calculations of
other physical quantities.

\subsection{glueballs}
\label{subsection_glueballerr}

The problems with calculating heavier glueball masses are much the
same as for $k$-strings and the solution will be similar. It is
perhaps less pressing here because there is little demand at present
for theoretical comparisons that require accuracy at the percent
level. (Except perhaps in recent work on the Pomeron
\cite{pomeron}.)
So, just as for the strings, we will construct improved operators and
will provide a calculation on anistropic lattices. For basic
improvements in the Monte Carlo algorithm we refer to
\cite{meyer}.

The finite volume issue is much less pressing here because the leading
corrections at large volume are exponentially small. Moreover, in
contrast to the case of $k$-strings, checking for finite volume
effects is straightforward because glueballs do not become very heavy
on large volumes. So the cost of the calculation is merely linear in
the volume -- and less so if one calculates the energies corresponding
to the lowest non-zero momenta as well.

The one troublesome finite volume effect which occurs on the
intermediate spatial sizes that ones uses in practice, comes from a
pair of (mutually conjugate) flux loops which can have a non-zero
overlap onto glueballs states. In particular in our calculation the
mass of such a pair of loops is close to that of the $0^{++\star}$ and
we suspect that the frequently poor continuum fits for this state are
a sign of mixing between the true $0^{++\star}$ and this finite-volume
state (whose mass will diverge in the thermodynamic limit). This can
be dealt with by including explicitly operators for such two loop
states in the basis so as to identify states which mix with it.
Similarly, if one is interested in heavier glueballs and wishes to
exclude scattering states of the same quantum numbers, it would be
useful to include explicitly operators for such scattering states in
the basis. We do not do so in the present paper. We note that the
overlaps between such scattering states and the real glueballs should
vanish as $N\to\infty$ by the usual large-$N$ arguments.

Finally we note in Table~\ref{table_meffQ} that the scalar glueball
does not appear to be affected by constraining the total topological
charge $Q$ to be constant, so we can assume that our mass calculations
are not affected by this loss of ergodicity.

\section{Improved operators}
\label{section_ops}

\subsection{algorithms}
\label{subsection_algorithms}

In a gauge-invariant lattice calculation, glueball and string
operators are composed of ordered products of link matrices around
closed loops. If $a$ is small then we would expect that a good
wave-functional for any of the lightest states will be smooth on
scales of the order of an appropriate physical length scale and so
certainly smooth on the scale of $a$. One can achieve this by summing
over paths between sites and using these, rather than the original
link matrices, as the basic components of the operator. By iterating
the procedure one can efficiently sum vast numbers of paths, producing
operators smooth on physical length scales.

Traditionally there have been two common variants of this procedure,
often referred to as `blocking'
\cite{block}
and `smearing' 
\cite{smear}
respectively. We only consider loops and links that are space-like, so
that the positivity of correlation functions is preserved. Thus all
indices will run from 1 to 3.

The first step of the usual smearing algorithm sums up the five
shortest paths between neighbouring sites, i.e.~the link $l$ and the
`staples' (with some relative weighting $p_a$).  This produces an
$N\times N$ matrix, ${\tilde U}^{s=1}_l$, which we assign to that
link, after first projecting it to a `nearby' SU($N$) matrix,
$U^{s=1}_l$ . We iterate the procedure
\begin{eqnarray}
{\tilde U}^{s+1}_i(n) & = & U^s_i(n)
+  p_a
\sum_{j\not= i}
U^s_j(n) U^s_i(n+\hat{\jmath})
U^{s\dagger}_j(n+\hat{\imath}) \nonumber \\
&  & + p_a
\sum_{j\not= i}
U^{s\dagger}_j(n-\hat{\jmath}) U^s_i(n-\hat{\jmath})
U^s_j(n-\hat{\jmath}+\hat{\imath}), \nonumber \\
U^{s=0}_i(n) & = & U_i(n)   
\label{eqn_oldsmear}
\end{eqnarray}
with
\begin{eqnarray}
U^{s+1}_i(n) = {\cal{U}}\Bigl\{ {\tilde U}^{s+1}_i(n)\Bigr\}
\label{eqn_sunitarise}
\end{eqnarray}
representing the procedure by which the matrix is transformed into a
special unitary one. We can now form operators by multiplying these
smeared link matrices around closed loops $c$
\begin{eqnarray}
\Phi({\cal C})
=
\sum_{c\in {\cal C} } \prod_{l\in c} U^s_l
\label{eqn_phi}
\end{eqnarray}
where it is understood that if we go backward along a link $l$ then we
use $U^{s\dagger}_l$ in place of $U^s_l$. The collection of loops
${\cal C}$ can be chosen so as to give the wave-function $\Phi({\cal
C})$ the desired quantum numbers.

Smearing produces SU($N$) matrices on the original links of the
lattice.  The parameter $p_a$ determines how rapidly the link field
spreads outwards as the procedure is iterated. Choosing $p_a$ small
means that very many smearing steps will be needed to produce links
smeared on the desired physical length scale and this will be
expensive. On the other hand this will produce operators that extend,
with a fine resolution, over all important length scales, so that good
overlaps are more likely to be achieved.
 
With blocking, the matrices live on `superlinks' joining sites that
are $2^b$ lattice spacing apart, where $b$ is the number of blocking
iterations. Each iteration involves adding the direct path to 4
elongated staples,
\begin{eqnarray}
{\tilde U}^{b+1}_i(n) & = & U^b_i(n)U^b_i(n+2^b\hat{\imath})
\nonumber \\
&  & +  p_a \sum_{j\not= i}
U^b_j(n) U^b_i(n+2^b\hat{\jmath})
U^b_i(n+2^b\hat{\jmath}+2^b\hat{\imath})
U^{b\dagger}_j(n+2^{b+1}\hat{\imath}) \nonumber \\
&  & + p_a
\sum_{j\not= i}
U^{b\dagger}_j(n-2^b\hat{\jmath}) U^b_i(n-2^b\hat{\jmath})
U^b_i(n-2^b\hat{\jmath}+2^b\hat{\imath})
U^b_j(n-2^b\hat{\jmath}+2^{b+1}\hat{\imath}), \\
U^{b=0}_i(n) & = & U_i(n)
\label{eqn_oldblock}
\end{eqnarray}
where again one projects ${\tilde U}^{b+1}_i(n)$ into SU($N$),
\begin{eqnarray}
U^{b+1}_i(n) = {\cal{U}}\Bigl\{ {\tilde U}^{b+1}_i(n)\Bigr\},
\label{eqn_unitarise}
\end{eqnarray}
and we form wavefunctions by multiplying these matrices around closed
loops of the corresponding superlinks. Here one typically chooses
$p_a=O(1)$ so that the width of the blocked link increases at least in
step with its length.

Blocking is faster because one smoothens by a factor of two in size at
each step. On the other hand the same factor of two gives it a rather
crude resolution. If the size of the wave-functional of the state of
interest falls between two blocking levels then the overlap might well
be suppressed leading to a poorer calculation than with a fine-grained
smearing.

An obvious way to try and improve the above smearing is to make it
more symmetric about the axis of the link.  In eqn(\ref{eqn_oldsmear})
we add to the link just the four nearest parallel links going outwards
along the lattice axes. (The staples are just these link matrices
parallel transported to the link of interest.)  After a number of
successive smearings this can lead to an operator with peculiar and
unnatural axial rotation properties. A first step to alleviate this is
to include the next set of parallel links, those which are a distance
$\surd 2a$ away, diagonally across the square lattice.  To parallel
transport these requires a minimum product of 2 link matrices at each
end, giving the algorithm
\begin{eqnarray}
 &  & {\tilde U}^{s+1}_i(n) =  U^s_i(n)
+  p_a
\sum_{j\not= i}
U^s_j(n) U^s_i(n+\hat{\jmath})
U^{s\dagger}_j(n+\hat{\imath}) \nonumber \\
&  & + p_a
\sum_{j\not= i}
U^{s\dagger}_j(n-\hat{\jmath}) U^s_i(n-\hat{\jmath})
U^s_j(n-\hat{\jmath}+\hat{\imath}) \nonumber \\
&  & + p_d
\sum_{j\not= i}\sum_{k\not= i,j}
U^s_j(n) U^s_k(n+\hat{\jmath})
U^s_i(n+\hat{\jmath}+\hat{k})
\sum_{j^\prime\not= i}\sum_{k^\prime\not= i,j^\prime}
U^{s\dagger}_{j^\prime}(n+\hat{\imath}+\hat{k^\prime}) 
U^{s\dagger}_{k^\prime}(n+\hat{\imath}) \nonumber \\
&  & + p_d \{rotations\} 
\label{eqn_newsmear}
\end{eqnarray}
with ${\tilde U}^{s+1}_i(n)$ then being unitarised to
${U}^{s+1}_i(n)$.  The term labelled `rotations' refers to three
further terms like the previous one, but rotated by a multiple of
$\pi/2$ around the $i$-axis (so that either $j$ or $k$ or both go
backwards).  So now the smearing consists of the direct path, the 4
staples and 16 `wiggly' staples. We now have two parameters, $p_a$ and
$p_d$, to choose (see below).

As for blocking, an obvious variant of the algorithm in
eqn(\ref{eqn_oldblock}), which is both more elegant and probably
better (because it includes more paths), is to simply multiply two
smeared links together
\begin{eqnarray}
U^{b=1}_i(n) & = & U^s_i(n) U^s_i(n+\hat{\imath}), \nonumber \\
U^{b=1,s}_i(n) & = & {\cal{S}}^s\{U^{b=1}_i(n)\}, \nonumber \\
U^{b+1}_i(n) & = & U^{b,s}_i(n) U^{b,s}_i(n+2^b\hat{\imath}). 
\label{eqn_newblock}
\end{eqnarray}
Here $U^s$ is a link that has been smeared $s$ times, and ${\cal S}$
denotes the smearing operation generalised to apply to blocked
links. That is to say, in eqn(\ref{eqn_oldsmear}) or
eqn(\ref{eqn_newsmear}) the matrices are replaced by blocked matrices
and the links are replaced by the appropriate superlinks. While each
step of the old blocking iteration summed just 5 paths (at the
previous blocking level) this improved blocking (using improved
smearing) will sum $21^2$ paths (ignoring double counting and
cancellations) and will be more axially symmetric (with an appropriate
choice of $p_a,p_d$).

Clearly these algorithms can be applied in various 
combinations. We focus on three simple strategies. 
{\hfil\break}
A) Iterated smearing using eqn(\ref{eqn_newsmear})
with small values of the parameters designed to give 
operators spanning all relevant scales with a fine
resolution.{\hfil\break}
B) Iterated blocking using eqn(\ref{eqn_newblock}) and 
eqn(\ref{eqn_newsmear}) with larger values of the parameters,
designed to rapidly go from ultraviolet to physical length
scales.{\hfil\break}
C) A hybrid where one initially blocks as in (B) and then
multiply smears the blocked links as in (A). The (iterated) 
blocking is designed to rapidly take us to the lower limit 
of physically interesting length scales, with the
subsequent smearing providing a fine-resolution exploration
of the larger length scales.{\hfil\break}
To determine the best choice of parameters we will calculate
the desired correlator and find which parameters give the
best overlap onto the desired state.

\subsubsection{unitarisation}
\label{subsubsection_unitarity}

Before that, some comments. There is no fundamental reason why the
smeared or blocked matrices should be unitary.  However in practice
one finds that projecting back to SU($N$) produces eventual overlaps
that are as good as with any other normalisation (and better than
most). Thus it is convenient to unitarise. We unitarise by finding the
SU($N$) matrix $U^b$ that maximises
\be
\max_{U^b\in \text{SU}(N)} Re Tr \{{\tilde U}^{b\dagger} U^b\}
\label{eqn_unitarise2}
\ee
and similarly for smearing. This can be done by using the
$\beta=\infty$ limit of the Cabibbo-Marinari heat bath, just as one
does when `cooling' lattice fields to calculate their topological
charge
\cite{cooling}.
This is an iterative procedure which needs to start with some initial
value for $U^b$, call it $U_s^b$.  Typically one will choose $U_s^b$
to be some very crude unitarisation of ${\tilde U}^b$.  After one or
two iterations one obtains a good approximation to $U^b$ in
eqn(\ref{eqn_unitarise2}).  The fact that this is an approximation
means that one does not maintain exact gauge invariance. However all
this will do is to increase slightly the statistical noise in the
calculation of correlators. A second potential problem arises in the
choice of the crude starting point, $U_s^b$. Typically this will not
be the same for $U^b$ and for $U^{b\dagger}$. For example if one
obtains $U_s^b$ by orthonormalising the columns of ${\tilde U}^{b}$
one by one, as we do, one gets a different matrix than if one started
by orthonormalising the rows. This effectively breaks the rotational
symmetry and can undermine in a subtle way the assignment of $J^{PC}$
quantum numbers in the construction of glueball operators
\footnote{We are grateful to Harvey Meyer for pointing out to us this
potential flaw in the usual blocking/smearing algorithms.}. 
In practice after the one or two cooling iterations that we normally
perform in our calculations, the resulting matrix loses almost all
memory of the starting matrix and the symmetry breaking is
insignificant. If one needed to be more careful a simple remedy would
be to choose to orthogonalise rows or columns at random, so that the
symmetry would be restored in the ensemble.

\subsection{numerical tests}
\label{subsection_tests}

The improved blocking we shall focus upon performs the blocking by
multiplying together two once-smeared links as in
eqn(\ref{eqn_newblock}).

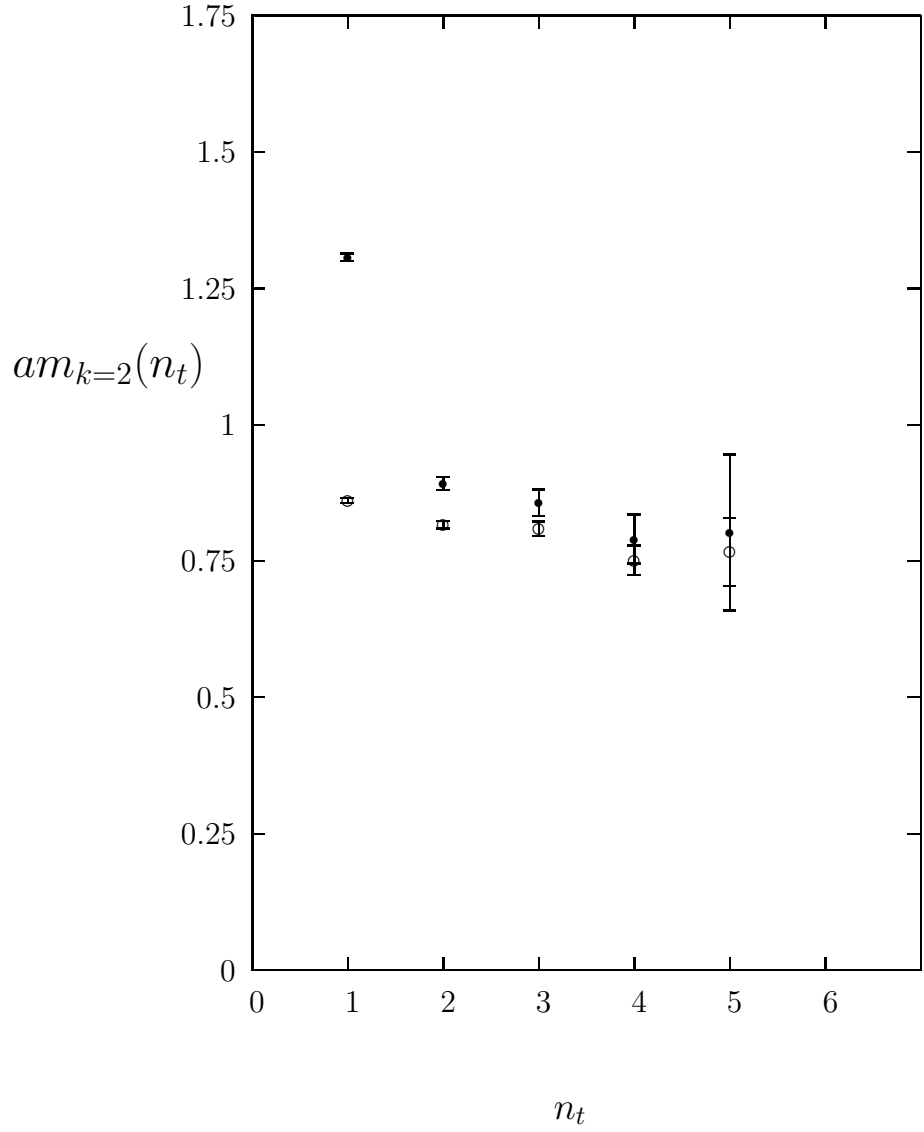
\begin	{figure}[p]
\begin	{center}
\leavevmode
\setlength{\unitlength}{0.240900pt}
\ifx\plotpoint\undefined\newsavebox{\plotpoint}\fi
\sbox{\plotpoint}{\rule[-0.200pt]{0.400pt}{0.400pt}}%
\begin{picture}(1500,1800)(0,0)
\font\gnuplot=cmr10 at 12pt
\gnuplot
\sbox{\plotpoint}{\rule[-0.200pt]{0.400pt}{0.400pt}}%
\put(375.0,250.0){\rule[-0.200pt]{4.818pt}{0.400pt}}
\put(350,250){\makebox(0,0)[r]{\ \ {$0$}}}
\put(1405.0,250.0){\rule[-0.200pt]{4.818pt}{0.400pt}}
\put(375.0,464.0){\rule[-0.200pt]{4.818pt}{0.400pt}}
\put(350,464){\makebox(0,0)[r]{\ \ {$0.25$}}}
\put(1405.0,464.0){\rule[-0.200pt]{4.818pt}{0.400pt}}
\put(375.0,679.0){\rule[-0.200pt]{4.818pt}{0.400pt}}
\put(350,679){\makebox(0,0)[r]{\ \ {$0.5$}}}
\put(1405.0,679.0){\rule[-0.200pt]{4.818pt}{0.400pt}}
\put(375.0,893.0){\rule[-0.200pt]{4.818pt}{0.400pt}}
\put(350,893){\makebox(0,0)[r]{\ \ {$0.75$}}}
\put(1405.0,893.0){\rule[-0.200pt]{4.818pt}{0.400pt}}
\put(375.0,1107.0){\rule[-0.200pt]{4.818pt}{0.400pt}}
\put(350,1107){\makebox(0,0)[r]{\ \ {$1$}}}
\put(1405.0,1107.0){\rule[-0.200pt]{4.818pt}{0.400pt}}
\put(375.0,1321.0){\rule[-0.200pt]{4.818pt}{0.400pt}}
\put(350,1321){\makebox(0,0)[r]{\ \ {$1.25$}}}
\put(1405.0,1321.0){\rule[-0.200pt]{4.818pt}{0.400pt}}
\put(375.0,1536.0){\rule[-0.200pt]{4.818pt}{0.400pt}}
\put(350,1536){\makebox(0,0)[r]{\ \ {$1.5$}}}
\put(1405.0,1536.0){\rule[-0.200pt]{4.818pt}{0.400pt}}
\put(375.0,1750.0){\rule[-0.200pt]{4.818pt}{0.400pt}}
\put(350,1750){\makebox(0,0)[r]{\ \ {$1.75$}}}
\put(1405.0,1750.0){\rule[-0.200pt]{4.818pt}{0.400pt}}
\put(375.0,250.0){\rule[-0.200pt]{0.400pt}{4.818pt}}
\put(375,200){\makebox(0,0){\ {$0$}}}
\put(375.0,1730.0){\rule[-0.200pt]{0.400pt}{4.818pt}}
\put(525.0,250.0){\rule[-0.200pt]{0.400pt}{4.818pt}}
\put(525,200){\makebox(0,0){\ {$1$}}}
\put(525.0,1730.0){\rule[-0.200pt]{0.400pt}{4.818pt}}
\put(675.0,250.0){\rule[-0.200pt]{0.400pt}{4.818pt}}
\put(675,200){\makebox(0,0){\ {$2$}}}
\put(675.0,1730.0){\rule[-0.200pt]{0.400pt}{4.818pt}}
\put(825.0,250.0){\rule[-0.200pt]{0.400pt}{4.818pt}}
\put(825,200){\makebox(0,0){\ {$3$}}}
\put(825.0,1730.0){\rule[-0.200pt]{0.400pt}{4.818pt}}
\put(975.0,250.0){\rule[-0.200pt]{0.400pt}{4.818pt}}
\put(975,200){\makebox(0,0){\ {$4$}}}
\put(975.0,1730.0){\rule[-0.200pt]{0.400pt}{4.818pt}}
\put(1125.0,250.0){\rule[-0.200pt]{0.400pt}{4.818pt}}
\put(1125,200){\makebox(0,0){\ {$5$}}}
\put(1125.0,1730.0){\rule[-0.200pt]{0.400pt}{4.818pt}}
\put(1275.0,250.0){\rule[-0.200pt]{0.400pt}{4.818pt}}
\put(1275,200){\makebox(0,0){\ {$6$}}}
\put(1275.0,1730.0){\rule[-0.200pt]{0.400pt}{4.818pt}}
\put(375.0,250.0){\rule[-0.200pt]{252.945pt}{0.400pt}}
\put(1425.0,250.0){\rule[-0.200pt]{0.400pt}{361.350pt}}
\put(375.0,1750.0){\rule[-0.200pt]{252.945pt}{0.400pt}}
\put(150,1200){\makebox(0,0){\Large{$am_{k=2}(n_t)$}}}
\put(875,25){\makebox(0,0){\large{$n_t$}}}
\put(375.0,250.0){\rule[-0.200pt]{0.400pt}{361.350pt}}
\put(525.0,1364.0){\rule[-0.200pt]{0.400pt}{2.891pt}}
\put(515.0,1364.0){\rule[-0.200pt]{4.818pt}{0.400pt}}
\put(515.0,1376.0){\rule[-0.200pt]{4.818pt}{0.400pt}}
\put(675.0,1004.0){\rule[-0.200pt]{0.400pt}{5.059pt}}
\put(665.0,1004.0){\rule[-0.200pt]{4.818pt}{0.400pt}}
\put(665.0,1025.0){\rule[-0.200pt]{4.818pt}{0.400pt}}
\put(825.0,964.0){\rule[-0.200pt]{0.400pt}{9.877pt}}
\put(815.0,964.0){\rule[-0.200pt]{4.818pt}{0.400pt}}
\put(815.0,1005.0){\rule[-0.200pt]{4.818pt}{0.400pt}}
\put(975.0,889.0){\rule[-0.200pt]{0.400pt}{18.549pt}}
\put(965.0,889.0){\rule[-0.200pt]{4.818pt}{0.400pt}}
\put(965.0,966.0){\rule[-0.200pt]{4.818pt}{0.400pt}}
\put(1125.0,815.0){\rule[-0.200pt]{0.400pt}{59.020pt}}
\put(1115.0,815.0){\rule[-0.200pt]{4.818pt}{0.400pt}}
\put(525,1370){\circle*{12}}
\put(675,1015){\circle*{12}}
\put(825,985){\circle*{12}}
\put(975,927){\circle*{12}}
\put(1125,937){\circle*{12}}
\put(1115.0,1060.0){\rule[-0.200pt]{4.818pt}{0.400pt}}
\put(525.0,984.0){\rule[-0.200pt]{0.400pt}{1.927pt}}
\put(515.0,984.0){\rule[-0.200pt]{4.818pt}{0.400pt}}
\put(515.0,992.0){\rule[-0.200pt]{4.818pt}{0.400pt}}
\put(675.0,944.0){\rule[-0.200pt]{0.400pt}{2.891pt}}
\put(665.0,944.0){\rule[-0.200pt]{4.818pt}{0.400pt}}
\put(665.0,956.0){\rule[-0.200pt]{4.818pt}{0.400pt}}
\put(825.0,932.0){\rule[-0.200pt]{0.400pt}{5.541pt}}
\put(815.0,932.0){\rule[-0.200pt]{4.818pt}{0.400pt}}
\put(815.0,955.0){\rule[-0.200pt]{4.818pt}{0.400pt}}
\put(975.0,871.0){\rule[-0.200pt]{0.400pt}{11.081pt}}
\put(965.0,871.0){\rule[-0.200pt]{4.818pt}{0.400pt}}
\put(965.0,917.0){\rule[-0.200pt]{4.818pt}{0.400pt}}
\put(1125.0,853.0){\rule[-0.200pt]{0.400pt}{25.776pt}}
\put(1115.0,853.0){\rule[-0.200pt]{4.818pt}{0.400pt}}
\put(525,988){\circle{18}}
\put(675,950){\circle{18}}
\put(825,943){\circle{18}}
\put(975,894){\circle{18}}
\put(1125,907){\circle{18}}
\put(1115.0,960.0){\rule[-0.200pt]{4.818pt}{0.400pt}}
\end{picture}
\end	{center}
\vskip 0.15in
\caption{Effective mass of the SU(4) $k=2$ string as a function 
of $n_t$ for a $16^4$ lattice with old blocking, $\bullet$,
and improved blocking, $\circ$. (The value of $a$ differs
by about 3\% in the two calculations.)}
\label{fig_oldnewn4}
\end 	{figure}

We illustrate the potential of operator improvement in
Fig.\ref{fig_oldnewn4}. Here we plot the effective mass versus $n_t$
of the $k=2$ string that winds around the spatial torus, from a
calculation
\cite{blmt-kstring}
on a $16^4$ lattice at $\beta=11.10$ in SU(4) using `old' blocking for
the (super)link construction.  Loops are composed of (super)links that
have been blocked up to four times, i.e. five blocking levels when we
include no blocking at all. The linear combination of these five
operators which minimises $am_{eff}(t=a)$ is chosen as the best
operator, and from its correlator we obtain $am_{eff}(t=n_t a)$ in
Fig.\ref{fig_oldnewn4}. Note that this corresponds to a variational
calculation of the ground state, where one maximises $\exp -aH$ in
this basis of five operators. We do the same with our improved
blocking, with $p_a=0.40$ and $p_d=0.16$, again on a $16^4$ lattice
for SU(4) and at the nearly identical coupling $\beta=11.085$.  We
contrast the two calculations in Fig.\ref{fig_oldnewn4} where we see a
dramatic improvement in the use of the new blocking technique: with
the same statistics the errors on the final mass estimate are reduced
by about a factor of two.

Clearly we need to determine the range of smearing/blocking parameters
that provide the best improvement and also how many smearing or
blocking steps need to be taken.  To do this we perform calculations
with a variety of different values of the parameters. Ideally we
should calculate the overlaps onto the lightest states in the various
channels of interest, so as to find the parameters that maximise these
overlaps, i.e. the $|c_n|^2$ in eqn(\ref{eqn_corrM}).  However usually
we do not have enough accuracy to do this, and instead we calculate
the effective mass from the values of the correlation function at
$t=0$ and $t=a$:
\be
am_{eff}(t=a) = -\ln {{C(t=a)}\over{C(t=0)}}.
\label{eqn_meff}
\ee
Because of the positivity of our correlation functions, the lower the
value of $am_{eff}(a)$ the better, in a variational sense, is the
operator.

We have performed a wide variety of tests and comparisons.  For the
sake of brevity we will choose to discuss only a subset of these here.

\subsubsection{blocking with smearing}
\label{subsubsection_blocksmear}

In this calculation we perform multiple smearing upon blocked
(super)links with $(p_a,p_d)=(0.10,0.0)$.  We block up to three times,
with the `old' blocking of eqn(\ref{eqn_oldblock}) with $p_a=1.0$. The
calculation is in SU(3) on a $12^4$ lattice at $\beta=5.90$.  The
purpose of this calculation is to see how many smearing and blocking
steps one needs to perform for a useful improvement.

In Table~\ref{table_blsmear} we show the values of $am_{eff}(t=a)$ for
the $0^{++}$ glueball operator based on $1\times 1$ loops
(`superplaquettes') made out of superlinks that have been blocked
$b=0,..,3$ times and then smeared up to 14 times. We see that if one
blocks too few times then one needs a large number of smearing steps
to obtain a useful improvement while if one blocks too many times the
operators are worse and cannot be improved by subsequent smearing.  In
this particular case the most efficient strategy appears to be to
block twice and then to smear about four times. If one smears too many
times, again the overlap gets worse.

For comparison we also show what one gets with the improved blocking
with parameters $(p_a,p_d)=(0.40,0.16)$.  Although it appears that we
could obtain a further small improvement in the overlap with further
multiple smearings, we shall not explore this possibility here and
shall from now on focus on simple improved blocking.

\subsubsection{improved blocking}
\label{subsubsection_impblock}

In this calculation we perform improved blocking with various values
of $(p_a,p_d)$ on a $12^4$ lattice at $\beta=10.90$ in SU(4).  We
perform $b=0,1,2,3$ blockings and use the various blocking levels to
construct our operators. So for each state we have a basis of
operators and within this basis we find the linear combination that
minimises $am_{eff}(t=a)$.  So $am_{eff}(t=a)$ is our best variational
estimate of the mass of the ground state in the channel of interest.

The channels we focus upon are the fundamental (i.e. k=1) string with
$\vec{p} = 0$, the same string with the lowest non-zero momentum, i.e.
$a\vec{p} = 2\pi/L_s$, the $k=2$ string and the lightest scalar and
tensor glueballs.

In Table~\ref{table_improvedbl} we list the values of $am_{eff}(t=a)$
that we obtain for the values of $p_a,p_d$ shown. We see that although
the best values of the parameters differ for different states, in
practice values like $(p_a,p_d)=(0.30,0.12)$ or
$(p_a,p_d)=(0.40,0.16)$ are a good compromise giving close to the best
overlaps for all the states.  These are therefore the values we shall
use in the rest of this paper.

The calculations we shall describe later on in this paper are with
high enough statistics that one can extract the overlap onto the
lightest state quite accurately. One can do the same for the
calculations in
\cite{blmt-kstring,blmt-glue}
and so compare the overlaps one obtains with `old' and improved
blocking. This we do in Table~\ref{table_ImpBl} for the lightest $k=1$
and $k=2$ strings and the lightest scalar glueball, for our various
SU(4) calculations. We observe a very large improvement in all cases
with our new techniques, particularly for the $k=2$ string.

\section{Some results}
\label{section_results}

To calculate the lightest mass in a given sector of string or glueball
states, we perform a standard variational calculation with different
blocking and/or smearing levels, and sometimes different loops,
providing the basis of operators.

In Tables~\ref{table_su2GK} to~\ref{table_su8GK} we list the masses of
the lightest and first excited $0^{++}$ glueballs, the mass of the
lightest $2^{++}$ glueball, and the tension of the fundamental ($k=1$)
string.  Tables~\ref{table_su46l} and ~\ref{table_su8l} contain our
values of the tensions of $k\geq 2$ strings.  All these calculations
use improved blocking, as in eqn(\ref{eqn_newsmear}) and
eqn(\ref{eqn_newblock}), with $(p_a,p_d)=(0.30,0.12)$ or
$(0.40,0.16)$.

The results of our SU(4) anisotropic lattice calculations are listed
in Tables~\ref{table_su4aM} and~\ref{table_su4aMK}.  Here we use one
or two blockings plus multiple smearing but, because these were
earlier calculations, the smearing was of the old variety, as in
eqn(\ref{eqn_oldsmear}). The bare anisotropy we chose was $\xi=0.5$
but, as we see from Table~\ref{table_su4aMK}, the renormalised
anisotropy (determined as described in Section~\ref{section_lattice})
is smaller by $\sim 8$\% and, as expected, $\xi_r \to \xi$ as $a\to
0$. Comparing the values of $a_s\sigma = a_t\sigma/\xi_r$ from
Table~\ref{table_su4aMK} to the values in Table~\ref{table_su4GK}, we
see that the tree level value of $\beta$, as in eqn(\ref{B6}), is also
significantly renormalised. For example our anistropic calculation at
$\beta=11.325$ has $a_s\sqrt\sigma$ nearly identical to that at
$\beta=11.085$ in the isotropic calculation.

\begin	{figure}[p]
\begin	{center}
\leavevmode
\input	{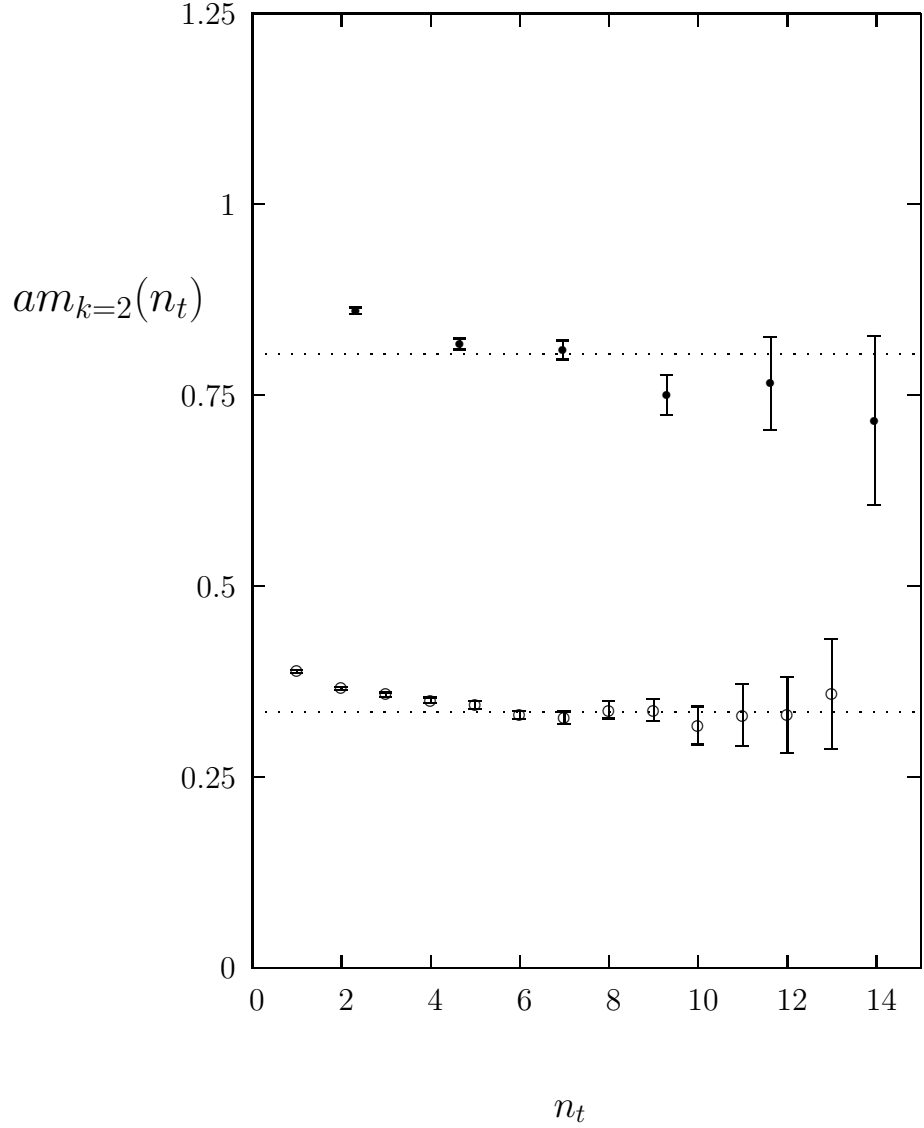}
\end	{center}
\vskip 0.15in
\caption{Effective mass of the SU(4) $k=2$ string as a function 
of $n_t$ for an anisotropic lattice, $\circ$, and an isotropic 
lattice, $\bullet$, both with the same value of $a_s$, 
$a_s\surd\sigma \simeq 0.198$, and $L_s=16$.
The value of $n_t$ is for the anisotropic lattice, and
the isotropic values of $n_t$ have been scaled up by the
calculated value of $a_s/a_t$.}
\label{fig_anison4}
\end 	{figure}

We can use the fact that the isotropic $\beta=11.085$ and anisotropic
$\beta=11.325$ calculations have essentially the same $a_s$ and
spatial size, $a_sL_s=16a_s$, and very similar statistics, to compare
directly the corresponding correlation functions. This we do in
Fig.\ref{fig_anison4} for the lightest $k=2$ periodic flux loop.  We
plot the effective mass, as defined in eqn(\ref{eqn_D3}), against
$n_t$. In the case of the isotropic calculation we scale $n_t$ up by a
factor of $1/\xi_r$ (although we leave the mass expressed in units of
the original lattice spacing, so that its value is about a factor of
$1/\xi_r$ larger). We also show the final mass estimates which we
obtain using exponential fits to the correlation functions.  We see
that although the $a_t$ we use only differs by a little more than a
factor of two between the two calculations, the extra resolution in
$t$ provides us with much more convincing evidence for an effective
mass plateau. This demonstrates the utility of such anisotropic
calculations.

The SU(8) calculations possess about half the statistics of the other
calculations, and the errors are probably underestimated. In
particular the mass estimates of heavier states, such as strings of
larger $k$, will suffer from the systematic errors spelt out earlier.

\subsection{glueballs}
\label{subsection_glueballs}

\begin	{figure}[p]
\begin	{center}
\leavevmode
\input	{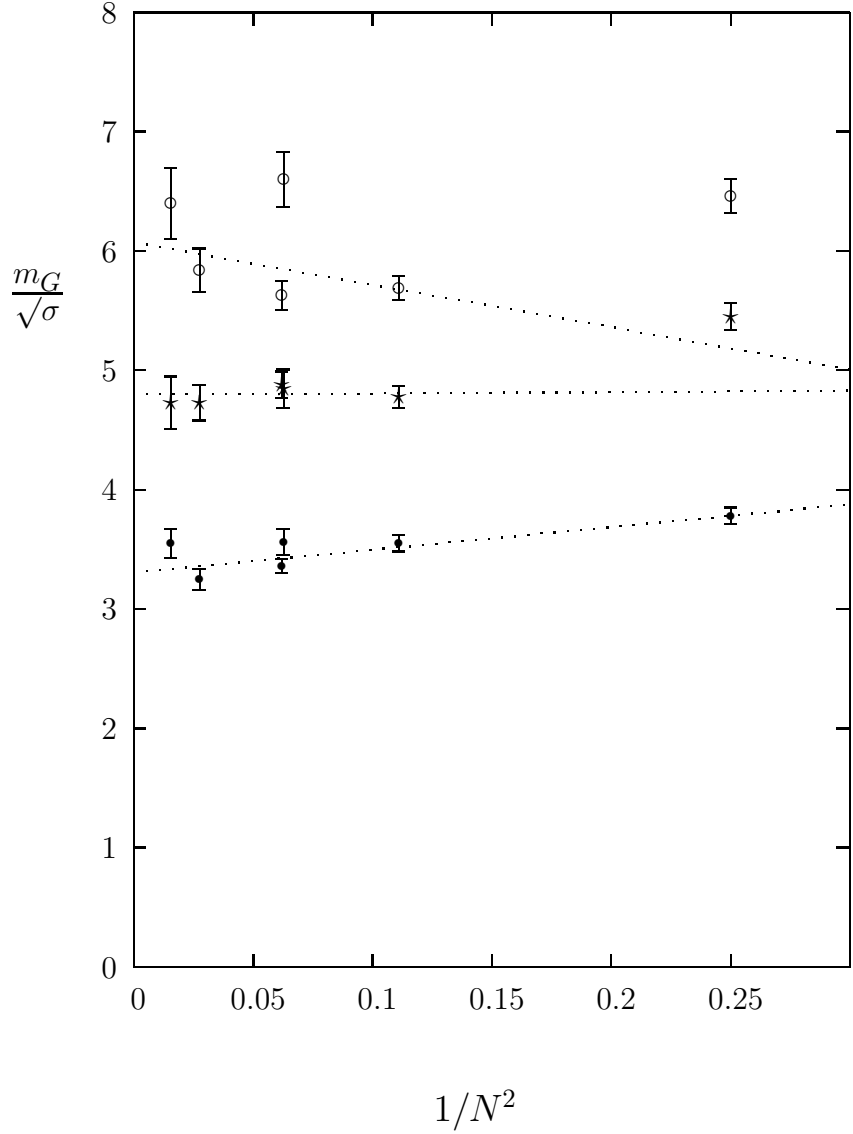}
\end	{center}
\vskip 0.15in
\caption{The lightest glueball masses expressed in units of
the fundamental string tension, in the continuum limit,
plotted against $1/N^2$. The $0^{++}$, $\bullet$, the
$2^{++}$, $\star$, and the first excited  $0^{++\star}$, $\circ$.
Dotted lines are extrapolations to $N=\infty$.}
\label{fig_gkNwa}
\end 	{figure}

We extrapolate to the continuum limit with an $O(a^2)$ correction:
\be
\frac{m_G}{\sqrt{\sigma}}(a) 
= 
\frac{m_G}{\sqrt{\sigma}}(0)
+
c a^2\sigma.
\label{eqn_contmk}
\ee
The results are listed in Table~\ref{table_continuumG}. These
continuum mass ratios can now be extrapolated to $N=\infty$ with an
$O(1/N^2)$ correction:
\be
\frac{m_G}{\sqrt{\sigma}}(N) 
= 
\frac{m_G}{\sqrt{\sigma}}(\infty)
+
\frac{c}{N^2}.
\label{eqn_Nmk}
\ee
This is shown in Fig.\ref{fig_gkNwa} and the $N=\infty$ limits are
listed in Table~\ref{table_continuumG}.

We make several observations. First, our $N=\infty$ mass ratios are
compatible with previous results
\cite{blmt-glue}
but the errors have been reduced by about a factor of two.  The
individual SU($N$) continuum limits of the two heavier states are
typically lower by about one standard deviation however, and we put
this down to the reduction in systematic errors arising from our
improved operators. The $0^{++\star}$ has a scatter which may arise
from the finite volume effects discussed earlier. Nonetheless, we see
that its mass is roughly twice that of the lightest scalar. Finally we
observe in Fig.\ref{fig_gkNwa} that with these more accurate
calculations, it has become impossible to include the SU(2) value of
the $2^{++}$, and perhaps the $ 0^{++\star}$, in a fit including just
the leading $O(1/N^2)$ correction.

\begin	{figure}[p]
\begin	{center}
\leavevmode
\input	{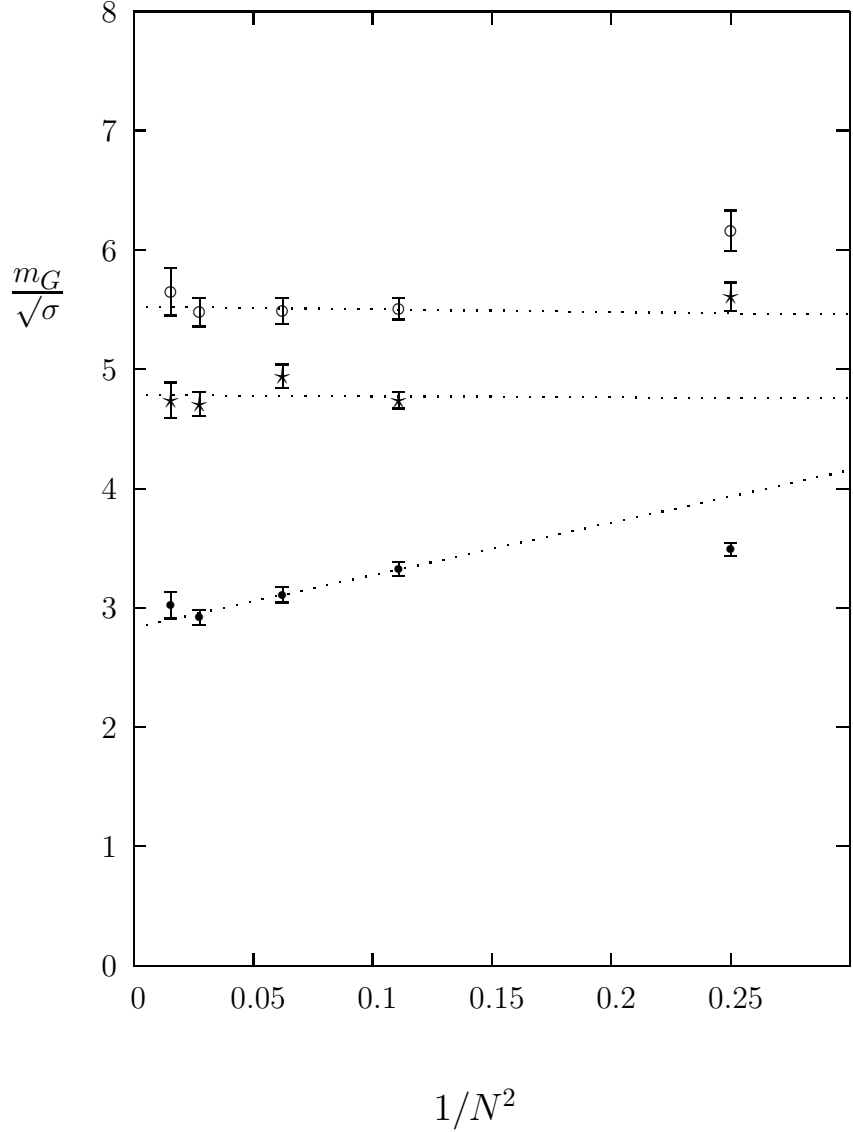}
\end	{center}
\vskip 0.15in
\caption{The lightest glueball masses expressed in units of
the fundamental string tension plotted against $1/N^2$.
The $0^{++}$, $\bullet$, the
$2^{++}$, $\star$, and the first excited  $0^{++}$, $\circ$.
Dotted lines are extrapolations to $N=\infty$.
On $16^4$ lattices with a nearly common value of the
lattice spacing, $a\sqrt{\sigma}\in[0.195,0.210]$, except for SU(2)
where $a\sqrt{\sigma}\simeq 0.24$.}
\label{fig_gkNl16}
\end 	{figure}

Our continuum extrapolations only involve a few lattice spacings and
one might worry that this may give the coarser lattice spacings too
much influence. It is therefore interesting to compare the mass ratios
calculated on the smallest value of $a$ that is `common' to all
$N$. In practice we mean by this that $a\sqrt{\sigma}\in
[0.195,0.210]$ for all $N$ except for SU(2) where
$a\sqrt{\sigma}\simeq 0.24$. This $a$ is well in the scaling regime
where lattice corrections are small.  All calculations are on $16^4$
lattices. We list the mass ratios in Table~\ref{table_l16G} and plot
them against $1/N^2$ in Fig.\ref{fig_gkNl16}, where we show large-$N$
extrapolations of the form in eqn(\ref{eqn_Nmk}). The errors are
smaller than for the continuum extrapolations, and it is now clear
that none of the SU(2) masses fall onto such a simple fit. It is
interesting that the $0^{++\star}$ is much better behaved than in
Fig.\ref{fig_gkNwa} and that at this (small) value of $a$ it is very
close to twice the mass gap.

\subsection{k-strings}
\label{subsection_kstrings}

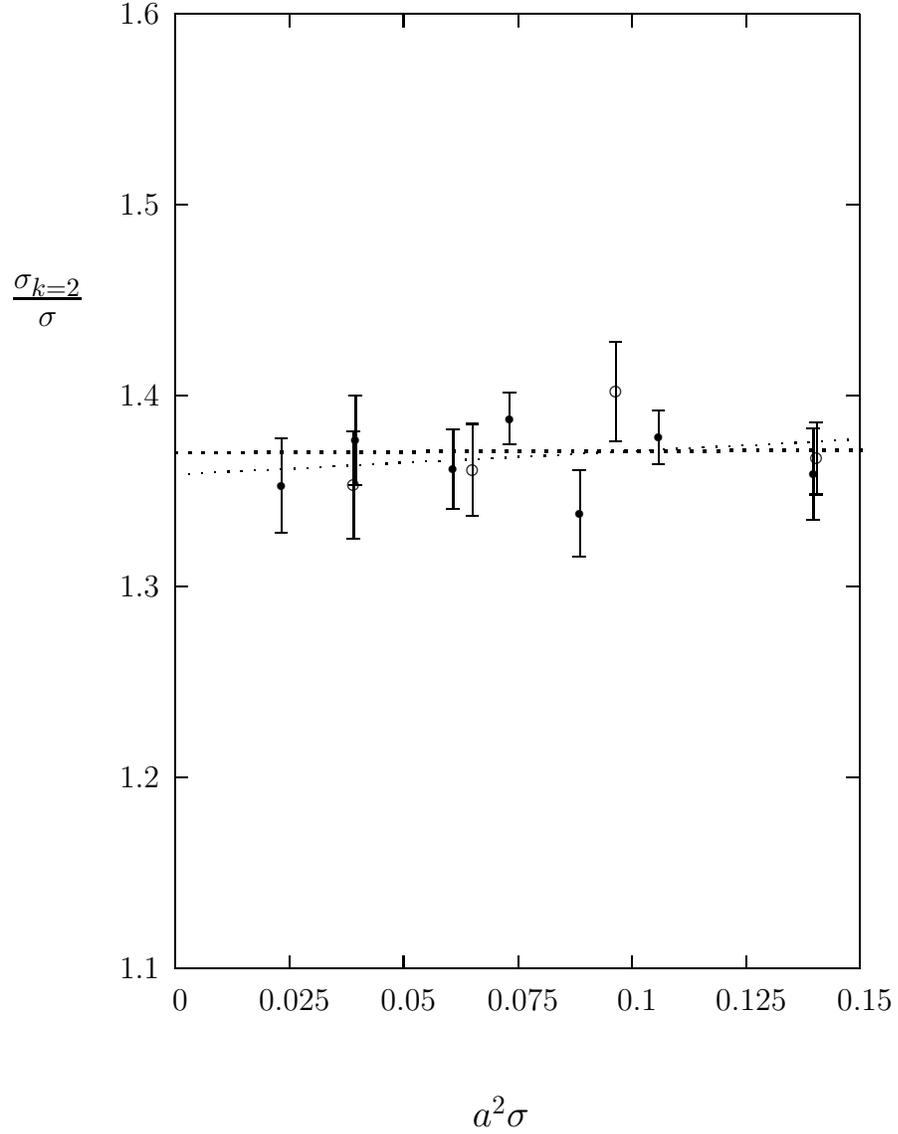
\begin	{figure}[p]
\begin	{center}
\leavevmode
\setlength{\unitlength}{0.240900pt}
\ifx\plotpoint\undefined\newsavebox{\plotpoint}\fi
\sbox{\plotpoint}{\rule[-0.200pt]{0.400pt}{0.400pt}}%
\begin{picture}(1500,1800)(0,0)
\font\gnuplot=cmr10 at 12pt
\gnuplot
\sbox{\plotpoint}{\rule[-0.200pt]{0.400pt}{0.400pt}}%
\put(350.0,250.0){\rule[-0.200pt]{4.818pt}{0.400pt}}
\put(325,250){\makebox(0,0)[r]{\ \ {$1.1$}}}
\put(1405.0,250.0){\rule[-0.200pt]{4.818pt}{0.400pt}}
\put(350.0,550.0){\rule[-0.200pt]{4.818pt}{0.400pt}}
\put(325,550){\makebox(0,0)[r]{\ \ {$1.2$}}}
\put(1405.0,550.0){\rule[-0.200pt]{4.818pt}{0.400pt}}
\put(350.0,850.0){\rule[-0.200pt]{4.818pt}{0.400pt}}
\put(325,850){\makebox(0,0)[r]{\ \ {$1.3$}}}
\put(1405.0,850.0){\rule[-0.200pt]{4.818pt}{0.400pt}}
\put(350.0,1150.0){\rule[-0.200pt]{4.818pt}{0.400pt}}
\put(325,1150){\makebox(0,0)[r]{\ \ {$1.4$}}}
\put(1405.0,1150.0){\rule[-0.200pt]{4.818pt}{0.400pt}}
\put(350.0,1450.0){\rule[-0.200pt]{4.818pt}{0.400pt}}
\put(325,1450){\makebox(0,0)[r]{\ \ {$1.5$}}}
\put(1405.0,1450.0){\rule[-0.200pt]{4.818pt}{0.400pt}}
\put(350.0,1750.0){\rule[-0.200pt]{4.818pt}{0.400pt}}
\put(325,1750){\makebox(0,0)[r]{\ \ {$1.6$}}}
\put(1405.0,1750.0){\rule[-0.200pt]{4.818pt}{0.400pt}}
\put(350.0,250.0){\rule[-0.200pt]{0.400pt}{4.818pt}}
\put(350,200){\makebox(0,0){\ {$0$}}}
\put(350.0,1730.0){\rule[-0.200pt]{0.400pt}{4.818pt}}
\put(529.0,250.0){\rule[-0.200pt]{0.400pt}{4.818pt}}
\put(529,200){\makebox(0,0){\ {$0.025$}}}
\put(529.0,1730.0){\rule[-0.200pt]{0.400pt}{4.818pt}}
\put(708.0,250.0){\rule[-0.200pt]{0.400pt}{4.818pt}}
\put(708,200){\makebox(0,0){\ {$0.05$}}}
\put(708.0,1730.0){\rule[-0.200pt]{0.400pt}{4.818pt}}
\put(888.0,250.0){\rule[-0.200pt]{0.400pt}{4.818pt}}
\put(888,200){\makebox(0,0){\ {$0.075$}}}
\put(888.0,1730.0){\rule[-0.200pt]{0.400pt}{4.818pt}}
\put(1067.0,250.0){\rule[-0.200pt]{0.400pt}{4.818pt}}
\put(1067,200){\makebox(0,0){\ {$0.1$}}}
\put(1067.0,1730.0){\rule[-0.200pt]{0.400pt}{4.818pt}}
\put(1246.0,250.0){\rule[-0.200pt]{0.400pt}{4.818pt}}
\put(1246,200){\makebox(0,0){\ {$0.125$}}}
\put(1246.0,1730.0){\rule[-0.200pt]{0.400pt}{4.818pt}}
\put(1425.0,250.0){\rule[-0.200pt]{0.400pt}{4.818pt}}
\put(1425,200){\makebox(0,0){\ {$0.15$}}}
\put(1425.0,1730.0){\rule[-0.200pt]{0.400pt}{4.818pt}}
\put(350.0,250.0){\rule[-0.200pt]{258.967pt}{0.400pt}}
\put(1425.0,250.0){\rule[-0.200pt]{0.400pt}{361.350pt}}
\put(350.0,1750.0){\rule[-0.200pt]{258.967pt}{0.400pt}}
\put(150,1300){\makebox(0,0){\Large{${{\sigma_{k=2}}\over{\sigma}}$}}}
\put(862,25){\makebox(0,0){\large{$a^2\sigma$}}}
\put(350.0,250.0){\rule[-0.200pt]{0.400pt}{361.350pt}}
\put(1352.0,954.0){\rule[-0.200pt]{0.400pt}{34.690pt}}
\put(1342.0,954.0){\rule[-0.200pt]{4.818pt}{0.400pt}}
\put(1342.0,1098.0){\rule[-0.200pt]{4.818pt}{0.400pt}}
\put(1109.0,1042.0){\rule[-0.200pt]{0.400pt}{20.236pt}}
\put(1099.0,1042.0){\rule[-0.200pt]{4.818pt}{0.400pt}}
\put(1099.0,1126.0){\rule[-0.200pt]{4.818pt}{0.400pt}}
\put(985.0,897.0){\rule[-0.200pt]{0.400pt}{32.521pt}}
\put(975.0,897.0){\rule[-0.200pt]{4.818pt}{0.400pt}}
\put(975.0,1032.0){\rule[-0.200pt]{4.818pt}{0.400pt}}
\put(875.0,1073.0){\rule[-0.200pt]{0.400pt}{19.513pt}}
\put(865.0,1073.0){\rule[-0.200pt]{4.818pt}{0.400pt}}
\put(865.0,1154.0){\rule[-0.200pt]{4.818pt}{0.400pt}}
\put(786.0,971.0){\rule[-0.200pt]{0.400pt}{30.353pt}}
\put(776.0,971.0){\rule[-0.200pt]{4.818pt}{0.400pt}}
\put(776.0,1097.0){\rule[-0.200pt]{4.818pt}{0.400pt}}
\put(633.0,1009.0){\rule[-0.200pt]{0.400pt}{33.967pt}}
\put(623.0,1009.0){\rule[-0.200pt]{4.818pt}{0.400pt}}
\put(623.0,1150.0){\rule[-0.200pt]{4.818pt}{0.400pt}}
\put(517.0,934.0){\rule[-0.200pt]{0.400pt}{35.653pt}}
\put(507.0,934.0){\rule[-0.200pt]{4.818pt}{0.400pt}}
\put(1352,1026){\circle*{12}}
\put(1109,1084){\circle*{12}}
\put(985,964){\circle*{12}}
\put(875,1113){\circle*{12}}
\put(786,1034){\circle*{12}}
\put(633,1080){\circle*{12}}
\put(517,1008){\circle*{12}}
\put(507.0,1082.0){\rule[-0.200pt]{4.818pt}{0.400pt}}
\put(1357.0,994.0){\rule[-0.200pt]{0.400pt}{27.463pt}}
\put(1347.0,994.0){\rule[-0.200pt]{4.818pt}{0.400pt}}
\put(1347.0,1108.0){\rule[-0.200pt]{4.818pt}{0.400pt}}
\put(1042.0,1078.0){\rule[-0.200pt]{0.400pt}{37.580pt}}
\put(1032.0,1078.0){\rule[-0.200pt]{4.818pt}{0.400pt}}
\put(1032.0,1234.0){\rule[-0.200pt]{4.818pt}{0.400pt}}
\put(817.0,961.0){\rule[-0.200pt]{0.400pt}{34.690pt}}
\put(807.0,961.0){\rule[-0.200pt]{4.818pt}{0.400pt}}
\put(807.0,1105.0){\rule[-0.200pt]{4.818pt}{0.400pt}}
\put(630.0,925.0){\rule[-0.200pt]{0.400pt}{40.471pt}}
\put(620.0,925.0){\rule[-0.200pt]{4.818pt}{0.400pt}}
\put(1357,1051){\circle{18}}
\put(1042,1156){\circle{18}}
\put(817,1033){\circle{18}}
\put(630,1009){\circle{18}}
\put(620.0,1093.0){\rule[-0.200pt]{4.818pt}{0.400pt}}
\sbox{\plotpoint}{\rule[-0.500pt]{1.000pt}{1.000pt}}%
\put(350,1060){\usebox{\plotpoint}}
\put(350.00,1060.00){\usebox{\plotpoint}}
\put(370.76,1060.00){\usebox{\plotpoint}}
\put(391.51,1060.00){\usebox{\plotpoint}}
\put(412.27,1060.00){\usebox{\plotpoint}}
\put(433.02,1060.00){\usebox{\plotpoint}}
\put(453.78,1060.00){\usebox{\plotpoint}}
\put(474.51,1060.50){\usebox{\plotpoint}}
\put(495.24,1061.00){\usebox{\plotpoint}}
\put(516.00,1061.00){\usebox{\plotpoint}}
\put(536.75,1061.00){\usebox{\plotpoint}}
\put(557.51,1061.00){\usebox{\plotpoint}}
\put(578.27,1061.00){\usebox{\plotpoint}}
\put(599.02,1061.00){\usebox{\plotpoint}}
\put(619.78,1061.00){\usebox{\plotpoint}}
\put(640.53,1061.00){\usebox{\plotpoint}}
\put(661.29,1061.00){\usebox{\plotpoint}}
\put(682.04,1061.00){\usebox{\plotpoint}}
\put(702.80,1061.00){\usebox{\plotpoint}}
\put(723.53,1061.41){\usebox{\plotpoint}}
\put(744.26,1062.00){\usebox{\plotpoint}}
\put(765.02,1062.00){\usebox{\plotpoint}}
\put(785.77,1062.00){\usebox{\plotpoint}}
\put(806.53,1062.00){\usebox{\plotpoint}}
\put(827.29,1062.00){\usebox{\plotpoint}}
\put(848.04,1062.00){\usebox{\plotpoint}}
\put(868.80,1062.00){\usebox{\plotpoint}}
\put(889.55,1062.00){\usebox{\plotpoint}}
\put(910.31,1062.00){\usebox{\plotpoint}}
\put(931.06,1062.00){\usebox{\plotpoint}}
\put(951.82,1062.00){\usebox{\plotpoint}}
\put(972.56,1062.32){\usebox{\plotpoint}}
\put(993.28,1063.00){\usebox{\plotpoint}}
\put(1014.04,1063.00){\usebox{\plotpoint}}
\put(1034.80,1063.00){\usebox{\plotpoint}}
\put(1055.55,1063.00){\usebox{\plotpoint}}
\put(1076.31,1063.00){\usebox{\plotpoint}}
\put(1097.06,1063.00){\usebox{\plotpoint}}
\put(1117.82,1063.00){\usebox{\plotpoint}}
\put(1138.57,1063.00){\usebox{\plotpoint}}
\put(1159.33,1063.00){\usebox{\plotpoint}}
\put(1180.08,1063.00){\usebox{\plotpoint}}
\put(1200.84,1063.00){\usebox{\plotpoint}}
\put(1221.58,1063.23){\usebox{\plotpoint}}
\put(1242.30,1064.00){\usebox{\plotpoint}}
\put(1263.06,1064.00){\usebox{\plotpoint}}
\put(1283.82,1064.00){\usebox{\plotpoint}}
\put(1304.57,1064.00){\usebox{\plotpoint}}
\put(1325.33,1064.00){\usebox{\plotpoint}}
\put(1346.08,1064.00){\usebox{\plotpoint}}
\put(1366.84,1064.00){\usebox{\plotpoint}}
\put(1387.59,1064.00){\usebox{\plotpoint}}
\put(1408.35,1064.00){\usebox{\plotpoint}}
\put(1425,1064){\usebox{\plotpoint}}
\sbox{\plotpoint}{\rule[-0.200pt]{0.400pt}{0.400pt}}%
\put(350,1026){\usebox{\plotpoint}}
\put(350.00,1026.00){\usebox{\plotpoint}}
\put(370.72,1026.88){\usebox{\plotpoint}}
\put(391.42,1027.84){\usebox{\plotpoint}}
\put(412.13,1029.00){\usebox{\plotpoint}}
\put(432.84,1030.00){\usebox{\plotpoint}}
\put(453.55,1031.00){\usebox{\plotpoint}}
\put(474.25,1032.00){\usebox{\plotpoint}}
\put(494.95,1033.36){\usebox{\plotpoint}}
\put(515.66,1034.24){\usebox{\plotpoint}}
\put(536.38,1035.14){\usebox{\plotpoint}}
\put(557.08,1036.10){\usebox{\plotpoint}}
\put(577.80,1037.00){\usebox{\plotpoint}}
\put(598.47,1038.86){\usebox{\plotpoint}}
\put(619.18,1039.82){\usebox{\plotpoint}}
\put(639.89,1040.72){\usebox{\plotpoint}}
\put(660.61,1041.60){\usebox{\plotpoint}}
\put(681.32,1042.48){\usebox{\plotpoint}}
\put(702.00,1044.00){\usebox{\plotpoint}}
\put(722.71,1045.00){\usebox{\plotpoint}}
\put(743.42,1046.00){\usebox{\plotpoint}}
\put(764.13,1047.00){\usebox{\plotpoint}}
\put(784.84,1048.08){\usebox{\plotpoint}}
\put(805.56,1049.00){\usebox{\plotpoint}}
\put(826.27,1050.00){\usebox{\plotpoint}}
\put(846.98,1051.00){\usebox{\plotpoint}}
\put(867.69,1052.00){\usebox{\plotpoint}}
\put(888.37,1053.58){\usebox{\plotpoint}}
\put(909.08,1054.46){\usebox{\plotpoint}}
\put(929.80,1055.38){\usebox{\plotpoint}}
\put(950.51,1056.32){\usebox{\plotpoint}}
\put(971.22,1057.20){\usebox{\plotpoint}}
\put(991.90,1059.00){\usebox{\plotpoint}}
\put(1012.60,1060.00){\usebox{\plotpoint}}
\put(1033.32,1060.94){\usebox{\plotpoint}}
\put(1054.03,1061.82){\usebox{\plotpoint}}
\put(1074.75,1062.70){\usebox{\plotpoint}}
\put(1095.44,1064.00){\usebox{\plotpoint}}
\put(1116.15,1065.00){\usebox{\plotpoint}}
\put(1136.86,1066.00){\usebox{\plotpoint}}
\put(1157.57,1067.00){\usebox{\plotpoint}}
\put(1178.27,1068.30){\usebox{\plotpoint}}
\put(1198.98,1069.18){\usebox{\plotpoint}}
\put(1219.70,1070.06){\usebox{\plotpoint}}
\put(1240.41,1071.04){\usebox{\plotpoint}}
\put(1261.12,1072.00){\usebox{\plotpoint}}
\put(1281.79,1073.80){\usebox{\plotpoint}}
\put(1302.51,1074.68){\usebox{\plotpoint}}
\put(1323.22,1075.66){\usebox{\plotpoint}}
\put(1343.94,1076.54){\usebox{\plotpoint}}
\put(1364.65,1077.42){\usebox{\plotpoint}}
\put(1385.33,1079.00){\usebox{\plotpoint}}
\put(1406.04,1080.00){\usebox{\plotpoint}}
\put(1425,1081){\usebox{\plotpoint}}
\end{picture}
\end	{center}
\vskip 0.15in
\caption{Ratio of the tension of the $k=2$ string 
to that of the fundamental ($k=1$) string in SU(4), for the
anisotropic ($\circ$) and isotropic ($\bullet$) calculations. 
Plotted against the square of the (spatial) lattice spacing.
Continuum extrapolations are shown.}
\label{fig_kkn4}
\end 	{figure}

The string tensions for $N=4,6,8$ are listed in
Tables~\ref{table_su46l} and \ref{table_su8l}.  The ratio of the $k=2$
string tension to $\sigma$ is plotted against $a_s^2\sigma$ in
Fig.\ref{fig_kkn4} for both the isotropic and the anisotropic SU(4)
lattice calculations. In the anisotropic calculation we take advantage
of the extended, finer resolution in $t$ to use exponential fit ranges
that are roughly constant in physical units. In this way we hope to
minimise the systematic errors discussed earlier, albeit at the price
of somewhat larger statistical errors. We show continuum fits of the
form in eqn(\ref{eqn_contmk}). In Fig.\ref{fig_kkn6} we do the same
for the $k=2,3$ strings in SU(6), and in Fig.\ref{fig_kkn8} for the
$k=2,3,4$ strings in SU(8), all on isotropic lattices.  The higher-$k$
SU(8) calculations, which are simultaneously more massive and are
performed with half the statistics of other calculations, no doubt
suffer rather severely from the systematic errors we emphasised
earlier, as suggested by the apparent increase in the ratio with
decreasing $a$.

Our continuum limits of $\sigma_k/\sigma$ are listed in
Table~\ref{table_theoryK} and compared with the values suggested by
`Casimir Scaling' and `MQCD'. We see that our values fall between
these two sets of predictions, except for the $k=3,4$ SU(8)
calculations -- but these we discount, because they are by far the
poorest.

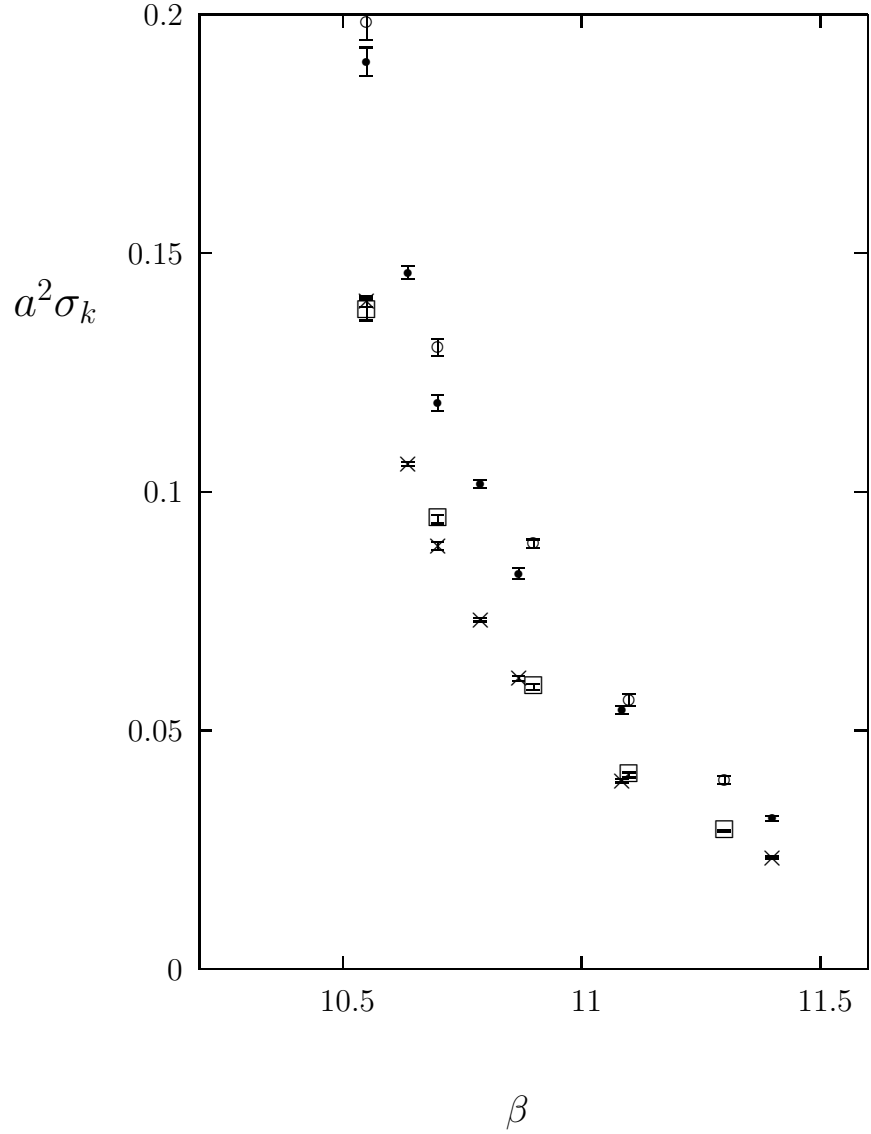
\begin	{figure}[p]
\begin	{center}
\leavevmode
\setlength{\unitlength}{0.240900pt}
\ifx\plotpoint\undefined\newsavebox{\plotpoint}\fi
\sbox{\plotpoint}{\rule[-0.200pt]{0.400pt}{0.400pt}}%
\begin{picture}(1500,1800)(0,0)
\font\gnuplot=cmr10 at 12pt
\gnuplot
\sbox{\plotpoint}{\rule[-0.200pt]{0.400pt}{0.400pt}}%
\put(375.0,250.0){\rule[-0.200pt]{4.818pt}{0.400pt}}
\put(350,250){\makebox(0,0)[r]{\ \ {$0$}}}
\put(1405.0,250.0){\rule[-0.200pt]{4.818pt}{0.400pt}}
\put(375.0,625.0){\rule[-0.200pt]{4.818pt}{0.400pt}}
\put(350,625){\makebox(0,0)[r]{\ \ {$0.05$}}}
\put(1405.0,625.0){\rule[-0.200pt]{4.818pt}{0.400pt}}
\put(375.0,1000.0){\rule[-0.200pt]{4.818pt}{0.400pt}}
\put(350,1000){\makebox(0,0)[r]{\ \ {$0.1$}}}
\put(1405.0,1000.0){\rule[-0.200pt]{4.818pt}{0.400pt}}
\put(375.0,1375.0){\rule[-0.200pt]{4.818pt}{0.400pt}}
\put(350,1375){\makebox(0,0)[r]{\ \ {$0.15$}}}
\put(1405.0,1375.0){\rule[-0.200pt]{4.818pt}{0.400pt}}
\put(375.0,1750.0){\rule[-0.200pt]{4.818pt}{0.400pt}}
\put(350,1750){\makebox(0,0)[r]{\ \ {$0.2$}}}
\put(1405.0,1750.0){\rule[-0.200pt]{4.818pt}{0.400pt}}
\put(600.0,250.0){\rule[-0.200pt]{0.400pt}{4.818pt}}
\put(600,200){\makebox(0,0){\ {$10.5$}}}
\put(600.0,1730.0){\rule[-0.200pt]{0.400pt}{4.818pt}}
\put(975.0,250.0){\rule[-0.200pt]{0.400pt}{4.818pt}}
\put(975,200){\makebox(0,0){\ {$11$}}}
\put(975.0,1730.0){\rule[-0.200pt]{0.400pt}{4.818pt}}
\put(1350.0,250.0){\rule[-0.200pt]{0.400pt}{4.818pt}}
\put(1350,200){\makebox(0,0){\ {$11.5$}}}
\put(1350.0,1730.0){\rule[-0.200pt]{0.400pt}{4.818pt}}
\put(375.0,250.0){\rule[-0.200pt]{252.945pt}{0.400pt}}
\put(1425.0,250.0){\rule[-0.200pt]{0.400pt}{361.350pt}}
\put(375.0,1750.0){\rule[-0.200pt]{252.945pt}{0.400pt}}
\put(150,1300){\makebox(0,0){\Large{$a^2\sigma_k$}}}
\put(875,25){\makebox(0,0){\large{$\beta$}}}
\put(375.0,250.0){\rule[-0.200pt]{0.400pt}{361.350pt}}
\put(638.0,1653.0){\rule[-0.200pt]{0.400pt}{10.840pt}}
\put(628.0,1653.0){\rule[-0.200pt]{4.818pt}{0.400pt}}
\put(628.0,1698.0){\rule[-0.200pt]{4.818pt}{0.400pt}}
\put(703.0,1334.0){\rule[-0.200pt]{0.400pt}{5.059pt}}
\put(693.0,1334.0){\rule[-0.200pt]{4.818pt}{0.400pt}}
\put(693.0,1355.0){\rule[-0.200pt]{4.818pt}{0.400pt}}
\put(750.0,1127.0){\rule[-0.200pt]{0.400pt}{6.022pt}}
\put(740.0,1127.0){\rule[-0.200pt]{4.818pt}{0.400pt}}
\put(740.0,1152.0){\rule[-0.200pt]{4.818pt}{0.400pt}}
\put(817.0,1006.0){\rule[-0.200pt]{0.400pt}{2.891pt}}
\put(807.0,1006.0){\rule[-0.200pt]{4.818pt}{0.400pt}}
\put(807.0,1018.0){\rule[-0.200pt]{4.818pt}{0.400pt}}
\put(877.0,863.0){\rule[-0.200pt]{0.400pt}{4.095pt}}
\put(867.0,863.0){\rule[-0.200pt]{4.818pt}{0.400pt}}
\put(867.0,880.0){\rule[-0.200pt]{4.818pt}{0.400pt}}
\put(1039.0,651.0){\rule[-0.200pt]{0.400pt}{2.891pt}}
\put(1029.0,651.0){\rule[-0.200pt]{4.818pt}{0.400pt}}
\put(1029.0,663.0){\rule[-0.200pt]{4.818pt}{0.400pt}}
\put(1275.0,483.0){\rule[-0.200pt]{0.400pt}{1.686pt}}
\put(1265.0,483.0){\rule[-0.200pt]{4.818pt}{0.400pt}}
\put(638,1675){\circle*{12}}
\put(703,1344){\circle*{12}}
\put(750,1140){\circle*{12}}
\put(817,1012){\circle*{12}}
\put(877,871){\circle*{12}}
\put(1039,657){\circle*{12}}
\put(1275,487){\circle*{12}}
\put(1265.0,490.0){\rule[-0.200pt]{4.818pt}{0.400pt}}
\put(638.0,1710.0){\rule[-0.200pt]{0.400pt}{9.636pt}}
\put(628.0,1710.0){\rule[-0.200pt]{4.818pt}{0.400pt}}
\put(628.0,1750.0){\rule[-0.200pt]{4.818pt}{0.400pt}}
\put(750.0,1213.0){\rule[-0.200pt]{0.400pt}{6.504pt}}
\put(740.0,1213.0){\rule[-0.200pt]{4.818pt}{0.400pt}}
\put(740.0,1240.0){\rule[-0.200pt]{4.818pt}{0.400pt}}
\put(900.0,912.0){\rule[-0.200pt]{0.400pt}{3.132pt}}
\put(890.0,912.0){\rule[-0.200pt]{4.818pt}{0.400pt}}
\put(890.0,925.0){\rule[-0.200pt]{4.818pt}{0.400pt}}
\put(1050.0,663.0){\rule[-0.200pt]{0.400pt}{4.577pt}}
\put(1040.0,663.0){\rule[-0.200pt]{4.818pt}{0.400pt}}
\put(1040.0,682.0){\rule[-0.200pt]{4.818pt}{0.400pt}}
\put(1200.0,541.0){\rule[-0.200pt]{0.400pt}{2.891pt}}
\put(1190.0,541.0){\rule[-0.200pt]{4.818pt}{0.400pt}}
\put(638,1738){\circle{18}}
\put(750,1227){\circle{18}}
\put(900,919){\circle{18}}
\put(1050,673){\circle{18}}
\put(1200,547){\circle{18}}
\put(1190.0,553.0){\rule[-0.200pt]{4.818pt}{0.400pt}}
\put(638.0,1290.0){\rule[-0.200pt]{0.400pt}{4.095pt}}
\put(628.0,1290.0){\rule[-0.200pt]{4.818pt}{0.400pt}}
\put(628.0,1307.0){\rule[-0.200pt]{4.818pt}{0.400pt}}
\put(703.0,1041.0){\rule[-0.200pt]{0.400pt}{1.445pt}}
\put(693.0,1041.0){\rule[-0.200pt]{4.818pt}{0.400pt}}
\put(693.0,1047.0){\rule[-0.200pt]{4.818pt}{0.400pt}}
\put(750.0,909.0){\rule[-0.200pt]{0.400pt}{2.891pt}}
\put(740.0,909.0){\rule[-0.200pt]{4.818pt}{0.400pt}}
\put(740.0,921.0){\rule[-0.200pt]{4.818pt}{0.400pt}}
\put(817.0,796.0){\rule[-0.200pt]{0.400pt}{1.445pt}}
\put(807.0,796.0){\rule[-0.200pt]{4.818pt}{0.400pt}}
\put(807.0,802.0){\rule[-0.200pt]{4.818pt}{0.400pt}}
\put(877.0,703.0){\rule[-0.200pt]{0.400pt}{1.686pt}}
\put(867.0,703.0){\rule[-0.200pt]{4.818pt}{0.400pt}}
\put(867.0,710.0){\rule[-0.200pt]{4.818pt}{0.400pt}}
\put(1039.0,543.0){\rule[-0.200pt]{0.400pt}{1.445pt}}
\put(1029.0,543.0){\rule[-0.200pt]{4.818pt}{0.400pt}}
\put(1029.0,549.0){\rule[-0.200pt]{4.818pt}{0.400pt}}
\put(1275.0,423.0){\rule[-0.200pt]{0.400pt}{0.964pt}}
\put(1265.0,423.0){\rule[-0.200pt]{4.818pt}{0.400pt}}
\put(638,1299){\makebox(0,0){$\times$}}
\put(703,1044){\makebox(0,0){$\times$}}
\put(750,915){\makebox(0,0){$\times$}}
\put(817,799){\makebox(0,0){$\times$}}
\put(877,707){\makebox(0,0){$\times$}}
\put(1039,546){\makebox(0,0){$\times$}}
\put(1275,425){\makebox(0,0){$\times$}}
\put(1265.0,427.0){\rule[-0.200pt]{4.818pt}{0.400pt}}
\put(638.0,1269.0){\rule[-0.200pt]{0.400pt}{7.950pt}}
\put(628.0,1269.0){\rule[-0.200pt]{4.818pt}{0.400pt}}
\put(628.0,1302.0){\rule[-0.200pt]{4.818pt}{0.400pt}}
\put(750.0,950.0){\rule[-0.200pt]{0.400pt}{3.373pt}}
\put(740.0,950.0){\rule[-0.200pt]{4.818pt}{0.400pt}}
\put(740.0,964.0){\rule[-0.200pt]{4.818pt}{0.400pt}}
\put(900.0,689.0){\rule[-0.200pt]{0.400pt}{2.168pt}}
\put(890.0,689.0){\rule[-0.200pt]{4.818pt}{0.400pt}}
\put(890.0,698.0){\rule[-0.200pt]{4.818pt}{0.400pt}}
\put(1050.0,551.0){\rule[-0.200pt]{0.400pt}{1.927pt}}
\put(1040.0,551.0){\rule[-0.200pt]{4.818pt}{0.400pt}}
\put(1040.0,559.0){\rule[-0.200pt]{4.818pt}{0.400pt}}
\put(1200.0,465.0){\rule[-0.200pt]{0.400pt}{0.964pt}}
\put(1190.0,465.0){\rule[-0.200pt]{4.818pt}{0.400pt}}
\put(638,1285){\raisebox{-.8pt}{\makebox(0,0){$\Box$}}}
\put(750,957){\raisebox{-.8pt}{\makebox(0,0){$\Box$}}}
\put(900,693){\raisebox{-.8pt}{\makebox(0,0){$\Box$}}}
\put(1050,555){\raisebox{-.8pt}{\makebox(0,0){$\Box$}}}
\put(1200,467){\raisebox{-.8pt}{\makebox(0,0){$\Box$}}}
\put(1190.0,469.0){\rule[-0.200pt]{4.818pt}{0.400pt}}
\end{picture}
\end	{center}
\vskip 0.15in
\caption{Comparison of the SU(4) $k=1$, $k=2$  string tensions 
obtained in this work ($\times$, $\bullet$ respectively) 
with those obtained in \cite{blmt-kstring} ($\Box$, $\circ$
respectively). Plotted against the inverse coupling
$\beta$.}
\label{fig_compOOKn4}
\end 	{figure}

\begin	{figure}[p]
\begin	{center}
\leavevmode
\setlength{\unitlength}{0.240900pt}
\ifx\plotpoint\undefined\newsavebox{\plotpoint}\fi
\sbox{\plotpoint}{\rule[-0.200pt]{0.400pt}{0.400pt}}%
\begin{picture}(1500,1800)(0,0)
\font\gnuplot=cmr10 at 12pt
\gnuplot
\sbox{\plotpoint}{\rule[-0.200pt]{0.400pt}{0.400pt}}%
\put(375.0,250.0){\rule[-0.200pt]{4.818pt}{0.400pt}}
\put(350,250){\makebox(0,0)[r]{\ \ {$0$}}}
\put(1405.0,250.0){\rule[-0.200pt]{4.818pt}{0.400pt}}
\put(375.0,550.0){\rule[-0.200pt]{4.818pt}{0.400pt}}
\put(350,550){\makebox(0,0)[r]{\ \ {$0.05$}}}
\put(1405.0,550.0){\rule[-0.200pt]{4.818pt}{0.400pt}}
\put(375.0,850.0){\rule[-0.200pt]{4.818pt}{0.400pt}}
\put(350,850){\makebox(0,0)[r]{\ \ {$0.1$}}}
\put(1405.0,850.0){\rule[-0.200pt]{4.818pt}{0.400pt}}
\put(375.0,1150.0){\rule[-0.200pt]{4.818pt}{0.400pt}}
\put(350,1150){\makebox(0,0)[r]{\ \ {$0.15$}}}
\put(1405.0,1150.0){\rule[-0.200pt]{4.818pt}{0.400pt}}
\put(375.0,1450.0){\rule[-0.200pt]{4.818pt}{0.400pt}}
\put(350,1450){\makebox(0,0)[r]{\ \ {$0.2$}}}
\put(1405.0,1450.0){\rule[-0.200pt]{4.818pt}{0.400pt}}
\put(375.0,1750.0){\rule[-0.200pt]{4.818pt}{0.400pt}}
\put(350,1750){\makebox(0,0)[r]{\ \ {$0.25$}}}
\put(1405.0,1750.0){\rule[-0.200pt]{4.818pt}{0.400pt}}
\put(450.0,250.0){\rule[-0.200pt]{0.400pt}{4.818pt}}
\put(450,200){\makebox(0,0){\ {$24.4$}}}
\put(450.0,1730.0){\rule[-0.200pt]{0.400pt}{4.818pt}}
\put(600.0,250.0){\rule[-0.200pt]{0.400pt}{4.818pt}}
\put(600,200){\makebox(0,0){\ {$24.6$}}}
\put(600.0,1730.0){\rule[-0.200pt]{0.400pt}{4.818pt}}
\put(750.0,250.0){\rule[-0.200pt]{0.400pt}{4.818pt}}
\put(750,200){\makebox(0,0){\ {$24.8$}}}
\put(750.0,1730.0){\rule[-0.200pt]{0.400pt}{4.818pt}}
\put(900.0,250.0){\rule[-0.200pt]{0.400pt}{4.818pt}}
\put(900,200){\makebox(0,0){\ {$25$}}}
\put(900.0,1730.0){\rule[-0.200pt]{0.400pt}{4.818pt}}
\put(1050.0,250.0){\rule[-0.200pt]{0.400pt}{4.818pt}}
\put(1050,200){\makebox(0,0){\ {$25.2$}}}
\put(1050.0,1730.0){\rule[-0.200pt]{0.400pt}{4.818pt}}
\put(1200.0,250.0){\rule[-0.200pt]{0.400pt}{4.818pt}}
\put(1200,200){\makebox(0,0){\ {$25.4$}}}
\put(1200.0,1730.0){\rule[-0.200pt]{0.400pt}{4.818pt}}
\put(1350.0,250.0){\rule[-0.200pt]{0.400pt}{4.818pt}}
\put(1350,200){\makebox(0,0){\ {$25.6$}}}
\put(1350.0,1730.0){\rule[-0.200pt]{0.400pt}{4.818pt}}
\put(375.0,250.0){\rule[-0.200pt]{252.945pt}{0.400pt}}
\put(1425.0,250.0){\rule[-0.200pt]{0.400pt}{361.350pt}}
\put(375.0,1750.0){\rule[-0.200pt]{252.945pt}{0.400pt}}
\put(150,1300){\makebox(0,0){\Large{$a^2\sigma_k$}}}
\put(875,25){\makebox(0,0){\large{$\beta$}}}
\put(375.0,250.0){\rule[-0.200pt]{0.400pt}{361.350pt}}
\put(525.0,1357.0){\rule[-0.200pt]{0.400pt}{11.563pt}}
\put(515.0,1357.0){\rule[-0.200pt]{4.818pt}{0.400pt}}
\put(515.0,1405.0){\rule[-0.200pt]{4.818pt}{0.400pt}}
\put(536.0,1374.0){\rule[-0.200pt]{0.400pt}{11.563pt}}
\put(526.0,1374.0){\rule[-0.200pt]{4.818pt}{0.400pt}}
\put(526.0,1422.0){\rule[-0.200pt]{4.818pt}{0.400pt}}
\put(653.0,1167.0){\rule[-0.200pt]{0.400pt}{6.504pt}}
\put(643.0,1167.0){\rule[-0.200pt]{4.818pt}{0.400pt}}
\put(643.0,1194.0){\rule[-0.200pt]{4.818pt}{0.400pt}}
\put(784.0,1022.0){\rule[-0.200pt]{0.400pt}{3.854pt}}
\put(774.0,1022.0){\rule[-0.200pt]{4.818pt}{0.400pt}}
\put(774.0,1038.0){\rule[-0.200pt]{4.818pt}{0.400pt}}
\put(938.0,860.0){\rule[-0.200pt]{0.400pt}{3.854pt}}
\put(928.0,860.0){\rule[-0.200pt]{4.818pt}{0.400pt}}
\put(928.0,876.0){\rule[-0.200pt]{4.818pt}{0.400pt}}
\put(1239.0,689.0){\rule[-0.200pt]{0.400pt}{2.409pt}}
\put(1229.0,689.0){\rule[-0.200pt]{4.818pt}{0.400pt}}
\put(525,1381){\circle*{12}}
\put(536,1398){\circle*{12}}
\put(653,1181){\circle*{12}}
\put(784,1030){\circle*{12}}
\put(938,868){\circle*{12}}
\put(1239,694){\circle*{12}}
\put(1229.0,699.0){\rule[-0.200pt]{4.818pt}{0.400pt}}
\put(618.0,1276.0){\rule[-0.200pt]{0.400pt}{8.672pt}}
\put(608.0,1276.0){\rule[-0.200pt]{4.818pt}{0.400pt}}
\put(608.0,1312.0){\rule[-0.200pt]{4.818pt}{0.400pt}}
\put(726.0,1156.0){\rule[-0.200pt]{0.400pt}{5.782pt}}
\put(716.0,1156.0){\rule[-0.200pt]{4.818pt}{0.400pt}}
\put(716.0,1180.0){\rule[-0.200pt]{4.818pt}{0.400pt}}
\put(942.0,892.0){\rule[-0.200pt]{0.400pt}{8.672pt}}
\put(932.0,892.0){\rule[-0.200pt]{4.818pt}{0.400pt}}
\put(932.0,928.0){\rule[-0.200pt]{4.818pt}{0.400pt}}
\put(1050.0,820.0){\rule[-0.200pt]{0.400pt}{5.782pt}}
\put(1040.0,820.0){\rule[-0.200pt]{4.818pt}{0.400pt}}
\put(1040.0,844.0){\rule[-0.200pt]{4.818pt}{0.400pt}}
\put(1266.0,703.0){\rule[-0.200pt]{0.400pt}{3.132pt}}
\put(1256.0,703.0){\rule[-0.200pt]{4.818pt}{0.400pt}}
\put(618,1294){\circle{18}}
\put(726,1168){\circle{18}}
\put(942,910){\circle{18}}
\put(1050,832){\circle{18}}
\put(1266,710){\circle{18}}
\put(1256.0,716.0){\rule[-0.200pt]{4.818pt}{0.400pt}}
\put(525.0,1531.0){\rule[-0.200pt]{0.400pt}{25.054pt}}
\put(515.0,1531.0){\rule[-0.200pt]{4.818pt}{0.400pt}}
\put(515.0,1635.0){\rule[-0.200pt]{4.818pt}{0.400pt}}
\put(536.0,1487.0){\rule[-0.200pt]{0.400pt}{24.090pt}}
\put(526.0,1487.0){\rule[-0.200pt]{4.818pt}{0.400pt}}
\put(526.0,1587.0){\rule[-0.200pt]{4.818pt}{0.400pt}}
\put(653.0,1294.0){\rule[-0.200pt]{0.400pt}{13.490pt}}
\put(643.0,1294.0){\rule[-0.200pt]{4.818pt}{0.400pt}}
\put(643.0,1350.0){\rule[-0.200pt]{4.818pt}{0.400pt}}
\put(784.0,1111.0){\rule[-0.200pt]{0.400pt}{9.877pt}}
\put(774.0,1111.0){\rule[-0.200pt]{4.818pt}{0.400pt}}
\put(774.0,1152.0){\rule[-0.200pt]{4.818pt}{0.400pt}}
\put(938.0,956.0){\rule[-0.200pt]{0.400pt}{6.986pt}}
\put(928.0,956.0){\rule[-0.200pt]{4.818pt}{0.400pt}}
\put(928.0,985.0){\rule[-0.200pt]{4.818pt}{0.400pt}}
\put(1239.0,739.0){\rule[-0.200pt]{0.400pt}{4.095pt}}
\put(1229.0,739.0){\rule[-0.200pt]{4.818pt}{0.400pt}}
\put(525,1583){\makebox(0,0){$\star$}}
\put(536,1537){\makebox(0,0){$\star$}}
\put(653,1322){\makebox(0,0){$\star$}}
\put(784,1131){\makebox(0,0){$\star$}}
\put(938,971){\makebox(0,0){$\star$}}
\put(1239,747){\makebox(0,0){$\star$}}
\put(1229.0,756.0){\rule[-0.200pt]{4.818pt}{0.400pt}}
\put(618.0,1402.0){\rule[-0.200pt]{0.400pt}{26.017pt}}
\put(608.0,1402.0){\rule[-0.200pt]{4.818pt}{0.400pt}}
\put(608.0,1510.0){\rule[-0.200pt]{4.818pt}{0.400pt}}
\put(726.0,1264.0){\rule[-0.200pt]{0.400pt}{11.563pt}}
\put(716.0,1264.0){\rule[-0.200pt]{4.818pt}{0.400pt}}
\put(716.0,1312.0){\rule[-0.200pt]{4.818pt}{0.400pt}}
\put(942.0,1000.0){\rule[-0.200pt]{0.400pt}{20.236pt}}
\put(932.0,1000.0){\rule[-0.200pt]{4.818pt}{0.400pt}}
\put(932.0,1084.0){\rule[-0.200pt]{4.818pt}{0.400pt}}
\put(1050.0,874.0){\rule[-0.200pt]{0.400pt}{17.345pt}}
\put(1040.0,874.0){\rule[-0.200pt]{4.818pt}{0.400pt}}
\put(1040.0,946.0){\rule[-0.200pt]{4.818pt}{0.400pt}}
\put(1266.0,772.0){\rule[-0.200pt]{0.400pt}{8.672pt}}
\put(1256.0,772.0){\rule[-0.200pt]{4.818pt}{0.400pt}}
\put(618,1456){\raisebox{-.8pt}{\makebox(0,0){$\Diamond$}}}
\put(726,1288){\raisebox{-.8pt}{\makebox(0,0){$\Diamond$}}}
\put(942,1042){\raisebox{-.8pt}{\makebox(0,0){$\Diamond$}}}
\put(1050,910){\raisebox{-.8pt}{\makebox(0,0){$\Diamond$}}}
\put(1266,790){\raisebox{-.8pt}{\makebox(0,0){$\Diamond$}}}
\put(1256.0,808.0){\rule[-0.200pt]{4.818pt}{0.400pt}}
\put(413.0,1148.0){\rule[-0.200pt]{0.400pt}{3.854pt}}
\put(403.0,1148.0){\rule[-0.200pt]{4.818pt}{0.400pt}}
\put(403.0,1164.0){\rule[-0.200pt]{4.818pt}{0.400pt}}
\put(525.0,943.0){\rule[-0.200pt]{0.400pt}{3.373pt}}
\put(515.0,943.0){\rule[-0.200pt]{4.818pt}{0.400pt}}
\put(515.0,957.0){\rule[-0.200pt]{4.818pt}{0.400pt}}
\put(536.0,931.0){\rule[-0.200pt]{0.400pt}{3.132pt}}
\put(526.0,931.0){\rule[-0.200pt]{4.818pt}{0.400pt}}
\put(526.0,944.0){\rule[-0.200pt]{4.818pt}{0.400pt}}
\put(653.0,813.0){\rule[-0.200pt]{0.400pt}{2.168pt}}
\put(643.0,813.0){\rule[-0.200pt]{4.818pt}{0.400pt}}
\put(643.0,822.0){\rule[-0.200pt]{4.818pt}{0.400pt}}
\put(784.0,717.0){\rule[-0.200pt]{0.400pt}{1.927pt}}
\put(774.0,717.0){\rule[-0.200pt]{4.818pt}{0.400pt}}
\put(774.0,725.0){\rule[-0.200pt]{4.818pt}{0.400pt}}
\put(938.0,625.0){\rule[-0.200pt]{0.400pt}{1.445pt}}
\put(928.0,625.0){\rule[-0.200pt]{4.818pt}{0.400pt}}
\put(928.0,631.0){\rule[-0.200pt]{4.818pt}{0.400pt}}
\put(1239.0,512.0){\rule[-0.200pt]{0.400pt}{0.964pt}}
\put(1229.0,512.0){\rule[-0.200pt]{4.818pt}{0.400pt}}
\put(413,1156){\makebox(0,0){$\times$}}
\put(525,950){\makebox(0,0){$\times$}}
\put(536,938){\makebox(0,0){$\times$}}
\put(653,817){\makebox(0,0){$\times$}}
\put(784,721){\makebox(0,0){$\times$}}
\put(938,628){\makebox(0,0){$\times$}}
\put(1239,514){\makebox(0,0){$\times$}}
\put(1229.0,516.0){\rule[-0.200pt]{4.818pt}{0.400pt}}
\put(618.0,876.0){\rule[-0.200pt]{0.400pt}{1.445pt}}
\put(608.0,876.0){\rule[-0.200pt]{4.818pt}{0.400pt}}
\put(608.0,882.0){\rule[-0.200pt]{4.818pt}{0.400pt}}
\put(726.0,779.0){\rule[-0.200pt]{0.400pt}{0.723pt}}
\put(716.0,779.0){\rule[-0.200pt]{4.818pt}{0.400pt}}
\put(716.0,782.0){\rule[-0.200pt]{4.818pt}{0.400pt}}
\put(942.0,633.0){\rule[-0.200pt]{0.400pt}{0.964pt}}
\put(932.0,633.0){\rule[-0.200pt]{4.818pt}{0.400pt}}
\put(932.0,637.0){\rule[-0.200pt]{4.818pt}{0.400pt}}
\put(1050.0,588.0){\rule[-0.200pt]{0.400pt}{0.964pt}}
\put(1040.0,588.0){\rule[-0.200pt]{4.818pt}{0.400pt}}
\put(1040.0,592.0){\rule[-0.200pt]{4.818pt}{0.400pt}}
\put(1266.0,514.0){\rule[-0.200pt]{0.400pt}{0.482pt}}
\put(1256.0,514.0){\rule[-0.200pt]{4.818pt}{0.400pt}}
\put(618,879){\raisebox{-.8pt}{\makebox(0,0){$\Box$}}}
\put(726,780){\raisebox{-.8pt}{\makebox(0,0){$\Box$}}}
\put(942,635){\raisebox{-.8pt}{\makebox(0,0){$\Box$}}}
\put(1050,590){\raisebox{-.8pt}{\makebox(0,0){$\Box$}}}
\put(1266,515){\raisebox{-.8pt}{\makebox(0,0){$\Box$}}}
\put(1256.0,516.0){\rule[-0.200pt]{4.818pt}{0.400pt}}
\end{picture}
\end	{center}
\vskip 0.15in
\caption{Comparison of the SU(6) $k=1$, $k=2$ and $k=3$ string tensions 
obtained in this work ($\times$, $\bullet$, $\star$ 
respectively) with those obtained in \cite{PisaK} ($\Box$, $\circ$, 
$\diamond$ respectively). Plotted against the inverse coupling
$\beta$.}
\label{fig_compKn6}
\end 	{figure}
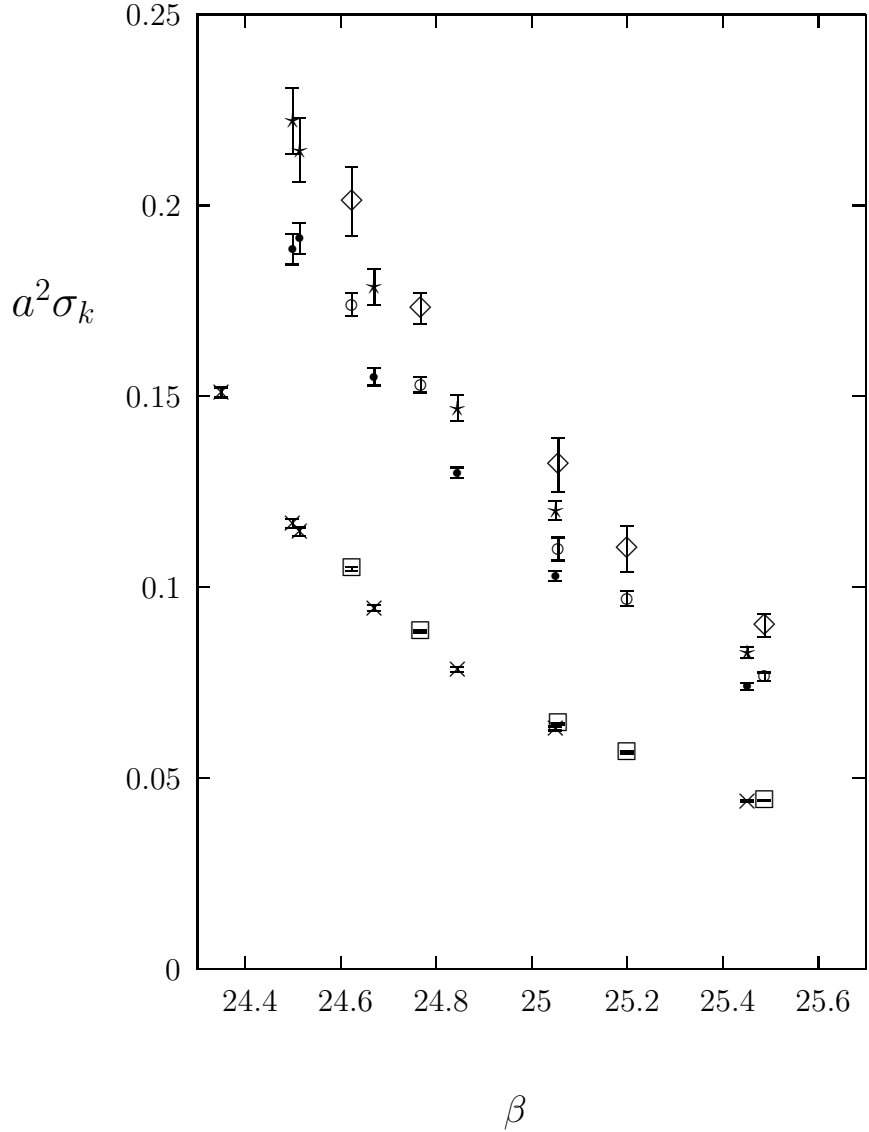

It is interesting to see by how much our calculation of $\sigma_k$
improves upon previous calculations.  In Fig.\ref{fig_compOOKn4} we
plot the SU(4) values of $a^2\sigma$ and $a^2\sigma_{k=2}$ as obtained
in the present calculation and in that of
\cite{blmt-kstring}.
We observe that our values are slightly lower with the difference
larger for $k=2$ than for $k=1$ -- precisely as we would expect from
the systematic error described earlier.  Because of the positivity of
our correlators, if the extracted mass is lower then the calculation
is unambiguously better.  In Fig.\ref{fig_compKn6} we compare our
SU(6) calculations with those of
\cite{PisaK}.
Here the masses are even heavier and the improvement much more
marked. Although our statistics is an order of magnitude less than
that of
\cite{PisaK}
our better operators clearly provide an unambiguously better
calculation.

\subsection{unstable strings}
\label{subsection_unstable}

\subsubsection{background}
\label{subsubsection_background}

At finite $N$ the only strings that are absolutely stable are the
$k$-strings. However for each value of $k$ there are sources in many
representations, and for each such source there will be a string (or
strings) carrying its flux.  At finite $N$ such strings can be
screened by the gluons in the vacuum, down to the lightest string of
the same value of $k$. However this charge screening/string breaking
typically vanishes as $N\to\infty$, so that the string become
increasingly stable and well-defined in that limit.

Typically the source arises from a direct product of $k+j$ fundamental
and $j$ conjugate fundamental sources.  Thus if screening is
suppressed one would expect that over long enough distances the flux
will travel through $k+2j$ fundamental flux tubes, i.e.
\be
\sigma_{k+j,j} = (k+2j)\sigma
\label{eqn_Kunstable}
\ee
unless these strings form a bound state of lower tension. Of course it
is only for $N\to\infty$ that screening can be neglected for arbitrary
separations, and in that limit we expect string binding to vanish, so
that the string that becomes stable has a tension that is precisely
$(k+2j)\sigma$.  The classic example of this is the $k=0$ adjoint
string, where we expect $\lim_{N\to\infty} \sigma_A = 2\sigma$.  What
would be interesting is not so much the verification of
eqn(\ref{eqn_Kunstable}), but the observation of a bound state or
resonant string at finite $N$. Of course, since the large-$N$ counting
tells us to expect the binding to be of the order of the decay width
(per unit length), it is not at all obvious that there is anything
unambiguous to find.

Recent calculations of the tension of the adjoint string and of
several other unstable strings in SU(3)
\cite{Deldar,Bali}
have turned out to be consistent with Casimir Scaling.  That is to
say, they typically find string tensions that are larger than in
eqn(\ref{eqn_Kunstable}).  These calculations have been performed with
explicit sources (i.e.~using Wilson rather than Polyakov loops) and
have exploited the fact that in such a set-up any flux tube will
remain completely stable up to some finite distance. This distance is
determined by the trade-off between the extra mass the source acquires
if it is to be screened, and the decrease in the string energy induced
by the screening. In practice the distance between the sources for
which the potential energy can be calculated is limited to about
$1\mathrm{fm}$, and some fraction of this distance falls into the
Coulomb region around the sources -- so that the actual strings are
probably no more than $\sim 0.5\mathrm{fm}$ in length. This is too
short a length for us to be entirely confident that we are seeing the
asymptotic string-like properties of the flux tube, and in particular
where we would expect eqn(\ref{eqn_Kunstable}) to give a lighter state
of this system.  Moreover, since the Coulomb interaction that
dominates the potential at shorter distances trivially satisfies
Casimir Scaling, one might worry that what we are seeing at
separations $\leq 1 \mathrm{fm}$ is some subasymptotic dynamics that
manifests approximate Casimir Scaling by continuity with one gluon
exchange. Alternatively it might be a genuine resonant multi-string
state, whose break-up into the separate fundamental strings requires
tunnelling through some barrier.

It would clearly be interesting to investigate such unstable strings
using the formalism where the string winds around the torus so that
there are no sources, and there is no Coulomb piece with Casimir
Scaling.  Moreover, in our typical calculation the length of the
string is $a_sL_s \sim 3.2 \sim 1.5\mathrm{fm}$, if we introduce `fm'
units through $\sqrt\sigma \simeq 0.45\mathrm{fm}$ (the value in QCD).
This is long enough that it might be asymptotic.  The downside is that
there is no length below which such an unstable string will be stable;
its approximate stability will depend entirely on $N$ being large
enough, and we do not know in advance what that will be.

We will focus here on the totally symmetric and antisymmetric $k=2$
and $k=3$ strings, and in addition the mixed $k=3$ string, which we
refer to as $kS,kA$ and $kM$ respectively.  (See Appendix A of
\cite{blmt-kstring}
for the details of the operator construction.) We recall that in
similar calculations in
\cite{blmt-kstring}
there was no good evidence found for such strings in $D=3+1$ SU($N$)
gauge theories, although some evidence was found in the case of
$D=2+1$. This might be because in
\cite{blmt-kstring}
the operators for strings of higher $k$ were much poorer in $D=3+1$
than in $D=2+1$. With the improved operators of the present paper we
can hope to do better.

In this Section we will compare our results to Casimir Scaling not so
much because the theoretical case is at all compelling, but rather
because there is at least a formula to compare with -- unlike the case
of `MQCD' which (as far as we know) makes no predictions for such
unstable strings. It may be useful to reproduce the expressions
relevant to our calculations:
\begin{equation}
{{\sigma_k}\over{\sigma}}
\stackrel{CS}{=} 
\begin{cases}
 k(N-k)/(N-1) & : \quad \text{anti-symmetric} \\
 k(N+k)/(N+1) & : \quad \text{symmetric} \\
 3(N^2-3)/(N^2-1) & : \quad \text{k=3,~mixed}.
\end{cases}
\label{eqn_CSall}
\end{equation}
(See Appendix A of
\cite{blmt-kstring}
for the derivation.)

\subsubsection{results}
\label{subsubsection_results}

We will begin with our results from the calculation on the SU(4)
anisotropic lattice with the smallest $a_s$ and $a_t$, i.e.~the $16^3
40$ lattice. We would hope that the fine resolution in $t$ will
provide us with our best chance of seeing some kind of effective mass
plateau (albeit temporary in $t$) for the heavy unstable string of
interest. On the other hand $N=4$ might be too small for these strings
to be sufficiently stable and well defined.

\begin	{figure}[p]
\begin	{center}
\leavevmode
\input	{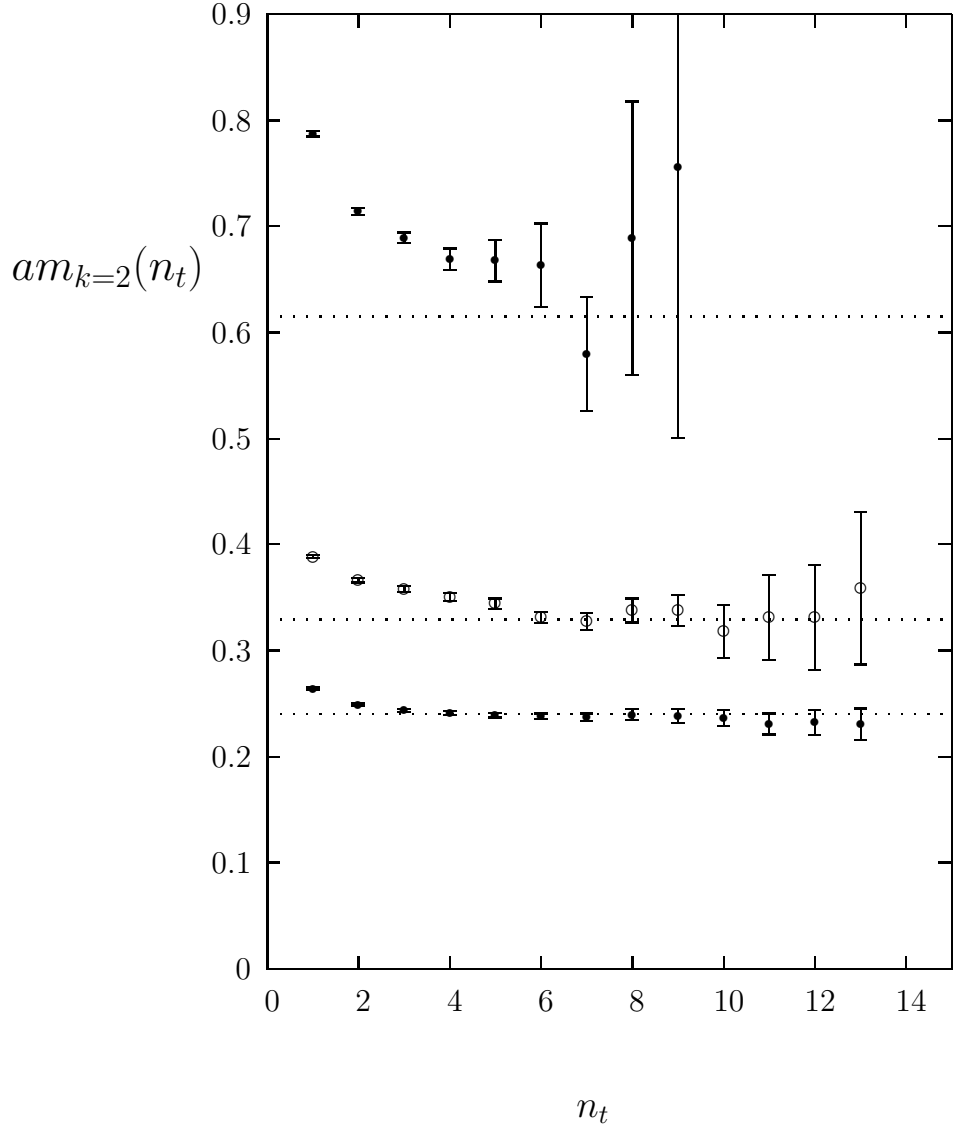}
\end	{center}
\vskip 0.15in
\caption{Effective mass of the SU(4) $k$-string as a function 
of $n_t$ from the $16^3 40$ anisotropic lattice calculation
with the smallest lattice spacing. The mass of the lightest 
string, in the fundamental representation, $\bullet$, and the
lightest masses in the $k=2$ totally antisymmetric, $\circ$, and the 
$k=2$ symmetric, $\bullet$, representations, are shown. Dashed lines
indicate the best mass estimate of the $k=1$ string, and
the masses that one would then expect for the (anti)symmetric 
$k=2$ strings if one scales up the $k=1$ tension by the Casimir 
Scaling factor.}
\label{fig_anisok2npn4}
\end 	{figure}

In Fig.\ref{fig_anisok2npn4} we show the effective mass as a function
of $n_t$ for the lightest $k=1$, $k=2A$ and $k=2S$ string states, as
obtained from a variational calculation
\cite{blmt-kstring}
using the basis of multiply-smeared blocked Polyakov loop operators
described at the beginning of this Section.  
Some observations:
\hfil\break 
$\bullet$ 
The lightest $k=2$ string is in the totally antisymmetric
representation, $2A$. This is true for all our $k$ and $N$ (and was
already noted in
\cite{blmt-kstring}). \hfil\break
$\bullet$
The lightest string in the $2S$ representation appears to be much
heavier, and we have significant evidence for an effective mass
plateau. \hfil\break
$\bullet$
The apparent absence of the lightest $2A$ state in the mass spectrum
obtained with the $2S$ operators, tells us that the overlap of this
state onto this basis must be very small; at most at the percent
level.  This is presumably the expected large-$N$ suppression and
suggests that for the present purposes $N=4$ is `large'.  \hfil\break
$\bullet$
We show in Fig.\ref{fig_anisok2npn4} the masses we would obtain for
the $k=2A$ and $k=2S$ strings if we assumed Casimir Scaling, and
scaled up from the $k=1$ mass using eqn(\ref{eqn_D4}) and
eqn(\ref{eqn_CSall}).  We see that Casimir Scaling works quite well
here.

On our isotropic lattices we cannot hope to go to large enough $n_t$
to obtain serious evidence for any effective mass plateaux for these
heavy strings.  We will assume on the basis of our above anisotropic
lattice calculation that these approximate plateaux do exist.  We will
not attempt continuum extrapolations, for obvious reasons, but will
simply work at the smallest $a$ at each $N$ i.e.~on the $16^4$
lattices for $N=4,6,8$.  We will typically extract the mass from $t=a$
to $t=2a$.  (We note that this corresponds to $m_{eff}(n_t\geq 3)$ in
our anisotropic lattice calculation, and we see from
Fig.\ref{fig_anisok2npn4} that such masses already lie, more-or-less,
on the corresponding plateaux.)  In Table~\ref{table_unstableK} we
list the lightest $2A,S$ masses for $N=4,6,8$ and the lightest
$3A,S,M$ masses for $N=6,8$.  We also list the masses one would expect
from Casimir Scaling (scaled up from the $k=1$ masses using
eqn(\ref{eqn_D4})).  Overall we find qualitative compatibility with
the Casimir Scaling formula in eqn(\ref{eqn_CSall}), and also some
sign that the agreement improves as $N$ increases. However there are a
number of other features of the spectrum that are both interesting and
puzzling, and which need to be understood before the comparison in
Table~\ref{table_unstableK} can be taken too seriously.

\subsubsection{puzzles}
\label{subsubsection_puzzles}
 
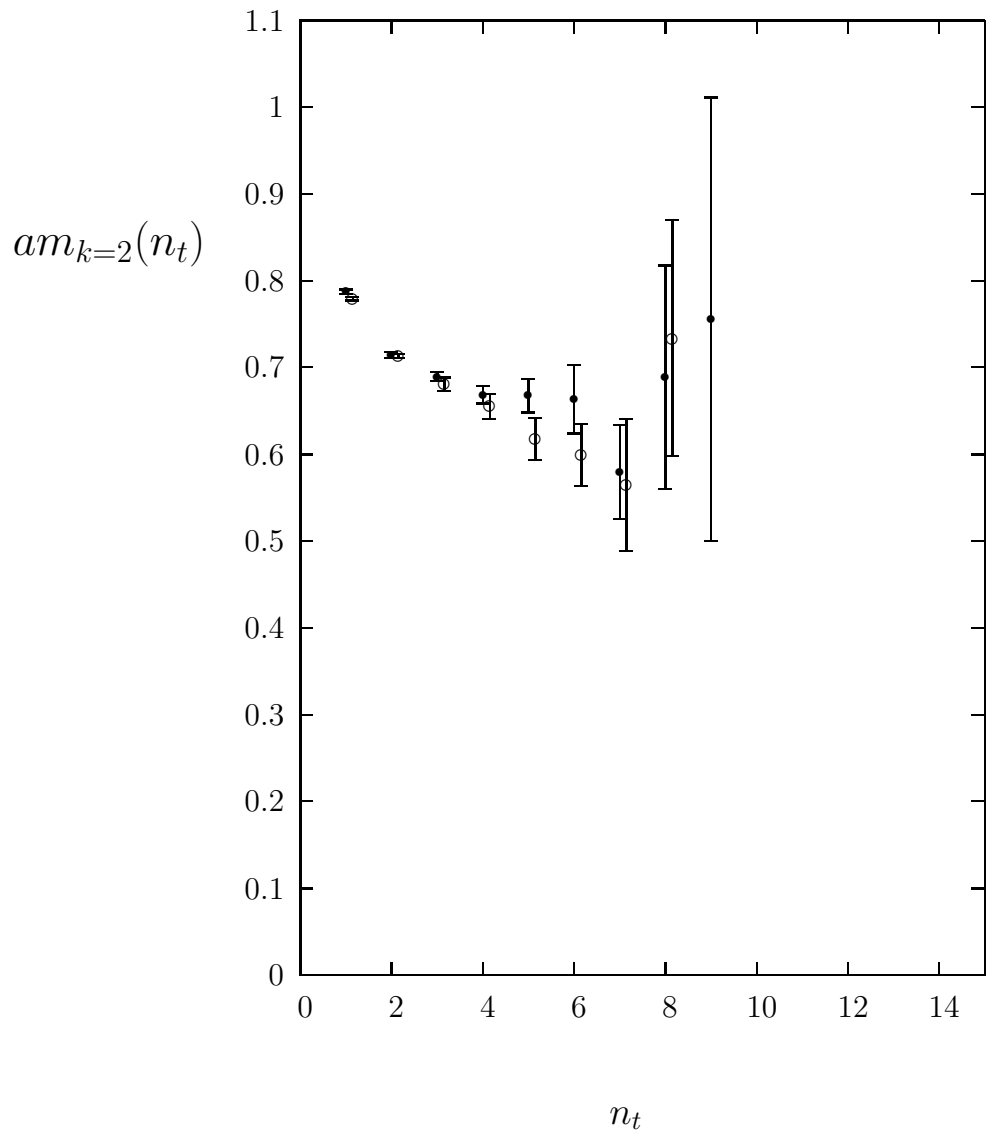
\begin	{figure}[p]
\begin	{center}
\leavevmode
\setlength{\unitlength}{0.240900pt}
\ifx\plotpoint\undefined\newsavebox{\plotpoint}\fi
\sbox{\plotpoint}{\rule[-0.200pt]{0.400pt}{0.400pt}}%
\begin{picture}(1500,1800)(0,0)
\font\gnuplot=cmr10 at 12pt
\gnuplot
\sbox{\plotpoint}{\rule[-0.200pt]{0.400pt}{0.400pt}}%
\put(350.0,250.0){\rule[-0.200pt]{4.818pt}{0.400pt}}
\put(325,250){\makebox(0,0)[r]{\ \ {$0$}}}
\put(1405.0,250.0){\rule[-0.200pt]{4.818pt}{0.400pt}}
\put(350.0,386.0){\rule[-0.200pt]{4.818pt}{0.400pt}}
\put(325,386){\makebox(0,0)[r]{\ \ {$0.1$}}}
\put(1405.0,386.0){\rule[-0.200pt]{4.818pt}{0.400pt}}
\put(350.0,523.0){\rule[-0.200pt]{4.818pt}{0.400pt}}
\put(325,523){\makebox(0,0)[r]{\ \ {$0.2$}}}
\put(1405.0,523.0){\rule[-0.200pt]{4.818pt}{0.400pt}}
\put(350.0,659.0){\rule[-0.200pt]{4.818pt}{0.400pt}}
\put(325,659){\makebox(0,0)[r]{\ \ {$0.3$}}}
\put(1405.0,659.0){\rule[-0.200pt]{4.818pt}{0.400pt}}
\put(350.0,795.0){\rule[-0.200pt]{4.818pt}{0.400pt}}
\put(325,795){\makebox(0,0)[r]{\ \ {$0.4$}}}
\put(1405.0,795.0){\rule[-0.200pt]{4.818pt}{0.400pt}}
\put(350.0,932.0){\rule[-0.200pt]{4.818pt}{0.400pt}}
\put(325,932){\makebox(0,0)[r]{\ \ {$0.5$}}}
\put(1405.0,932.0){\rule[-0.200pt]{4.818pt}{0.400pt}}
\put(350.0,1068.0){\rule[-0.200pt]{4.818pt}{0.400pt}}
\put(325,1068){\makebox(0,0)[r]{\ \ {$0.6$}}}
\put(1405.0,1068.0){\rule[-0.200pt]{4.818pt}{0.400pt}}
\put(350.0,1205.0){\rule[-0.200pt]{4.818pt}{0.400pt}}
\put(325,1205){\makebox(0,0)[r]{\ \ {$0.7$}}}
\put(1405.0,1205.0){\rule[-0.200pt]{4.818pt}{0.400pt}}
\put(350.0,1341.0){\rule[-0.200pt]{4.818pt}{0.400pt}}
\put(325,1341){\makebox(0,0)[r]{\ \ {$0.8$}}}
\put(1405.0,1341.0){\rule[-0.200pt]{4.818pt}{0.400pt}}
\put(350.0,1477.0){\rule[-0.200pt]{4.818pt}{0.400pt}}
\put(325,1477){\makebox(0,0)[r]{\ \ {$0.9$}}}
\put(1405.0,1477.0){\rule[-0.200pt]{4.818pt}{0.400pt}}
\put(350.0,1614.0){\rule[-0.200pt]{4.818pt}{0.400pt}}
\put(325,1614){\makebox(0,0)[r]{\ \ {$1$}}}
\put(1405.0,1614.0){\rule[-0.200pt]{4.818pt}{0.400pt}}
\put(350.0,1750.0){\rule[-0.200pt]{4.818pt}{0.400pt}}
\put(325,1750){\makebox(0,0)[r]{\ \ {$1.1$}}}
\put(1405.0,1750.0){\rule[-0.200pt]{4.818pt}{0.400pt}}
\put(350.0,250.0){\rule[-0.200pt]{0.400pt}{4.818pt}}
\put(350,200){\makebox(0,0){\ {$0$}}}
\put(350.0,1730.0){\rule[-0.200pt]{0.400pt}{4.818pt}}
\put(493.0,250.0){\rule[-0.200pt]{0.400pt}{4.818pt}}
\put(493,200){\makebox(0,0){\ {$2$}}}
\put(493.0,1730.0){\rule[-0.200pt]{0.400pt}{4.818pt}}
\put(637.0,250.0){\rule[-0.200pt]{0.400pt}{4.818pt}}
\put(637,200){\makebox(0,0){\ {$4$}}}
\put(637.0,1730.0){\rule[-0.200pt]{0.400pt}{4.818pt}}
\put(780.0,250.0){\rule[-0.200pt]{0.400pt}{4.818pt}}
\put(780,200){\makebox(0,0){\ {$6$}}}
\put(780.0,1730.0){\rule[-0.200pt]{0.400pt}{4.818pt}}
\put(923.0,250.0){\rule[-0.200pt]{0.400pt}{4.818pt}}
\put(923,200){\makebox(0,0){\ {$8$}}}
\put(923.0,1730.0){\rule[-0.200pt]{0.400pt}{4.818pt}}
\put(1067.0,250.0){\rule[-0.200pt]{0.400pt}{4.818pt}}
\put(1067,200){\makebox(0,0){\ {$10$}}}
\put(1067.0,1730.0){\rule[-0.200pt]{0.400pt}{4.818pt}}
\put(1210.0,250.0){\rule[-0.200pt]{0.400pt}{4.818pt}}
\put(1210,200){\makebox(0,0){\ {$12$}}}
\put(1210.0,1730.0){\rule[-0.200pt]{0.400pt}{4.818pt}}
\put(1353.0,250.0){\rule[-0.200pt]{0.400pt}{4.818pt}}
\put(1353,200){\makebox(0,0){\ {$14$}}}
\put(1353.0,1730.0){\rule[-0.200pt]{0.400pt}{4.818pt}}
\put(350.0,250.0){\rule[-0.200pt]{258.967pt}{0.400pt}}
\put(1425.0,250.0){\rule[-0.200pt]{0.400pt}{361.350pt}}
\put(350.0,1750.0){\rule[-0.200pt]{258.967pt}{0.400pt}}
\put(50,1400){\makebox(0,0){\Large{$am_{k=2}(n_t)$}}}
\put(862,25){\makebox(0,0){\large{$n_t$}}}
\put(350.0,250.0){\rule[-0.200pt]{0.400pt}{361.350pt}}
\put(422.0,1320.0){\rule[-0.200pt]{0.400pt}{1.686pt}}
\put(412.0,1320.0){\rule[-0.200pt]{4.818pt}{0.400pt}}
\put(412.0,1327.0){\rule[-0.200pt]{4.818pt}{0.400pt}}
\put(493.0,1220.0){\rule[-0.200pt]{0.400pt}{2.168pt}}
\put(483.0,1220.0){\rule[-0.200pt]{4.818pt}{0.400pt}}
\put(483.0,1229.0){\rule[-0.200pt]{4.818pt}{0.400pt}}
\put(565.0,1183.0){\rule[-0.200pt]{0.400pt}{3.373pt}}
\put(555.0,1183.0){\rule[-0.200pt]{4.818pt}{0.400pt}}
\put(555.0,1197.0){\rule[-0.200pt]{4.818pt}{0.400pt}}
\put(637.0,1148.0){\rule[-0.200pt]{0.400pt}{6.745pt}}
\put(627.0,1148.0){\rule[-0.200pt]{4.818pt}{0.400pt}}
\put(627.0,1176.0){\rule[-0.200pt]{4.818pt}{0.400pt}}
\put(708.0,1134.0){\rule[-0.200pt]{0.400pt}{12.768pt}}
\put(698.0,1134.0){\rule[-0.200pt]{4.818pt}{0.400pt}}
\put(698.0,1187.0){\rule[-0.200pt]{4.818pt}{0.400pt}}
\put(780.0,1101.0){\rule[-0.200pt]{0.400pt}{25.776pt}}
\put(770.0,1101.0){\rule[-0.200pt]{4.818pt}{0.400pt}}
\put(770.0,1208.0){\rule[-0.200pt]{4.818pt}{0.400pt}}
\put(852.0,966.0){\rule[-0.200pt]{0.400pt}{35.653pt}}
\put(842.0,966.0){\rule[-0.200pt]{4.818pt}{0.400pt}}
\put(842.0,1114.0){\rule[-0.200pt]{4.818pt}{0.400pt}}
\put(923.0,1014.0){\rule[-0.200pt]{0.400pt}{84.556pt}}
\put(913.0,1014.0){\rule[-0.200pt]{4.818pt}{0.400pt}}
\put(913.0,1365.0){\rule[-0.200pt]{4.818pt}{0.400pt}}
\put(995.0,932.0){\rule[-0.200pt]{0.400pt}{167.907pt}}
\put(985.0,932.0){\rule[-0.200pt]{4.818pt}{0.400pt}}
\put(422,1324){\circle*{12}}
\put(493,1224){\circle*{12}}
\put(565,1190){\circle*{12}}
\put(637,1162){\circle*{12}}
\put(708,1161){\circle*{12}}
\put(780,1155){\circle*{12}}
\put(852,1040){\circle*{12}}
\put(923,1189){\circle*{12}}
\put(995,1281){\circle*{12}}
\put(985.0,1629.0){\rule[-0.200pt]{4.818pt}{0.400pt}}
\put(432.0,1310.0){\rule[-0.200pt]{0.400pt}{1.204pt}}
\put(422.0,1310.0){\rule[-0.200pt]{4.818pt}{0.400pt}}
\put(422.0,1315.0){\rule[-0.200pt]{4.818pt}{0.400pt}}
\put(504.0,1219.0){\rule[-0.200pt]{0.400pt}{1.686pt}}
\put(494.0,1219.0){\rule[-0.200pt]{4.818pt}{0.400pt}}
\put(494.0,1226.0){\rule[-0.200pt]{4.818pt}{0.400pt}}
\put(576.0,1168.0){\rule[-0.200pt]{0.400pt}{5.059pt}}
\put(566.0,1168.0){\rule[-0.200pt]{4.818pt}{0.400pt}}
\put(566.0,1189.0){\rule[-0.200pt]{4.818pt}{0.400pt}}
\put(647.0,1124.0){\rule[-0.200pt]{0.400pt}{9.395pt}}
\put(637.0,1124.0){\rule[-0.200pt]{4.818pt}{0.400pt}}
\put(637.0,1163.0){\rule[-0.200pt]{4.818pt}{0.400pt}}
\put(719.0,1059.0){\rule[-0.200pt]{0.400pt}{15.899pt}}
\put(709.0,1059.0){\rule[-0.200pt]{4.818pt}{0.400pt}}
\put(709.0,1125.0){\rule[-0.200pt]{4.818pt}{0.400pt}}
\put(791.0,1018.0){\rule[-0.200pt]{0.400pt}{23.608pt}}
\put(781.0,1018.0){\rule[-0.200pt]{4.818pt}{0.400pt}}
\put(781.0,1116.0){\rule[-0.200pt]{4.818pt}{0.400pt}}
\put(862.0,916.0){\rule[-0.200pt]{0.400pt}{50.107pt}}
\put(852.0,916.0){\rule[-0.200pt]{4.818pt}{0.400pt}}
\put(852.0,1124.0){\rule[-0.200pt]{4.818pt}{0.400pt}}
\put(934.0,1065.0){\rule[-0.200pt]{0.400pt}{89.374pt}}
\put(924.0,1065.0){\rule[-0.200pt]{4.818pt}{0.400pt}}
\put(432,1312){\circle{18}}
\put(504,1222){\circle{18}}
\put(576,1179){\circle{18}}
\put(647,1144){\circle{18}}
\put(719,1092){\circle{18}}
\put(791,1067){\circle{18}}
\put(862,1020){\circle{18}}
\put(934,1250){\circle{18}}
\put(924.0,1436.0){\rule[-0.200pt]{4.818pt}{0.400pt}}
\end{picture}
\end	{center}
\vskip 0.15in
\caption{Effective mass of the SU(4) $k=2$ string as a function 
of $n_t$ for the $16^3 40$ anisotropic lattice with the smallest 
lattice spacing. We plot the lightest string mass in the
totally symmetric $k=2S$ representation, $\bullet$,
and the first excited mass in the totally
antisymmetric $k=2A^\star$ representation, $\circ$.}
\label{fig_anisok2npn4b}
\end 	{figure}

The main puzzle is an unexpected (to us) pattern of degeneracies.  A
striking illustration is provided in Fig.\ref{fig_anisok2npn4b}, where
we plot $am_{eff}(n_t)$ for the first excited state of the $2A$ string
and compare it to the ground state of the $2S$ string (as plotted in
Fig.\ref{fig_anisok2npn4}).  We see that the two states are, within
errors, degenerate with each other. Although this degeneracy is most
convincing on an anisotropic lattice, because of the finer resolution
in $t$, we see good evidence for similar degeneracies on the isotropic
lattice calculations, as shown in Table~\ref{table_meffK2}.
Interestingly we see there some evidence that the near-degeneracy
becomes less precise with increasing $N$. When we look at the $k=3$
strings we observe analogous degeneracies.  We observe in
Tables~\ref{table_meffK3N6} and ~\ref{table_meffK3N8} that the first
excited $3A$ string has a similar mass to the ground state $3M$
string, and that the first excited $3M$ and, more approximately, the
second excited $3A$, have the same mass as the lightest $3S$
string. In all these cases these are real near-degeneracies -- they
are not the same states appearing from variational analyses using
different but overlapping sets of operators. We can verify this (see
below) by performing variational calculations with the full basis of
$k=2$ or $k=3$ operators, whereupon we find these near-degenerate
states appearing there as separate states, with approximately the same
masses.

These near-degeneracies are, of course, only significant to the extent
that the energy differences are much less than typical splittings
between excited states. To demonstrate this in the case of SU(4) we
list in Table~\ref{table_mstarN4} the values of $am_{eff}(t=a)$ for
the string ground state and the next few excited states.  We do so for
the $k=1$, $k=2$ and for the $k=2A,S$ representations, for both the
anisotropic and isotropic lattice calculations with the smallest
values of $a$. Of course $am_{eff}(a)$ is not ideal; it would be
better to use $am_{eff}(t)$ at a larger value of $t$. However in the
case of the highly excited states, values of $am_{eff}(t>a)$ are too
imprecise to be useful. So to keep the treatment uniform we use this
measure of the energy for all states. Note that our variational
calculation is based on maximising $\exp -aH$, so $am_{eff}(a)$ is
precisely the mass estimate provided by the variational
calculation. One first calculates the ground state using all
operators, then the first excited state in the basis orthogonal to
this ground state, and so on. This means that the higher excited
states are determined using an ever-smaller basis, and are less
reliable. In practice our $k=1$, $k=2A,S$ and $k=3A,S,M$ bases contain
5 or 10 operators each for the isotropic and anisotropic calculations
respectively, while the $k=2,3$ bases are 2 or 3 times as large.

Focussing first on the anisotropic lattice calculation, we see in
Table~\ref{table_mstarN4} that there is an extensive pattern of
near-degeneracies
\begin{equation}
m_{2S} \simeq m^{\star}_{2A} \qquad ;  \qquad
m^{\star}_{2S} \simeq m^{\star\star}_{2A} \qquad ;  \qquad
m^{\star\star}_{2S} \simeq m^{\star\star\star}_{2A} \qquad ;  \qquad
\hdots
\label{eqn_k2degen}
\end{equation}
These states appear as separate states in the $k=2$ spectrum (also
shown). Other than these near-degeneracies, the typical splittings in
the $k=1$, $k=2A,2S$ and $k=3A,3S,3M$ spectra, are very much larger,
$a_t\Delta E \sim 0.4$, justifying our use of the term `nearly
degenerate'. We also note some evidence that the degeneracy becomes
less exact for the higher excitations.  Finally we note that the same
pattern appears in the isotropic lattice calculation.

Analogous results for $k=2$ and $k=3$ strings in SU(6) and SU(8) are
listed in Table~\ref{table_mstarN6} and Table~\ref{table_mstarN8}. In
the $k=2$ sector we only have a clear near-degeneracy between $m_{2S}$
and $m^{\star}_{2A}$ and even that is weaker than for SU(4).  In the
$k=3$ sector we note
\begin{equation}
m_{3M} \simeq m^{\star}_{3A} \quad ;  \quad
m_{3S} \simeq m^{\star}_{3M}  \sim m^{\star\star}_{3A}
\label{eqn_k3degen}
\end{equation}
with the relation denoted by `$\sim$' being less certain.  Indeed in
SU(8) the evidence is more for $m^{\star}_{3S} \sim
m^{\star\star}_{3A}$.

It is interesting to ask what happens in SU(2) and SU(3).
Unfortunately in most of those calculations we did not include $k>1$
strings (for obvious reasons) and we cannot present results at the
same $a$ and on the same volume as for $N\geq 4$. Instead we show an
example for each of $N=2$ and $N=3$ in Table~\ref{table_mstarN23}. In
SU(2) the $k=2A$ operator is just a constant and this also renders the
$k=2$ diagonalisation singular. In the $k=2S$ sector we see
non-trivial masses, although the effective mass plateaux are often not
well defined. We note that the lightest $k=2S$ state is significantly
below the Casimir Scaling prediction. In the case of SU(3) two
fundamental fluxes can become a single (conjugate) fundamental flux
and we see this in the fact that the $k=1$ loop mass appears as the
lightest $k=2A$ (and $k=2$) state. This is also what one expects from
Casimir Scaling. We also see that just as in SU($N\geq 4$) the
lightest $k=2S$ state is nearly degenerate with the first excited
$k=2A$. Both states appear in the $k=2$ spectrum showing that they are
indeed different states.  An extra oddity in SU(3) is that the first
excited $k=1$ state has the same mass. We might expect it to appear in
the $k=2A$ spectrum just as the lightest $k=1$ state does. On the
other hand we have the same $k=2S, 2A^\star $ near-degeneracy in SU(4)
where we do not expect the $k=1$ states to appear in the $k=2$
spectrum. Despite this, in SU(4) the first excited $k=1$ state does
indeed have a mass that is not very different from that of these $k=2$
states. All this accentuates the puzzle of these spectra.

As a first step towards understanding these apparent
near-degeneracies, it might help to consider what kind of `excited'
states we might expect to find. \\ 
$\bullet$ 
The $k=2(3)$ operators should project onto scattering states of two
(three) $k=1$ loops.  Where will such states lie? We observe that the
relevant $k=2$ operator is a product of two $k=1$ lines at the same
position; thus it projects onto all relative momenta. (The total
momentum being zero of course.) On a $L_s=16$ lattice one can estimate
that the correlator of an operator that projects equally onto all
lattice momenta will lead to $m^{k=2}_{eff}(a) \sim 3$ with the
precise value depending on the lattice corrections to the
energy-momentum dispersion relation. Thus the scattering states should
appear at the upper end of the mass range we are considering. On the
other hand because the string operators are highly smeared in the
directions transverse to the string axis, one might expect the overlap
onto higher momenta to be suppressed. If this suppression were quite
radical then it might be that the spectra we are seeing could be
related to the lightest 2 or 3 string scattering states (including
excitations of the latter). To resolve this issue definitively one
would need to include explicit scattering state operators in the basis
and examine overlaps. These states would need to include a large range
of relative momenta as well as excitations of the $k=1$ strings.  Such
a calculation goes well beyond what we are attempting here. \\
$\bullet$ 
In our basis we do not include the conjugate operators which, if we
did, would lead to trivial degeneracies (even in the $k=1$
sector). However in SU(4) the $k=2$ and conjugate, $k=-2$, operators
`mix' because $z^2=z^{\star 2}, \quad \forall z \in Z_4$.  Thus all
$k=2$ excitations might form $C=+$ and $C=-$ linear combinations
leading to an approximate doubling in the $k=2$ spectrum. This might
be a factor in the more extensive degeneracy structure that we see in
SU(4). (The same argument would apply to $k=3$ in SU(6).)  On the
other hand it should also apply to the lightest $k=2A$ state for which
there is no degeneracy. So this seems unlikely.  \\ 
$\bullet$ 
We note that our operators are maximally symmetric, not only so as to
have $\vec{p}=0$, but also with respect to translations along the axis
of the loop and rotations around the axis. Nonetheless since we only
have lattice symmetries, somewhere amongst the excited states of the
strings will be states that in the continuum become states with spin
$J=4$ around the string axis.  In a simple string model, the pattern
of fluctuations of $J\not= 0$ and excited $J=0$ states is related and
perhaps leads to some of the degeneracy structure.  This is also
something to consider. \\ 
$\bullet$ 
There will also be states involving compression waves along the flux
tube, but we have no idea of the mass scale of these.

None of the above provides an immediate explanation for the pattern of
degeneracies we observe.  This makes our study of the unstable strings
much more ambiguous than that of the stable strings.

\section{Mass calculations in the $\mathrm{\bf N\to\infty}$ limit}
\label{section_massatlargeN}

We expect correlators of gauge invariant operators to factorise at
large $N$, e.g.
\be
\langle \Phi^{\dagger}(t)\Phi(0)\rangle
\stackrel{N\to\infty}{=} 
|\langle \Phi \rangle|^2 \{1+O(\frac{1}{N^2})\}
\label{eqn_factcorr}
\ee
where $\Phi$ is some typical glueball or string single-trace trial
wavenfunctional.  This means that the connected piece of the
correlator, from which we extract masses, should vanish rapidly at
large $N$:
\be
\sum_{n{\not =}\Omega} 
|\langle\Omega|\Phi^{\dagger}|n\rangle|^2 e^{-E_n t}
\stackrel{N\to\infty}{=} 
O(\frac{1}{N^2})\times {|\langle\Omega|\Phi^{\dagger}|\Omega\rangle|^2}
\label{eqn_factcorr2}
\ee
where $|\Omega\rangle$ is the vacuum state.  If this is really so,
does it mean that mass calculations become increasingly difficult as
$N\to\infty$?

To see if the factorisation in eqn(\ref{eqn_factcorr}) does indeed
occur in the range of $N$ that we study, we calculate the correlation
function
\be
C_2(t)
=
{{\langle \Phi(t)\Phi(0)\rangle}
\over
{\langle \Phi\rangle}^2}
- 1
=
{{\langle \Phi_v(t)\Phi_v(0)\rangle}
\over
{\langle \Phi\rangle}^2}
\label{eqn_factcorr3}
\ee
where
\be
\Phi_v(t)
=
\Phi(t) - \langle \Phi \rangle
\label{eqn_factcorr6}
\ee
and $\Phi$ is the thrice-blocked (super-)plaquette $0^{++}$ glueball
operator (which is real so we drop the conjugation from now
on). $C_2(t)$ is the vacuum subtracted correlator normalised by the
disconnected piece and, if calculated at fixed $a$, it should be
$\propto 1/N^2$ as $N\to\infty$. We use our calculations on $10^4$
lattices at $a\simeq 1/5T_c$, all of which use $10^5$ Monte Carlo
sweeps for the averages. Thus the calculations are as nearly identical
as possible, apart from the variation with $N$. (We have checked that
if one fixes $a$ using the string tension rather than $T_c$ it makes
no difference to our conclusions.)  We note that at this value of $a$
the above operator has a very large overlap onto the lightest scalar
glueball.

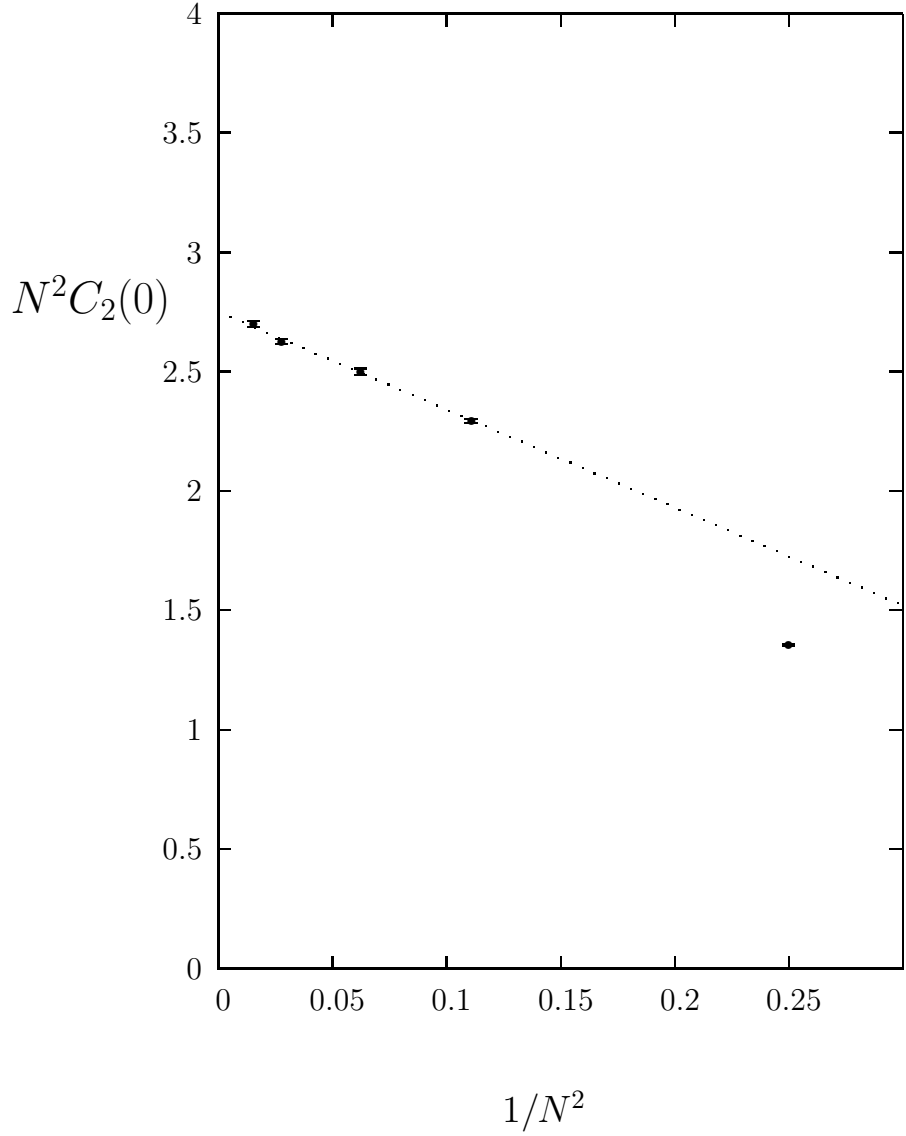
\begin	{figure}[p]
\begin	{center}
\leavevmode
\setlength{\unitlength}{0.240900pt}
\ifx\plotpoint\undefined\newsavebox{\plotpoint}\fi
\sbox{\plotpoint}{\rule[-0.200pt]{0.400pt}{0.400pt}}%
\begin{picture}(1500,1800)(0,0)
\font\gnuplot=cmr10 at 12pt
\gnuplot
\sbox{\plotpoint}{\rule[-0.200pt]{0.400pt}{0.400pt}}%
\put(350.0,250.0){\rule[-0.200pt]{4.818pt}{0.400pt}}
\put(325,250){\makebox(0,0)[r]{\ \ {$0$}}}
\put(1405.0,250.0){\rule[-0.200pt]{4.818pt}{0.400pt}}
\put(350.0,438.0){\rule[-0.200pt]{4.818pt}{0.400pt}}
\put(325,438){\makebox(0,0)[r]{\ \ {$0.5$}}}
\put(1405.0,438.0){\rule[-0.200pt]{4.818pt}{0.400pt}}
\put(350.0,625.0){\rule[-0.200pt]{4.818pt}{0.400pt}}
\put(325,625){\makebox(0,0)[r]{\ \ {$1$}}}
\put(1405.0,625.0){\rule[-0.200pt]{4.818pt}{0.400pt}}
\put(350.0,813.0){\rule[-0.200pt]{4.818pt}{0.400pt}}
\put(325,813){\makebox(0,0)[r]{\ \ {$1.5$}}}
\put(1405.0,813.0){\rule[-0.200pt]{4.818pt}{0.400pt}}
\put(350.0,1000.0){\rule[-0.200pt]{4.818pt}{0.400pt}}
\put(325,1000){\makebox(0,0)[r]{\ \ {$2$}}}
\put(1405.0,1000.0){\rule[-0.200pt]{4.818pt}{0.400pt}}
\put(350.0,1188.0){\rule[-0.200pt]{4.818pt}{0.400pt}}
\put(325,1188){\makebox(0,0)[r]{\ \ {$2.5$}}}
\put(1405.0,1188.0){\rule[-0.200pt]{4.818pt}{0.400pt}}
\put(350.0,1375.0){\rule[-0.200pt]{4.818pt}{0.400pt}}
\put(325,1375){\makebox(0,0)[r]{\ \ {$3$}}}
\put(1405.0,1375.0){\rule[-0.200pt]{4.818pt}{0.400pt}}
\put(350.0,1563.0){\rule[-0.200pt]{4.818pt}{0.400pt}}
\put(325,1563){\makebox(0,0)[r]{\ \ {$3.5$}}}
\put(1405.0,1563.0){\rule[-0.200pt]{4.818pt}{0.400pt}}
\put(350.0,1750.0){\rule[-0.200pt]{4.818pt}{0.400pt}}
\put(325,1750){\makebox(0,0)[r]{\ \ {$4$}}}
\put(1405.0,1750.0){\rule[-0.200pt]{4.818pt}{0.400pt}}
\put(350.0,250.0){\rule[-0.200pt]{0.400pt}{4.818pt}}
\put(350,200){\makebox(0,0){\ {$0$}}}
\put(350.0,1730.0){\rule[-0.200pt]{0.400pt}{4.818pt}}
\put(529.0,250.0){\rule[-0.200pt]{0.400pt}{4.818pt}}
\put(529,200){\makebox(0,0){\ {$0.05$}}}
\put(529.0,1730.0){\rule[-0.200pt]{0.400pt}{4.818pt}}
\put(708.0,250.0){\rule[-0.200pt]{0.400pt}{4.818pt}}
\put(708,200){\makebox(0,0){\ {$0.1$}}}
\put(708.0,1730.0){\rule[-0.200pt]{0.400pt}{4.818pt}}
\put(888.0,250.0){\rule[-0.200pt]{0.400pt}{4.818pt}}
\put(888,200){\makebox(0,0){\ {$0.15$}}}
\put(888.0,1730.0){\rule[-0.200pt]{0.400pt}{4.818pt}}
\put(1067.0,250.0){\rule[-0.200pt]{0.400pt}{4.818pt}}
\put(1067,200){\makebox(0,0){\ {$0.2$}}}
\put(1067.0,1730.0){\rule[-0.200pt]{0.400pt}{4.818pt}}
\put(1246.0,250.0){\rule[-0.200pt]{0.400pt}{4.818pt}}
\put(1246,200){\makebox(0,0){\ {$0.25$}}}
\put(1246.0,1730.0){\rule[-0.200pt]{0.400pt}{4.818pt}}
\put(350.0,250.0){\rule[-0.200pt]{258.967pt}{0.400pt}}
\put(1425.0,250.0){\rule[-0.200pt]{0.400pt}{361.350pt}}
\put(350.0,1750.0){\rule[-0.200pt]{258.967pt}{0.400pt}}
\put(150,1300){\makebox(0,0){\Large{$N^2C_2(0)$}}}
\put(862,25){\makebox(0,0){\large{$1/N^2$}}}
\put(350.0,250.0){\rule[-0.200pt]{0.400pt}{361.350pt}}
\put(1246.0,757.0){\rule[-0.200pt]{0.400pt}{0.723pt}}
\put(1236.0,757.0){\rule[-0.200pt]{4.818pt}{0.400pt}}
\put(1236.0,760.0){\rule[-0.200pt]{4.818pt}{0.400pt}}
\put(748.0,1107.0){\rule[-0.200pt]{0.400pt}{1.686pt}}
\put(738.0,1107.0){\rule[-0.200pt]{4.818pt}{0.400pt}}
\put(738.0,1114.0){\rule[-0.200pt]{4.818pt}{0.400pt}}
\put(574.0,1183.0){\rule[-0.200pt]{0.400pt}{2.409pt}}
\put(564.0,1183.0){\rule[-0.200pt]{4.818pt}{0.400pt}}
\put(564.0,1193.0){\rule[-0.200pt]{4.818pt}{0.400pt}}
\put(450.0,1231.0){\rule[-0.200pt]{0.400pt}{1.927pt}}
\put(440.0,1231.0){\rule[-0.200pt]{4.818pt}{0.400pt}}
\put(440.0,1239.0){\rule[-0.200pt]{4.818pt}{0.400pt}}
\put(406.0,1258.0){\rule[-0.200pt]{0.400pt}{2.168pt}}
\put(396.0,1258.0){\rule[-0.200pt]{4.818pt}{0.400pt}}
\put(1246,759){\circle*{12}}
\put(748,1111){\circle*{12}}
\put(574,1188){\circle*{12}}
\put(450,1235){\circle*{12}}
\put(406,1262){\circle*{12}}
\put(396.0,1267.0){\rule[-0.200pt]{4.818pt}{0.400pt}}
\put(350,1282){\usebox{\plotpoint}}
\put(350.00,1282.00){\usebox{\plotpoint}}
\put(368.90,1273.41){\usebox{\plotpoint}}
\put(388.04,1265.48){\usebox{\plotpoint}}
\put(406.94,1256.93){\usebox{\plotpoint}}
\put(426.09,1248.96){\usebox{\plotpoint}}
\put(445.24,1241.00){\usebox{\plotpoint}}
\put(464.13,1232.43){\usebox{\plotpoint}}
\put(483.29,1224.51){\usebox{\plotpoint}}
\put(502.19,1215.93){\usebox{\plotpoint}}
\put(521.42,1208.17){\usebox{\plotpoint}}
\put(540.42,1199.83){\usebox{\plotpoint}}
\put(559.41,1191.45){\usebox{\plotpoint}}
\put(578.65,1183.71){\usebox{\plotpoint}}
\put(597.54,1175.12){\usebox{\plotpoint}}
\put(616.68,1167.16){\usebox{\plotpoint}}
\put(635.61,1158.69){\usebox{\plotpoint}}
\put(654.74,1150.66){\usebox{\plotpoint}}
\put(673.91,1142.76){\usebox{\plotpoint}}
\put(692.77,1134.12){\usebox{\plotpoint}}
\put(711.93,1126.21){\usebox{\plotpoint}}
\put(731.17,1118.47){\usebox{\plotpoint}}
\put(750.07,1109.88){\usebox{\plotpoint}}
\put(769.20,1101.90){\usebox{\plotpoint}}
\put(788.15,1093.49){\usebox{\plotpoint}}
\put(807.26,1085.43){\usebox{\plotpoint}}
\put(826.46,1077.56){\usebox{\plotpoint}}
\put(845.29,1068.86){\usebox{\plotpoint}}
\put(864.46,1060.97){\usebox{\plotpoint}}
\put(883.40,1052.49){\usebox{\plotpoint}}
\put(902.59,1044.64){\usebox{\plotpoint}}
\put(921.70,1036.56){\usebox{\plotpoint}}
\put(940.55,1027.93){\usebox{\plotpoint}}
\put(959.79,1020.19){\usebox{\plotpoint}}
\put(978.69,1011.60){\usebox{\plotpoint}}
\put(997.93,1003.85){\usebox{\plotpoint}}
\put(1016.79,995.26){\usebox{\plotpoint}}
\put(1035.88,987.14){\usebox{\plotpoint}}
\put(1055.09,979.33){\usebox{\plotpoint}}
\put(1074.02,970.81){\usebox{\plotpoint}}
\put(1093.11,962.68){\usebox{\plotpoint}}
\put(1112.07,954.25){\usebox{\plotpoint}}
\put(1131.24,946.34){\usebox{\plotpoint}}
\put(1150.37,938.32){\usebox{\plotpoint}}
\put(1169.20,929.64){\usebox{\plotpoint}}
\put(1188.44,921.89){\usebox{\plotpoint}}
\put(1207.33,913.30){\usebox{\plotpoint}}
\put(1226.57,905.56){\usebox{\plotpoint}}
\put(1245.46,897.01){\usebox{\plotpoint}}
\put(1264.53,888.85){\usebox{\plotpoint}}
\put(1283.76,881.09){\usebox{\plotpoint}}
\put(1302.66,872.52){\usebox{\plotpoint}}
\put(1321.75,864.38){\usebox{\plotpoint}}
\put(1340.73,856.01){\usebox{\plotpoint}}
\put(1359.89,848.05){\usebox{\plotpoint}}
\put(1379.03,840.08){\usebox{\plotpoint}}
\put(1397.84,831.34){\usebox{\plotpoint}}
\put(1417.08,823.60){\usebox{\plotpoint}}
\put(1425,820){\usebox{\plotpoint}}
\end{picture}
\end	{center}
\vskip 0.15in
\caption{The $N$-dependence of the two point correlation function,
$C_2(t)$, defined in eqn(\ref{eqn_factcorr3}). The calculations are 
on $10^4$ lattices at fixed lattice spacing, $a \simeq 1/5T_c$.} 
\label{fig_Ncor}
\end 	{figure}

We plot the values of $C_2(t=0)$ multiplied by $N^2$ in
Fig.\ref{fig_Ncor}. We see that for $N\geq 3$ we obtain a good fit to
the form
\be
N^2 C_2(t=0)
=
2.751(10) - \frac{4.11(17)}{N^2}.
\label{eqn_factcorr4}
\ee
This demonstrates that the correlators we use for our mass
calculations do indeed show the expected suppression with increasing
$N$, even at the smallest values of $N$. That is to say our `measured
signal' -- the connected piece of the two-point correlator when
normalised to the disconnected piece -- does indeed vanish rapidly
with increasing $N$.

How does all this affect the accuracy of our mass calculations?  This
will be determined by how the fluctuations of $C_2(t=0)$ vary with
$N$. We calculate these from
\begin{eqnarray}
\sigma^2_c(t)
& = &
\langle \{\Phi_v(t)\Phi_v(0)
- \langle \Phi_v(t)\Phi_v(0) \rangle\}^2 \rangle \nonumber \\
& = &
\langle \Phi_v(t)\Phi_v(0)\Phi_v(t)\Phi_v(0) \rangle
-
\langle \Phi_v(t)\Phi_v(0) \rangle^2
\label{eqn_factcorr5}
\end{eqnarray}
and the $N$-dependence of the fluctations on the correlator $C_2(t=0)$
will be given by
\be
\sigma[C_2(t)]
=
\frac{\sigma_c(t)}
{{\langle \Phi\rangle}^2}.
\label{eqn_factcorr7}
\ee
We plot the ratio of this quantity to $C_2(t=0)$ in
Fig.\ref{fig_Nfluctcor}. We see that this ratio becomes rapidly
independent of $N$ as $N$ increases. That is to say, the fluctuations
of the connected 2-point correlator decrease with $N$ in the same way
as the correlator itself and masses can, in principle, be extracted
with more-or-less the same precision at all $N$.

What we observe numerically is in fact what one expects from the usual
large-$N$ counting. The 4-point correlator in eqn(\ref{eqn_factcorr5})
contains (despite first appearances) $\langle \Phi_v \Phi_v \rangle
\langle \Phi_v \Phi_v \rangle$ disconnected pieces which will dominate
the correlator at large $N$.  Thus $\sigma_c(t)$ will have the same
$N$-dependence as our connected two point correlator, $\langle
\Phi_v(t) \Phi_v(0) \rangle$, and our above results are in accordance
with this expectation.

As an aside we recall (rearranging the fields) that
\begin{eqnarray}
\sigma^2_c(t)
& = &
\langle \Phi_v(t)\Phi_v(t)\Phi_v(0)\Phi_v(0) \rangle
-
\langle \Phi_v(t)\Phi_v(0) \rangle^2 \nonumber \\
& = &
\sum_{n} |\langle \Omega | \Phi_v\Phi_v |n \rangle|^2 e^{-E_n t}
- 
\{\sum_{n} |\langle\Omega | \Phi_v  |n \rangle|^2 e^{-E_n t}\}^2 \nonumber \\
& \stackrel{t\to\infty}{=} &
|\langle \Omega | \Phi_v\Phi_v | \Omega \rangle|^2.
\label{eqn_factcorr8}
\end{eqnarray}
That is to say, the errors become independent of $t$ at larger $t$, so
that the error-to-signal ratio grows $\propto \exp{+mt}$ if $m$ is the
mass we are trying to calculate. Moreover as $N$ decreases this
behaviour will set in at smaller $t$ since the disconnected piece
becomes increasingly dominant. As is well known, for hadrons that are
not flavour singlets the correlator explicitly excludes such a
disconnected piece, and the $t$-dependence of the error-to-signal
ratio is much more favourable.

\begin	{figure}[p]
\begin	{center}
\leavevmode
\setlength{\unitlength}{0.240900pt}
\ifx\plotpoint\undefined\newsavebox{\plotpoint}\fi
\sbox{\plotpoint}{\rule[-0.200pt]{0.400pt}{0.400pt}}%
\begin{picture}(1500,1800)(0,0)
\font\gnuplot=cmr10 at 12pt
\gnuplot
\sbox{\plotpoint}{\rule[-0.200pt]{0.400pt}{0.400pt}}%
\put(400.0,250.0){\rule[-0.200pt]{4.818pt}{0.400pt}}
\put(375,250){\makebox(0,0)[r]{\ \ {$0$}}}
\put(1405.0,250.0){\rule[-0.200pt]{4.818pt}{0.400pt}}
\put(400.0,591.0){\rule[-0.200pt]{4.818pt}{0.400pt}}
\put(375,591){\makebox(0,0)[r]{\ \ {$0.005$}}}
\put(1405.0,591.0){\rule[-0.200pt]{4.818pt}{0.400pt}}
\put(400.0,932.0){\rule[-0.200pt]{4.818pt}{0.400pt}}
\put(375,932){\makebox(0,0)[r]{\ \ {$0.01$}}}
\put(1405.0,932.0){\rule[-0.200pt]{4.818pt}{0.400pt}}
\put(400.0,1273.0){\rule[-0.200pt]{4.818pt}{0.400pt}}
\put(375,1273){\makebox(0,0)[r]{\ \ {$0.015$}}}
\put(1405.0,1273.0){\rule[-0.200pt]{4.818pt}{0.400pt}}
\put(400.0,1614.0){\rule[-0.200pt]{4.818pt}{0.400pt}}
\put(375,1614){\makebox(0,0)[r]{\ \ {$0.02$}}}
\put(1405.0,1614.0){\rule[-0.200pt]{4.818pt}{0.400pt}}
\put(400.0,250.0){\rule[-0.200pt]{0.400pt}{4.818pt}}
\put(400,200){\makebox(0,0){\ {$0$}}}
\put(400.0,1730.0){\rule[-0.200pt]{0.400pt}{4.818pt}}
\put(605.0,250.0){\rule[-0.200pt]{0.400pt}{4.818pt}}
\put(605,200){\makebox(0,0){\ {$2$}}}
\put(605.0,1730.0){\rule[-0.200pt]{0.400pt}{4.818pt}}
\put(810.0,250.0){\rule[-0.200pt]{0.400pt}{4.818pt}}
\put(810,200){\makebox(0,0){\ {$4$}}}
\put(810.0,1730.0){\rule[-0.200pt]{0.400pt}{4.818pt}}
\put(1015.0,250.0){\rule[-0.200pt]{0.400pt}{4.818pt}}
\put(1015,200){\makebox(0,0){\ {$6$}}}
\put(1015.0,1730.0){\rule[-0.200pt]{0.400pt}{4.818pt}}
\put(1220.0,250.0){\rule[-0.200pt]{0.400pt}{4.818pt}}
\put(1220,200){\makebox(0,0){\ {$8$}}}
\put(1220.0,1730.0){\rule[-0.200pt]{0.400pt}{4.818pt}}
\put(1425.0,250.0){\rule[-0.200pt]{0.400pt}{4.818pt}}
\put(1425,200){\makebox(0,0){\ {$10$}}}
\put(1425.0,1730.0){\rule[-0.200pt]{0.400pt}{4.818pt}}
\put(400.0,250.0){\rule[-0.200pt]{246.922pt}{0.400pt}}
\put(1425.0,250.0){\rule[-0.200pt]{0.400pt}{361.350pt}}
\put(400.0,1750.0){\rule[-0.200pt]{246.922pt}{0.400pt}}
\put(200,1425){\makebox(0,0){\Large{$\frac{error}{signal}$}}}
\put(887,25){\makebox(0,0){\large{$N$}}}
\put(400.0,250.0){\rule[-0.200pt]{0.400pt}{361.350pt}}
\put(605,502){\circle*{12}}
\put(708,550){\circle*{12}}
\put(810,598){\circle*{12}}
\put(1015,509){\circle*{12}}
\put(1220,557){\circle*{12}}
\put(605,1068){\circle{18}}
\put(708,1477){\circle{18}}
\put(810,1273){\circle{18}}
\put(1015,1136){\circle{18}}
\put(1220,1409){\circle{18}}
\end{picture}
\end	{center}
\vskip 0.15in
\caption{The $N$-dependence of the error to signal ratio for the 
two point correlation function,  $C_2(t)$, defined in 
eqns(\ref{eqn_factcorr3} - \ref{eqn_factcorr8}) after $10^5$ sweeps
on $10^4$ lattices at fixed lattice spacing, $a \simeq 1/5T_c$.
For $t=0$ ($\bullet$) and $t=a$ ($\circ$).} 
\label{fig_Nercor}
\end 	{figure}
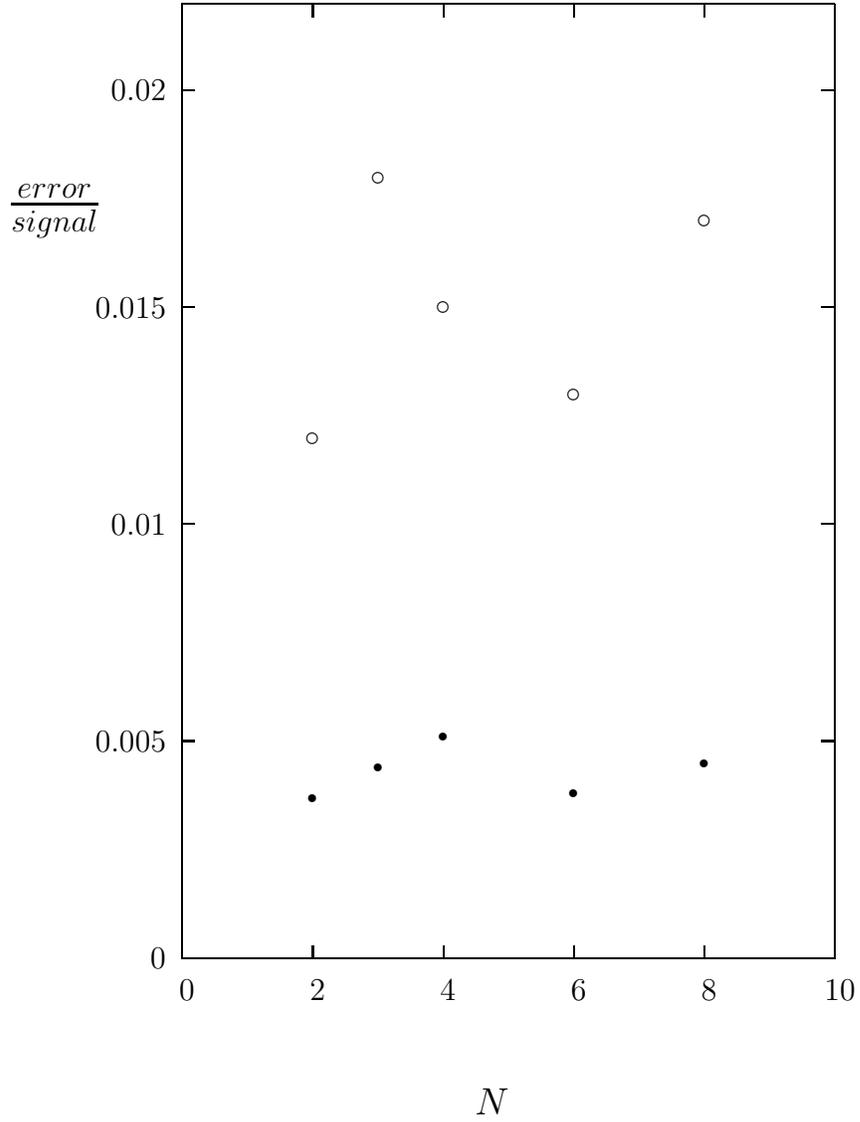

We have seen that the correlator and its fluctuations decrease in the
same way with increasing $N$ so that in principle the error on the
`same' Monte Carlo calculation can be the same at all $N$.  Do our
calculations achieve this ideal in practice? To answer this question
we plot in Fig.\ref{fig_Nercor} the error-to-signal ratio on our
connected 2-point correlator at both $t=0$ and $t=a$ as a function of
$N$. Recall that the same number of lattice fields are used in each
calculation. We observe that this ratio is independent of $N$ up to
fluctuations that are presumably within the error on the errors (which
are not shown because they have not been calculated). Thus large-$N$
factorisation does not make mass calculations any more difficult.

\section{Discussion}
\label{section_discuss}

In this paper we introduced some improved algorithms for constructing
operators with very good overlaps onto the physical states of SU($N$)
gauge theories. We performed a variety of tests and comparisons and
identified a range of parameters that work well.

Using these techniques we performed calculations of the lightest
glueballs and $k$-strings for $N=2,3,4,6,8$.  The glueball results are
consistent with earlier work, although with improved accuracy and a
greater lever arm in $N$. Our accuracy has improved sufficiently that
we are beginning to find that the SU(2) masses cannot be described by
just a $1/N^2$ correction to $N=\infty$, even though the corrections
are small. We confirm that SU(3) $\simeq$ SU($\infty$).

The focus of our calculations was on $k$-string tensions.  We
emphasised the main systematic errors. The first simply arises from
the fact that the mass of the string increases with $k$ and the
natural systematic bias is for its mass to be extracted at too small a
value of $t$ and so to be increasingly overestimated for larger $k$
and larger $N$. The step we made towards solving this problem was to
construct operators with very good overlaps onto the lightest
$k$-strings. Comparing our results with earlier SU(4) and SU(6)
calculations we saw that our calculations improved significantly upon
that earlier work.  (The criterion of `improvement' is unambiguous
because of the positivity of the lattice action.) In particular it is
clear that the string tension ratios are below the predictions of
`MQCD'. Our current calculations suggest that they are above `Casimir
Scaling' but it is possible that more precise calculations will alter
this conclusion.

The main caveat on our $k$-string calculation concerns the leading
string correction that we (and others) use. Previous work suggests
that it is very accurate for spatial sizes $l\sqrt{\sigma} \geq 3$,
but this has not been investigated with great accuracy for larger $k$
(and $N$). We showed that a natural physical picture suggested that
this would be an increasing underestimate as $N$ grows.  This is an
important source of potential systematic error which could, in
principle, lift the ratios up to the `MQCD' values. Although this
seems unlikely for our range of $N$, this judgement is based on
simple-minded estimates that might be incorrect. This is a problem
that needs to be addressed.

We also attempted to find evidence for other, unstable, strings. Using
a calculation with an anisotropic lattice, $a_t \ll a_s$, we found
significant evidence for a string of this type. This calculation adds
to previous work on this issue, in that our string winds around the
torus and so is without the sources which might otherwise confuse the
issue with their Coulomb interaction. We found in addition an
intriguing pattern of near-degeneracies for which we could produce no
simple explanation. All this needs to be better understood -- in
particular the role of multi-string scattering states -- before one
can claim support for one theoretical picture or another.

\begin	{figure}[p]
\begin	{center}
\leavevmode
\input	{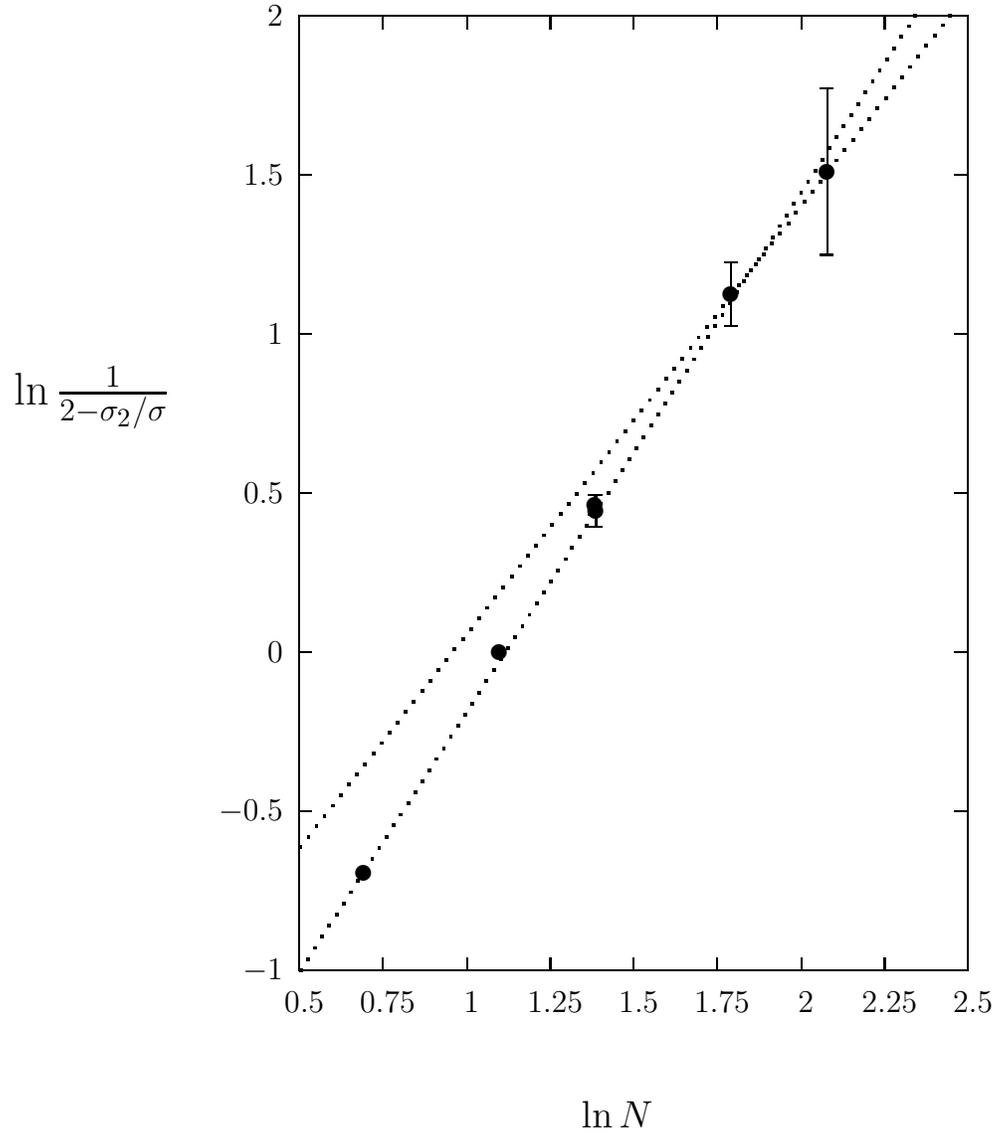}
\end	{center}
\vskip 0.15in
\caption{The calculated ratio $\sigma_{k=2}/\sigma$ for 
$N=4,6,8$, plotted so that the slope gives the power, $\alpha$, 
of the correction to the $N=\infty$ limit if that is parameterised 
as $1/N^\alpha$. (We also show the `trivial' values for $N=2,3$.) 
The best fits for  $N\geq 4$ and  $N\geq 6$ are shown.}
\label{fig_npower}
\end 	{figure}

Although the `MQCD' and `Casimir Scaling' formulae have some
theoretical motivation, they are also very interesting in that while
the former has the conventional $O(1/N^2)$ correction to the large-$N$
limit, Casimir Scaling has an unconventional $O(1/N)$ correction. We
recall that the argument for $O(1/N^2)$ is from diagrams to all
orders, and one might speculate that the result might be different for
completely non-perturbative objects like strings. Whether this makes
the question interesting (see e.g.
\cite{mt-tempe})
or tells us that `Casimir Scaling' is excluded (see e.g.
\cite{shifman})
is controversial. In any case it is interesting to see if our
calculation tells us something about the correction in a
model-independent way. To investigate this we take our $k=2$ string
results, which are reasonably accurate for all $N$, and we
parameterise them as
\begin{equation}
2-\frac{\sigma_k}{\sigma} = \frac{c}{N^\alpha}.
\label{eqn_npower}
\end{equation}
In Fig.\ref{fig_npower} we plot our results for this on a logarithmic
plot, and we see that the best fits give
\begin{equation}
\alpha = \begin{cases}
1.82 \pm 0.35 & : \quad N\geq 4 \\
1.36 \pm 0.98 & : \quad N\geq 6 \\
\end{cases}
\label{eqn_npower2}
\end{equation}
It is clear that to have something really useful on this question we
require calculations that are an order of magnitude more precise.

Finally we asked whether the vanishing of fluctations in the
$N=\infty$ limit presents an obstruction to the calculation of masses
at large $N$ -- given that we calculate masses from the correlation
between such fluctuations.  We saw that the connected 2-point
correlators we use for our mass calculations do in fact show the
expected rapid suppression with increasing $N$, even at small
$N$. However we also saw that the fluctuation of such a connected
2-point correlator about its average decreases with $N$ in the same
way, so that the `error-to-signal' ratio is more-or-less independent
of $N$.  We conclude that correlators encode masses with much the same
precision at all values of $N$, even if at $N=\infty$ there are in
fact no fluctuations to correlate.

\section*{Acknowledgements}

We have had useful conversations with many colleagues, including Adi
Armoni, Chris Korthals-Altes and Harvey Meyer.  Our lattice
calculations were carried out on PPARC and EPSRC funded Alpha Compaq
workstations in Oxford Theoretical Physics, and on a desktop funded by
All Souls College.  During the course of most of this research, UW was
supported by a PPARC SPG fellowship, and BL by a EU Marie
Sk{\l}odowska-Curie postdoctoral fellowship.

\vfill\eject

\begin{table}
\begin{center}
\begin{tabular}{|cc||ccc|c|}\hline
\multicolumn{6}{|c|}{$am_{eff}(n_t)$} \\ \hline
state & $n_t$ & $Q\simeq 0$ & $|Q|=1$ & $|Q|\simeq 2$ & $all$ \\ \hline
$G_{0^{++}}$  & 1 &  0.81(3) & 0.81(1) & 0.82(3) & 0.80(1) \\
              & 2 &  0.71(5) & 0.78(3) & 0.79(6) & 0.76(2) \\ \hline
$l_{k=1}$ & 1 &  0.75(1) & 0.75(1) & 0.76(1) & 0.75(1) \\
          & 2 &  0.71(2) & 0.71(1) & 0.76(2) & 0.72(1) \\ \hline
$l_{k=2}$ & 1 &  1.43(2) & 1.41(1) & 1.41(2)  & 1.42(1) \\
          & 2 &  1.39(7) & 1.32(3) & 1.38(12) & 1.34(3)  \\ \hline
\end{tabular}
\caption{\label{table_meffQ}
Effective masses at $t=n_t a$ for the lightest $0^{++}$ glueball
and the lightest $k=1$ and $k=2$ winding strings, on a $12^4$
lattice at $\beta=44.85$ in SU(8). The masses have been 
calculated separately for sequences of fields with the
indicated topological charges.} 
\end{center}
\end{table}

\begin{table}
\begin{center}
\begin{tabular}{|cc|c||cc|c||cc|c||cc|c|}\hline
\multicolumn{12}{|c|}{blocking + smearing : SU(3) at $\beta=5.9$} \\ \hline
$b$ & $s$ & $m(a)$ & $b$ & $s$ & $m(a)$ &  $b$ & $s$ & $m(a)$ &  $b$ & $s$ & $m(a)$ \\ \hline
0 & 0  & 2.28 & 1 & 0  & 1.40 & 2 & 0  & 0.99 & 3 & 0  & 1.05 \\
0 & 2  & 1.72 & 1 & 2  & 1.15 & 2 & 2  & 0.92 & 3 & 2  & 1.09 \\
0 & 4  & 1.49 & 1 & 4  & 1.04 & 2 & 4  & 0.91 & 3 & 4  & 1.12 \\
0 & 6  & 1.35 & 1 & 6  & 0.97 & 2 & 6  & 0.93 & 3 & 6  & 1.14 \\
0 & 8  & 1.25 & 1 & 8  & 0.94 & 2 & 8  & 0.96 & 3 & 8  & 1.15 \\ 
0 & 10 & 1.19 & 1 & 10 & 0.93 & 2 & 10 & 1.00 & 3 & 10 & 1.16 \\
0 & 12 & 1.14 & 1 & 12 & 0.93 & 2 & 12 & 1.04 & 3 & 12 & 1.17 \\
0 & 14 & 1.10 & 1 & 14 & 0.93 & 2 & 14 & 1.09 & 3 & 14 & 1.17 \\ \hline
0 & 0  & 2.26 & 1 & 0  & 1.38 & 2 & 0  & 0.95 & 3 & 0  & 1.07 \\ \hline
\end{tabular}
\caption{\label{table_blsmear}
Effective $0^{++}$ glueball masses at $t=a$ for operators that
are $1\times 1$ loops of (super)links blocked $b$ times and
then smeared $s$ times. Parameters $p_a=1.0$ for blocking and
$p_a=0.10$ for smearing. Last row is from a separate calculation
with improved blocking with $p_a=0.40$ and $p_d=0.16$.
The calculation is for SU(3) on a $12^4$ lattice at $\beta=5.9$.}  
\end{center}
\end{table}

\begin{table}
\begin{center}
\begin{tabular}{|c|c||c|c|c|c|c|}\hline
\multicolumn{7}{|c|}{Improved blocking : SU(4) at $\beta=10.9$} \\ \hline
\multicolumn{2}{|c||}{parameters} & 
\multicolumn{5}{|c|}{lightest effective masses at $t=a$ :}  \\ \hline
$p_a$ & $p_d$ & $l_{k=1}(p=0)$ & $l_{k=1}(p=1)$ & $l_{k=2}(p=0)$ & 
$G_{0^{++}}$ & $G_{2^{++}}$ \\ \hline
0.15  & 0.03 & 0.707(16)  & 1.147(7) & 1.068(34) & 0.915(17) & 1.395(25) \\  
0.20  & 0.04 & 0.659(15)  & 1.076(6) & 0.973(30) & 0.820(17) & 1.249(23) \\  
0.25  & 0.05 & 0.639(15)  & 1.048(6) & 0.936(32) & 0.780(19) & 1.188(22) \\  
0.30  & 0.06 & 0.630(15)  & 1.036(6) & 0.919(31) & 0.762(19) & 1.163(22) \\  
0.35  & 0.07 & 0.626(15)  & 1.031(6) & 0.909(31) & 0.754(19) & 1.153(22) \\  
0.40  & 0.08 & 0.623(15)  & 1.029(6) & 0.903(31) & 0.749(19) & 1.149(22) \\  
0.50  & 0.10 & 0.620(15)  & 1.027(6) & 0.897(31) & 0.744(19) & 1.149(22) \\  
0.60  & 0.12 & 0.619(15)  & 1.027(6) & 0.894(31) & 0.742(19) & 1.150(22) \\  
0.75  & 0.15 & 0.619(15)  & 1.029(6) & 0.892(31) & 0.741(19) & 1.154(22) \\  \hline 
0.15  & 0.06 & 0.632(9)   & 1.029(7)  & 0.937(20) & 0.830(23) & 1.258(37) \\ 
0.20  & 0.08 & 0.616(9)   & 1.010(7)  & 0.904(20) & 0.796(23) & 1.196(40) \\ 
0.25  & 0.10 & 0.611(10)  & 1.006(8)  & 0.893(20) & 0.786(23) & 1.180(39) \\ 
0.30  & 0.12 & 0.609(10)  & 1.006(8)  & 0.888(19) & 0.782(23) & 1.176(37) \\ 
0.35  & 0.14 & 0.608(10)  & 1.007(8)  & 0.886(19) & 0.781(23) & 1.175(34) \\ 
0.40  & 0.16 & 0.608(10)  & 1.008(8)  & 0.885(19) & 0.780(23) & 1.175(32) \\ 
0.50  & 0.20 & 0.608(10)  & 1.011(8)  & 0.884(19) & 0.779(23) & 1.176(30) \\ 
0.60  & 0.24 & 0.608(10)  & 1.013(8)  & 0.883(19) & 0.779(23) & 1.177(29) \\ 
0.75  & 0.30 & 0.608(10)  & 1.015(8)  & 0.883(19) & 0.780(23) & 1.178(28) \\  \hline
0.25  & 0.15 & 0.608(10)  & 1.009(8)  & 0.884(19) & 0.781(23) & 1.175(34) \\ 
0.35  & 0.21 & 0.608(10)  & 1.012(8)  & 0.882(19) & 0.780(22) & 1.175(30) \\ 
0.50  & 0.30 & 0.609(10)  & 1.016(8)  & 0.883(19) & 0.780(22) & 1.177(28) \\ 
0.75  & 0.45 & 0.610(10)  & 1.022(8)  & 0.884(19) & 0.781(22) & 1.179(28) \\  \hline
\end{tabular}
\caption{\label{table_improvedbl}
How improved blocking varies with parameters. 
Effective masses at $t=a$ of periodic $k=1$ and $k=2$ strings, and
the  $t=a$ energy of the $k=1$ string with lowest non-zero momentum.
Also the lightest scalar and tensor glueballs. Calculations
are in SU(4) on a $12^4$ lattice at $\beta=10.9$. From the
best operators formed out of (super)links using improved blocking 
and the various blocking parameters shown.}
\end{center}
\end{table}


\begin{table}
\begin{center}
\begin{tabular}{|c|c|c|c||c|c|c|c|}\hline
\multicolumn{8}{|c|}{Best Overlaps in SU(4)} \\ \hline
\multicolumn{4}{|c||}{Old blocking \cite{blmt-kstring}} & 
\multicolumn{4}{|c|}{Improved blocking}  \\ \hline
$\beta$ & $l_{k=1}$ &  $l_{k=2}$ & $G_{0^{++}}$ &
$\beta$ & $l_{k=1}$ &  $l_{k=2}$ & $G_{0^{++}}$ \\ \hline
10.55 & 0.69 & 0.59 & 0.90 & 10.55  & 0.968 & 0.945 & 0.963 \\  
10.70 & 0.81 & 0.67 & 0.89 & 10.70  & 0.967 & 0.937 & 0.973 \\
10.90 & 0.84 & 0.71 & 0.76 & 10.87  & 0.975 & 0.951 & 0.963 \\
11.10 & 0.77 & 0.59 & 0.84 & 11.085 & 0.965 & 0.941 & 0.963 \\
11.30 & 0.79 & 0.62 & 0.86 & 11.40  & 0.981 & 0.962 & 0.965 \\ \hline
\end{tabular}
\caption{\label{table_ImpBl}
Best overlaps for the lightest $k=1$ and $k=2$ loops and the $0^{++}$
glueball. From the old blocking calculation of \cite{blmt-kstring}
and from the improved blocking calculation of this paper.}
\end{center}
\end{table}

\begin{table}
\begin{center}
\begin{tabular}{|c|c|c|c|c|c|}\hline
\multicolumn{6}{|c|}{SU(2)}  \\ \hline
$\beta$ & $L$ & $a\surd\sigma$ & $am_{0^{++}}$ & $am_{0^{++\star}}$ & $am_{2^{++}}$ \\ \hline
2.1768  & 8  & 0.5149(77) & 1.626(57) & 2.65(48) & 2.73(45) \\
2.2986  & 10 & 0.3667(18) & 1.212(22) & 2.00(11) & 1.95(9) \\
2.3715  & 12 & 0.2879(13) & 0.994(18) & 1.64(7)  & 1.56(5) \\
2.3726  & 12 & 0.2879(10) & 1.015(13) & 1.60(5)  & 1.562(29) \\
2.4265  & 16 & 0.2388(9)  & 0.834(13) & 1.47(4)  & 1.340(27) \\
2.5115  & 20 & 0.1768(8)  & 0.658(10) & 1.100(17) & 0.954(13) \\ \hline
\end{tabular}
\caption{\label{table_su2GK}
The string tension, the lightest and first excited $J^{PC}=0^{++}$
glueball masses and the lightest $2^{++}$ glueball mass: calculated 
at the indicated values of $\beta$ on $L^4$ lattices in SU(2).}
\end{center}
\end{table}

\begin{table}
\begin{center}
\begin{tabular}{|c|c|c|c|c|c|}\hline
\multicolumn{6}{|c|}{SU(3)}  \\ \hline
$\beta$ & $L$ & $a\surd\sigma$ & $am_{0^{++}}$ & $am_{0^{++\star}}$ & $am_{2^{++}}$ \\ \hline
5.6925 & 8  & 0.3970(19)  & 0.941(25) & 1.99(12)  & 2.01(14) \\
5.6993 & 8  & 0.3933(16)  & 0.969(18) & 2.12(14)  & 1.97(12) \\
5.7995 & 10 & 0.3148(12)  & 0.909(15) & 1.65(6)   & 1.52(5)  \\
5.8000 & 10 & 0.3133(13)  & 0.945(21) & 1.58(8)   & 1.57(6) \\
5.8945 & 12 & 0.2607(11)  & 0.795(13) & 1.42(4)   & 1.279(22) \\
6.0625 & 16 & 0.19466(73) & 0.648(11) & 1.073(16) & 0.922(13) \\
6.3380 & 24 & 0.12930(69) & 0.448(11) & 0.731(13) & 0.636(12) \\ \hline
 \end{tabular}
\caption{\label{table_su3GK}
The string tension, the lightest and first excited $J^{PC}=0^{++}$
glueball masses and the lightest $2^{++}$ glueball mass: calculated 
at the indicated values of $\beta$ on $L^4$ lattices in SU(3).}
\end{center}
\end{table}

\begin{table}
\begin{center}
\begin{tabular}{|c|c|c|c|c|c|}\hline
\multicolumn{6}{|c|}{SU(4)}  \\ \hline
$\beta$ & $L$ & $a\surd\sigma$ & $am_{0^{++}}$ & $am_{0^{++\star}}$ & $am_{2^{++}}$ \\ \hline
10.550 &  8 & 0.3739(15)  & 0.819(21) & 1.626(73) & 1.66(7) \\
10.637 & 10 & 0.3254(6)   & 0.864(9)  & 1.652(44) & 1.575(40) \\
10.700 & 10 & 0.2977(13)  & 0.839(17) & 1.425(56) & 1.427(50) \\
10.789 & 12 & 0.2706(8)   & 0.788(12) & 1.440(35) & 1.305(35) \\
10.870 & 12 & 0.2467(11)  & 0.746(16) & 1.357(33) & 1.185(28) \\
11.085 & 16 & 0.19868(83) & 0.618(13) & 1.093(19) & 0.981(19) \\
11.400 & 20 & 0.15277(72) & 0.476(11) & 0.825(15) & 0.736(15) \\ \hline
\end{tabular}
\caption{\label{table_su4GK}
The string tension, the lightest and first excited $J^{PC}=0^{++}$
glueball masses and the lightest $2^{++}$ glueball mass: calculated 
at the indicated values of $\beta$ on $L^4$ lattices in SU(4).}
\end{center}
\end{table}

\begin{table}
\begin{center}
\begin{tabular}{|c|c|c|c|c|c|}\hline
\multicolumn{6}{|c|}{SU(6)}  \\ \hline
$\beta$ & $L$ & $a\surd\sigma$ & $am_{0^{++}}$ & $am_{0^{++\star}}$ & $am_{2^{++}}$ \\ \hline
24.350 & 8  & 0.3886(18)  & 0.764(24) & 1.77(11)  & 1.83(11) \\
24.500 & 10 & 0.3416(18)  & 0.837(21) & 1.731(73) & 1.60(6) \\
24.515 & 10 & 0.3385(15)  & 0.856(20) & 1.643(75) & 1.57(6) \\
24.670 & 10 & 0.3075(14)  & 0.812(19) & 1.509(65) & 1.438(47) \\
24.845 & 12 & 0.2801(13)  & 0.774(13) & 1.514(50) & 1.315(35) \\
25.050 & 12 & 0.2510(10)  & 0.727(16) & 1.316(54) & 1.181(35) \\
25.452 & 16 & 0.20992(82) & 0.613(12) & 1.150(24) & 0.989(20) \\ \hline
\end{tabular}
\caption{\label{table_su6GK}
The string tension, the lightest and first excited $J^{PC}=0^{++}$
glueball masses and the lightest $2^{++}$ glueball mass: calculated 
at the indicated values of $\beta$ on $L^4$ lattices in SU(6).}
\end{center}
\end{table}

\begin{table}
\begin{center}
\begin{tabular}{|c|c|c|c|c|c|}\hline
\multicolumn{6}{|c|}{SU(8)}  \\ \hline
$\beta$ & $L$ & $a\surd\sigma$ & $am_{0^{++}}$ & $am_{0^{++\star}}$ & $am_{2^{++}}$ \\ \hline
 43.85 & 8  & 0.3630(21) & 0.835(30) & 1.63(16) & 1.64(12) \\
 44.00 & 10 & 0.3413(13) & 0.829(20) & 1.59(7) & 1.64(6) \\
 44.35 & 10 & 0.3008(14) & 0.804(23) & 1.36(7) & 1.39(6) \\
 44.85 & 12 & 0.2598(12) & 0.776(19) & 1.46(5) & 1.22(3) \\
 45.70 & 16 & 0.2090(13) & 0.632(23) & 1.18(4) & 0.99(3) \\ \hline
\end{tabular}
\caption{\label{table_su8GK}
The string tension, the lightest and first excited $J^{PC}=0^{++}$
glueball masses and the lightest $2^{++}$ glueball mass: calculated 
at the indicated values of $\beta$ on $L^4$ lattices in SU(8).}
\end{center}
\end{table}

\begin{table}
\begin{center}
\begin{tabular}{|c|c|c||c|c|c|c|}\hline
\multicolumn{3}{|c||}{SU(4)} & \multicolumn{4}{|c|}{SU(6)}  \\ \hline
$\beta$ & $a^2\sigma_{k=1}$ &  $a^2\sigma_{k=2}$ & $\beta$ &
$a^2\sigma_{k=1}$ &  $a^2\sigma_{k=2}$ &  $a^2\sigma_{k=3}$\\ \hline
10.550 & 0.1398(11)  & 0.1900(30)   & 24.350 & 0.1510(14)  & 0.2511(70)  & 0.264(12) \\
10.637 & 0.10588(38) & 0.1459(14)   & 24.500 & 0.1167(12)  & 0.1885(40)  & 0.2222(87) \\
10.700 & 0.08863(77) & 0.1186(17)   & 24.515 & 0.1146(11)  & 0.1913(40)  & 0.2145(84) \\
10.789 & 0.07321(40) & 0.10160(83)  & 24.670 & 0.09453(76) & 0.1551(23)  & 0.1787(47) \\
10.870 & 0.06087(51) & 0.08286(108) & 24.845 & 0.07846(65) & 0.1300(13)  & 0.1469(34) \\  
11.085 & 0.03947(34) & 0.05433(80)  & 25.050 & 0.06299(50) & 0.1030(13)  & 0.1201(24) \\
11.400 & 0.02334(22) & 0.03157(49)  & 25.452 & 0.04407(34) & 0.07398(77) & 0.0829(14) \\ \hline    
\end{tabular}
\caption{\label{table_su46l}
Tensions of $k$-strings in our SU(4) and SU(6) calculations.}
\end{center}
\end{table}

\begin{table}
\begin{center}
\begin{tabular}{|c|c|c|c|c|}\hline
\multicolumn{5}{|c|}{SU(8)}  \\ \hline
$\beta$ & $a^2\sigma_{k=1}$ &  $a^2\sigma_{k=2}$ &  $a^2\sigma_{k=3}$ &
$a^2\sigma_{k=4}$ \\ \hline
 43.85 & 0.1318(15)  & 0.2317(56) & 0.276(18)  & 0.296(32) \\
 44.00 & 0.11646(91) & 0.2107(54) & 0.256(14)  & 0.271(31) \\
 44.35 & 0.09049(83) & 0.1580(26) & 0.2012(68) & 0.215(12) \\
 44.85 & 0.06750(64) & 0.1184(25) & 0.1541(40) & 0.1548(60) \\
 45.70 & 0.04370(50) & 0.0782(12) & 0.1003(19) & 0.1135(38) \\ \hline    
\end{tabular}
\caption{\label{table_su8l}
Tensions of $k$-strings in our SU(8) calculations.}
\end{center}
\end{table}

\begin{table}
\begin{center}
\begin{tabular}{|c|c|c|c|c|c|c|}\hline
\multicolumn{7}{|c|}{SU(4) anisotropic: $\xi=0.5$}  \\ \hline
$\beta$ & lattice & $a_t m_{k=1}$ &  $a_t m_{k=2}$ & $a_tm_{0^{++}}$ & 
$a_t m_{0^{++\star}}$ & $a_t m_{2^{++}}$ \\ \hline
10.700 & $8^3 20$  & 0.4021(36) & 0.5692(68) & 0.3920(73) & 0.827(14) & 0.750(25) \\ 
10.850 & $10^3 26$ & 0.3516(43) & 0.5100(81) & 0.3697(55) & 0.724(13) & 0.630(15)  \\ 
11.050 & $12^3 30$ & 0.2878(26) & 0.4048(64) & 0.3234(84) & 0.616(17) & 0.507(13) \\ 
11.325 & $16^3 40$ & 0.2400(21) & 0.3347(69) & 0.2767(82) & 0.558(18) & 0.417(8) \\  \hline
\end{tabular}
\caption{\label{table_su4aM}
Masses of the lightest $k=1$ and $k=2$ winding strings, of 
the lightest and first excited $J^{PC}=0^{++}$
glueball masses and of the lightest $2^{++}$ glueball. Calculated
in SU(4) with a bare anisotropy  $\xi=a_t/a_s=0.5$   
at the indicated values of $\beta$ on the lattices shown.}
\end{center}
\end{table}

\begin{table}
\begin{center}
\begin{tabular}{|c|c|c|c|c|c|c|}\hline
\multicolumn{7}{|c|}{SU(4) anisotropic : $\xi=0.5$}  \\ \hline
$\beta$ & $\xi_r$  & $a_t\surd\sigma$ & 
$\sigma_{k=2}/\sigma_{k=1}$ & $m_{0^{++}}/\surd\sigma$ & 
$m_{0^{++\star}}/\surd\sigma$ & $m_{2^{++}}/\surd\sigma$\\ \hline
10.700 & 0.405(5) & 0.1518(12)  & 1.367(19) & 2.582(52) & 5.45(10) & 4.94(17)\\ 
10.850 & 0.408(8) & 0.1268(15)  & 1.402(26) & 2.916(56) & 5.71(13) & 4.97(13) \\   
11.050 & 0.415(4) & 0.10586(67) & 1.361(24) & 3.055(82) & 5.82(17) & 4.79(13) \\ 
11.325 & 0.430(6) & 0.08489(68) & 1.353(28) & 3.260(99) & 6.57(22) & 4.91(10) \\ \hline
$\infty$ &        &             &  1.358(33) & 3.56(11) & 6.60(23)$^\star$ & 4.85(16) \\ \hline 
\end{tabular}
\caption{\label{table_su4aMK}
The renormalised anisotropy, $\xi_r$, the string tension and
mass ratios from the masses in Table~\ref{table_su4aM}. With
the results of the continuum extrapolations. (The $\star$
indicates a poor fit.)}
\end{center}
\end{table}

\begin{table}
\begin{center}
\begin{tabular}{|c||c|c|c|}\hline
\multicolumn{4}{|c|}{continuum limit} \\ \hline
 & $m_{0^{++}}/\surd\sigma$ & $m_{0^{++\star}}/\surd\sigma$
 & $m_{2^{++}}/\surd\sigma$  \\ \hline
SU(2) & 3.78(7)  & 6.46(14)  & 5.45(11) \\
SU(3) & 3.55(7)  & 5.69(10)  & 4.78(9)  \\ 
SU(4) & 3.36(6)  & 5.63(12)$^\star$  & 4.88(11) \\ 
SU(4)$_{an}$ & 3.56(11) & 6.60(23)$^\star$  & 4.85(16) \\ 
SU(6) & 3.25(9)  & 5.84(18)  & 4.73(15) \\ 
SU(8) & 3.55(12) & 6.40(30)$^\star$  & 4.73(22) \\  \hline 
SU($\infty$) & 3.307(53) & 6.07(17) & 4.80(14) \\  \hline 
\end{tabular}
\caption{\label{table_continuumG}
The continuum limit of the lightest scalar and tensor
glueball masses, and the first excited scalar mass,
all in units of the string tension $\sigma$. (The $^\star$ label
denotes a poor fit.) The extrapolation of these to $N=\infty$ is shown.}
\end{center}
\end{table}

\begin{table}
\begin{center}
\begin{tabular}{|c||c|c|c|}\hline
\multicolumn{4}{|c|}{$L=16$ ; $a \simeq 0.2/\surd\sigma$} \\ \hline
 & $m_{0^{++}}/\surd\sigma$ & $m_{0^{++\star}}/\surd\sigma$
 & $m_{2^{++}}/\surd\sigma$  \\ \hline
SU(2) & 3.493(56)  & 6.16(17)  & 5.61(12) \\
SU(3) & 3.329(58)  & 5.51(9)   & 4.74(7)   \\ 
SU(4) & 3.111(67)  & 5.49(11)  & 4.94(10) \\ 
SU(6) & 2.920(59)  & 5.48(12)  & 4.71(10) \\ 
SU(8) & 3.024(112) & 5.65(20)  & 4.74(15) \\  \hline 
SU($\infty$) & 2.84(7) & 5.52(13) & 4.78(11) \\  \hline 
\end{tabular}
\caption{\label{table_l16G}
The  masses of the lightest scalar and tensor
glueballs, and the first excited scalar,
in units of the string tension $\sigma$, as
calculated at the smallest value of $a$ that
is `common' to  all $N$. The value of  $a\surd\sigma$ 
lies in the range [0.195,0.210] except for SU(2)
where it is $\sim 0.24$. The
extrapolation of these to $N=\infty$ is shown.}
\end{center}
\end{table}

\begin{table}
\begin{center}
\begin{tabular}{|c||c|c|c|}\hline
\multicolumn{4}{|c|}{ $\sigma_{k}/\sigma$ } \\ \hline
 (N,k) & Casimir scaling & this paper & `MQCD' \\ \hline
(4,2) & 1.333 & 1.370(20) &  1.414 \\
(4,2) & 1.333 & 1.358(33) &  1.414 \\
(6,2) & 1.600 & 1.675(31) &  1.732 \\
(6,3) & 1.800 & 1.886(61) &  2.000 \\
(8,2) & 1.714 & 1.779(51) &  1.848 \\
(8,3) & 2.143 & 2.38(10)  &  2.414 \\
(8,4) & 2.286 & 2.69(17)  &  2.613 \\ \hline
\end{tabular}
\caption{\label{table_theoryK}
Predictions of `Casimir Scaling' and `MQCD' compared against
our calculated values of
the ratio of the tension of the lightest $k$-string
to that of the fundamental ($k=1$) string. The second
SU(4) calculation is the one on the anisotropic lattice.}
\end{center}
\end{table}

\clearpage

\begin{table}
\begin{center}
\begin{tabular}{|c||c|c||c|c||c|c|}\hline
\multicolumn{7}{|c|}{ string masses }
\\ \hline
\multicolumn{1}{|c||}{} & \multicolumn{2}{|c||}{SU(4)} & 
 \multicolumn{2}{|c||}{SU(6)} & \multicolumn{2}{|c|}{SU(8)}
\\ \hline
 $k$ & $am$ & CS &  $am$ & CS &  $am$ & CS  \\ \hline
 1  & 0.566(6)  & 0.566 & 0.640(6)  & 0.640 & 0.634(8)  & 0.634 \\
 2A & 0.804(10) & 0.777 & 1.118(13) & 1.063 & 1.186(18) & 1.133 \\ 
 2S & 1.235(20) & 1.451 & 1.411(21) & 1.547 & 1.443(23) & 1.488 \\ 
 3A &           &       & 1.281(16) & 1.204 & 1.542(29) & 1.433 \\ 
 3S &           &       & 2.16(10)  & 2.656 & 2.19(13)  & 2.498 \\ 
 3M &           &       & 1.863(50) & 1.930 & 1.97(6)   & 1.966 \\  \hline
\end{tabular}
\caption{\label{table_unstableK}
Our mass estimates for strings in various representations, 
compared to the Casimir Scaling prediction obtained by scaling
up the $k=1$ mass, assuming the leading string correction. 
All from the isotropic $16^4$ lattice calculations 
at the smallest value of $a$.} 
\end{center}
\end{table}

\begin{table}
\begin{center}
\begin{tabular}{|c||c|c|c||c|c|c||c|c|c|}\hline
\multicolumn{10}{|c|}{ $am_{k}(n_t)$}
\\ \hline
\multicolumn{1}{|c||}{} & \multicolumn{3}{|c||}{SU(4)} & 
\multicolumn{3}{|c||}{SU(6)} & \multicolumn{3}{|c|}{SU(8)}
\\ \hline
$n_t$ & $k=1$ & $k=2S$ & $k=2A^{\star}$ &  $k=1$ & $k=2S$ 
& $k=2A^{\star}$ & $k=1$ & $k=2S$ & $k=2A^{\star}$ \\ \hline
 1 & 0.60(1) & 1.40(1) & 1.40(1) & 0.67(1) & 1.54(1) & 1.67(1) & 0.66(1) & 1.51(1) & 1.69(1)  \\
 2 & 0.57(1) & 1.25(2) & 1.30(2) & 0.64(1) & 1.42(2) & 1.55(2) & 0.64(1) & 1.42(2) & 1.53(3)  \\
 3 & 0.56(1) & 1.22(4) & 1.25(7) & 0.63(1) & 1.39(6) & 1.57(8) & 0.63(1) & 1.56(7) & 1.54(12) \\ \hline
\end{tabular}
\caption{\label{table_meffK2}
Effective masses versus $t=an_t$ for $k=1$ and $k=2$ 
strings in the representations shown. (Errors rounded up.)
From isotropic $16^4$ lattices at our smallest lattice spacing.}
\end{center}
\end{table}

\begin{table}
\begin{center}
\begin{tabular}{|c||c|c|c|c|c|}\hline
\multicolumn{6}{|c|}{ SU(6) : $am_{k}(n_t)$}
\\ \hline
$n_t$ & $k=3S$ & $k=3M$ &  $k=3M^{\star}$ & $k=3A^{\star}$ 
& $k=3A^{\star\star}$ \\ \hline
 1 & 2.55(1)  & 2.03(2)  & 2.46(1) & 1.94(1) & 2.74(2) \\ 
 2 & 2.16(10) & 1.86(5)  & 2.21(9) & 1.77(4) & 2.38(19) \\ \hline
\end{tabular}
\caption{\label{table_meffK3N6}
Effective masses versus $t=an_t$ for $k=3$ 
strings in the representations shown. (Errors rounded up.)
From isotropic $16^4$ lattices at our smallest lattice spacing.}
\end{center}
\end{table}

\begin{table}
\begin{center}
\begin{tabular}{|c||c|c|c|c|c|}\hline
\multicolumn{6}{|c|}{ SU(8) : $am_{k}(n_t)$}
\\ \hline
$n_t$ & $k=3S$ & $k=3M$ &  $k=3M^{\star}$ & $k=3A^{\star}$ 
& $k=3A^{\star\star}$ \\ \hline
 1 & 2.50(2)  & 2.08(2) & 2.52(2)  & 2.15(1) & 3.17(3) \\
 2 & 2.19(13) & 1.97(6) & 2.17(12) & 1.93(7) & 2.53(28) \\ \hline
\end{tabular}
\caption{\label{table_meffK3N8}
Effective masses versus $t=an_t$ for $k=3$ 
strings in the representations shown. (Errors rounded up.)
From isotropic $16^4$ lattices at our smallest lattice spacing.}
\end{center}
\end{table}

\begin{table}
\begin{center}
\begin{tabular}{|c||c|c|c|c||c|c|c|c|}\hline
\multicolumn{9}{|c|}{ SU(4) : $am_{k}(a)$} \\ \hline
\multicolumn{1}{|c||}{} & \multicolumn{4}{|c||}{anisotropic}
 & \multicolumn{4}{|c|}{isotropic} \\ \hline
state  & $k=1$ & $k=2$ & $k=2A$ &  $k=2S$ 
 &  $k=1$ & $k=2$ & $k=2A$ &  $k=2S$ \\ \hline
1 & 0.320(1) & 0.388(2)  &  0.388(2) &  0.787(2) & 0.60(1) & 0.86(1) &  0.86(1) & 1.40(1)   \\
2 & 0.671(1) & 0.779(2)  &  0.779(2) &  1.258(3) & 1.27(1) & 1.36(1) &  1.40(1) & 1.87(1)   \\
3 & 1.063(2) & 0.787(2)  &  1.204(2) &  1.768(5) & 1.77(1) & 1.49(1) &  1.98(1) & 3.03(2)   \\
4 & 1.484(3) & 1.204(2)  &  1.649(5) &  2.299(6) & 3.31(2) & 1.91(1) &  4.01(6) & 6.6(5)    \\
5 & 1.929(5) & 1.258(3)  &  2.144(8) &  2.88(1)  &         & 2.00(1) &          &           \\ \hline
\end{tabular}
\caption{\label{table_mstarN4}
Effective masses at $t=a$ for strings in SU(4) in the representations
shown. The ground and excited states are listed in increasing order 
of mass. Shown separately are results from the anisotropic and
isotropic lattices with the smallest lattice spacing.}
\end{center}
\end{table}

\begin{table}
\begin{center}
\begin{tabular}{|c||c||c|c|c||c|c|c|}\hline
\multicolumn{8}{|c|}{ SU(6) : $am_{k}(a)$} \\ \hline
state  & $k=1$ & $k=2$ & $k=2A$ &  $k=2S$ 
 & $k=3A$ & $k=3M$ &  $k=3S$ \\ \hline
1 & 0.668(2) & 1.18(1) & 1.18(1)  & 1.54(1) & 1.37(1) & 2.03(2) & 2.55(1) \\  
2 & 1.332(3) & 1.53(1) & 1.67(1)  & 1.99(1) & 1.94(1) & 2.46(1) & 3.26(2) \\
3 & 1.894(5) & 1.72(1) & 2.44(1)  & 3.26(2) & 2.74(2) & 3.95(4) & 5.32(16) \\
4 & 3.61(3)  & 2.04(1) & 5.07(12) &         & 5.4(3)  & 6.5(6)  & 7.2(1.1) \\ \hline
\end{tabular}
\caption{\label{table_mstarN6}
Effective masses at $t=a$ for strings in SU(6) in the representations
shown. The ground and excited states are listed in increasing order 
of mass. From the $16^4$ lattice at the smallest lattice spacing.}
\end{center}
\end{table}

\begin{table}
\begin{center}
\begin{tabular}{|c||c||c|c|c||c|c|c|}\hline
\multicolumn{8}{|c|}{ SU(8) : $am_{k}(a)$} \\ \hline
state  & $k=1$ & $k=2$ & $k=2A$ &  $k=2S$ 
 & $k=3A$ & $k=3M$ &  $k=3S$ \\ \hline
1 & 0.658(3) & 1.24(1) & 1.25(1)  & 1.51(1) & 1.64(1)  & 2.08(2) & 2.50(2)  \\  
2 & 1.320(4) & 1.51(1) & 1.69(1)  & 1.95(1) & 2.15(1)  & 2.52(2) & 3.11(3)  \\
3 & 1.887(8) & 1.74(1) & 2.54(2)  & 3.22(3) & 3.17(3)  & 4.09(7) & 5.20(21) \\
4 & 3.56(4)  & 2.01(1) & 5.5(4)   & 6.5(8)  & 6.3(8)   & 6.8(9)  & 6.1(6)   \\ \hline
\end{tabular}
\caption{\label{table_mstarN8}
Effective masses at $t=a$ for strings in SU(8) in the representations
shown. The ground and excited states are listed in increasing order 
of mass. From the $16^4$ lattice at the smallest lattice spacing.}
\end{center}
\end{table}

\begin{table}
\begin{center}
\begin{tabular}{|c||c|c|c|c||c|c|c|c|}\hline
\multicolumn{9}{|c|}{ $am_{k}(a)$} \\ \hline
\multicolumn{1}{|c||}{} & \multicolumn{4}{|c||}{SU(2) : $a=1/8T_c$}
 & \multicolumn{4}{|c|}{SU(3) : $a=1/6T_c$} \\ \hline
state  & $k=1$ & $k=2$ & $k=2A$ &  $k=2S$ 
 &  $k=1$ & $k=2$ & $k=2A$ &  $k=2S$ \\ \hline
1 & 0.573(6)  & -- & -- &  1.20(6) & 0.727(7) & 0.726(8)  &  0.726(8) & 1.641(33) \\
2 & 1.100(12) & -- & -- &  1.77(4) & 1.56(4)  & 1.524(33) &  1.527(33) & 2.64(30) \\
3 & 1.40(10) & --  & -- &  2.9(4)  & 2.25(20) & 1.651(31) &  2.2(2) & --  \\
4 & 2.2(2)   & --  & -- &  --      &  --      &  2.8(4)   &     --  & --  \\ \hline
\end{tabular}
\caption{\label{table_mstarN23}
Effective masses for strings in SU(2) and in SU(3) in the representations
shown. The ground and excited states are listed in increasing order 
of mass.}
\end{center}
\end{table}

\clearpage

\begin	{figure}[p]
\begin	{center}
\leavevmode
\setlength{\unitlength}{0.240900pt}
\ifx\plotpoint\undefined\newsavebox{\plotpoint}\fi
\sbox{\plotpoint}{\rule[-0.200pt]{0.400pt}{0.400pt}}%
\begin{picture}(1500,1800)(0,0)
\font\gnuplot=cmr10 at 12pt
\gnuplot
\sbox{\plotpoint}{\rule[-0.200pt]{0.400pt}{0.400pt}}%
\put(375.0,250.0){\rule[-0.200pt]{4.818pt}{0.400pt}}
\put(350,250){\makebox(0,0)[r]{\ \ {$1.25$}}}
\put(1405.0,250.0){\rule[-0.200pt]{4.818pt}{0.400pt}}
\put(375.0,625.0){\rule[-0.200pt]{4.818pt}{0.400pt}}
\put(350,625){\makebox(0,0)[r]{\ \ {$1.5$}}}
\put(1405.0,625.0){\rule[-0.200pt]{4.818pt}{0.400pt}}
\put(375.0,1000.0){\rule[-0.200pt]{4.818pt}{0.400pt}}
\put(350,1000){\makebox(0,0)[r]{\ \ {$1.75$}}}
\put(1405.0,1000.0){\rule[-0.200pt]{4.818pt}{0.400pt}}
\put(375.0,1375.0){\rule[-0.200pt]{4.818pt}{0.400pt}}
\put(350,1375){\makebox(0,0)[r]{\ \ {$2$}}}
\put(1405.0,1375.0){\rule[-0.200pt]{4.818pt}{0.400pt}}
\put(375.0,1750.0){\rule[-0.200pt]{4.818pt}{0.400pt}}
\put(350,1750){\makebox(0,0)[r]{\ \ {$2.25$}}}
\put(1405.0,1750.0){\rule[-0.200pt]{4.818pt}{0.400pt}}
\put(375.0,250.0){\rule[-0.200pt]{0.400pt}{4.818pt}}
\put(375,200){\makebox(0,0){\ {$0$}}}
\put(375.0,1730.0){\rule[-0.200pt]{0.400pt}{4.818pt}}
\put(534.0,250.0){\rule[-0.200pt]{0.400pt}{4.818pt}}
\put(534,200){\makebox(0,0){\ {$0.025$}}}
\put(534.0,1730.0){\rule[-0.200pt]{0.400pt}{4.818pt}}
\put(693.0,250.0){\rule[-0.200pt]{0.400pt}{4.818pt}}
\put(693,200){\makebox(0,0){\ {$0.05$}}}
\put(693.0,1730.0){\rule[-0.200pt]{0.400pt}{4.818pt}}
\put(852.0,250.0){\rule[-0.200pt]{0.400pt}{4.818pt}}
\put(852,200){\makebox(0,0){\ {$0.075$}}}
\put(852.0,1730.0){\rule[-0.200pt]{0.400pt}{4.818pt}}
\put(1011.0,250.0){\rule[-0.200pt]{0.400pt}{4.818pt}}
\put(1011,200){\makebox(0,0){\ {$0.1$}}}
\put(1011.0,1730.0){\rule[-0.200pt]{0.400pt}{4.818pt}}
\put(1170.0,250.0){\rule[-0.200pt]{0.400pt}{4.818pt}}
\put(1170,200){\makebox(0,0){\ {$0.125$}}}
\put(1170.0,1730.0){\rule[-0.200pt]{0.400pt}{4.818pt}}
\put(1330.0,250.0){\rule[-0.200pt]{0.400pt}{4.818pt}}
\put(1330,200){\makebox(0,0){\ {$0.15$}}}
\put(1330.0,1730.0){\rule[-0.200pt]{0.400pt}{4.818pt}}
\put(375.0,250.0){\rule[-0.200pt]{252.945pt}{0.400pt}}
\put(1425.0,250.0){\rule[-0.200pt]{0.400pt}{361.350pt}}
\put(375.0,1750.0){\rule[-0.200pt]{252.945pt}{0.400pt}}
\put(150,1300){\makebox(0,0){\Large{${{\sigma_k}\over{\sigma}}$}}}
\put(875,25){\makebox(0,0){\large{$a^2\sigma$}}}
\put(375.0,250.0){\rule[-0.200pt]{0.400pt}{361.350pt}}
\put(1336.0,796.0){\rule[-0.200pt]{0.400pt}{35.412pt}}
\put(1326.0,796.0){\rule[-0.200pt]{4.818pt}{0.400pt}}
\put(1326.0,943.0){\rule[-0.200pt]{4.818pt}{0.400pt}}
\put(1118.0,741.0){\rule[-0.200pt]{0.400pt}{27.463pt}}
\put(1108.0,741.0){\rule[-0.200pt]{4.818pt}{0.400pt}}
\put(1108.0,855.0){\rule[-0.200pt]{4.818pt}{0.400pt}}
\put(1104.0,821.0){\rule[-0.200pt]{0.400pt}{27.944pt}}
\put(1094.0,821.0){\rule[-0.200pt]{4.818pt}{0.400pt}}
\put(1094.0,937.0){\rule[-0.200pt]{4.818pt}{0.400pt}}
\put(977.0,795.0){\rule[-0.200pt]{0.400pt}{19.995pt}}
\put(967.0,795.0){\rule[-0.200pt]{4.818pt}{0.400pt}}
\put(967.0,878.0){\rule[-0.200pt]{4.818pt}{0.400pt}}
\put(875.0,828.0){\rule[-0.200pt]{0.400pt}{15.658pt}}
\put(865.0,828.0){\rule[-0.200pt]{4.818pt}{0.400pt}}
\put(865.0,893.0){\rule[-0.200pt]{4.818pt}{0.400pt}}
\put(776.0,791.0){\rule[-0.200pt]{0.400pt}{17.586pt}}
\put(766.0,791.0){\rule[-0.200pt]{4.818pt}{0.400pt}}
\put(766.0,864.0){\rule[-0.200pt]{4.818pt}{0.400pt}}
\put(656.0,861.0){\rule[-0.200pt]{0.400pt}{15.658pt}}
\put(646.0,861.0){\rule[-0.200pt]{4.818pt}{0.400pt}}
\put(1336,869){\circle*{12}}
\put(1118,798){\circle*{12}}
\put(1104,879){\circle*{12}}
\put(977,836){\circle*{12}}
\put(875,860){\circle*{12}}
\put(776,828){\circle*{12}}
\put(656,893){\circle*{12}}
\put(646.0,926.0){\rule[-0.200pt]{4.818pt}{0.400pt}}
\put(1336.0,876.0){\rule[-0.200pt]{0.400pt}{58.539pt}}
\put(1326.0,876.0){\rule[-0.200pt]{4.818pt}{0.400pt}}
\put(1326.0,1119.0){\rule[-0.200pt]{4.818pt}{0.400pt}}
\put(1118.0,1116.0){\rule[-0.200pt]{0.400pt}{55.407pt}}
\put(1108.0,1116.0){\rule[-0.200pt]{4.818pt}{0.400pt}}
\put(1108.0,1346.0){\rule[-0.200pt]{4.818pt}{0.400pt}}
\put(1104.0,1069.0){\rule[-0.200pt]{0.400pt}{54.925pt}}
\put(1094.0,1069.0){\rule[-0.200pt]{4.818pt}{0.400pt}}
\put(1094.0,1297.0){\rule[-0.200pt]{4.818pt}{0.400pt}}
\put(977.0,1132.0){\rule[-0.200pt]{0.400pt}{37.580pt}}
\put(967.0,1132.0){\rule[-0.200pt]{4.818pt}{0.400pt}}
\put(967.0,1288.0){\rule[-0.200pt]{4.818pt}{0.400pt}}
\put(875.0,1114.0){\rule[-0.200pt]{0.400pt}{33.244pt}}
\put(865.0,1114.0){\rule[-0.200pt]{4.818pt}{0.400pt}}
\put(865.0,1252.0){\rule[-0.200pt]{4.818pt}{0.400pt}}
\put(776.0,1174.0){\rule[-0.200pt]{0.400pt}{29.631pt}}
\put(766.0,1174.0){\rule[-0.200pt]{4.818pt}{0.400pt}}
\put(766.0,1297.0){\rule[-0.200pt]{4.818pt}{0.400pt}}
\put(656.0,1144.0){\rule[-0.200pt]{0.400pt}{25.294pt}}
\put(646.0,1144.0){\rule[-0.200pt]{4.818pt}{0.400pt}}
\put(1336,997){\circle{18}}
\put(1118,1231){\circle{18}}
\put(1104,1183){\circle{18}}
\put(977,1210){\circle{18}}
\put(875,1183){\circle{18}}
\put(776,1236){\circle{18}}
\put(656,1197){\circle{18}}
\put(646.0,1249.0){\rule[-0.200pt]{4.818pt}{0.400pt}}
\sbox{\plotpoint}{\rule[-0.500pt]{1.000pt}{1.000pt}}%
\put(375,1204){\usebox{\plotpoint}}
\put(375.00,1204.00){\usebox{\plotpoint}}
\put(395.76,1204.00){\usebox{\plotpoint}}
\put(416.51,1204.00){\usebox{\plotpoint}}
\put(437.27,1204.00){\usebox{\plotpoint}}
\put(458.02,1204.00){\usebox{\plotpoint}}
\put(478.78,1204.00){\usebox{\plotpoint}}
\put(499.49,1205.00){\usebox{\plotpoint}}
\put(520.24,1205.00){\usebox{\plotpoint}}
\put(541.00,1205.00){\usebox{\plotpoint}}
\put(561.75,1205.00){\usebox{\plotpoint}}
\put(582.51,1205.00){\usebox{\plotpoint}}
\put(603.27,1205.00){\usebox{\plotpoint}}
\put(624.02,1205.00){\usebox{\plotpoint}}
\put(644.78,1205.00){\usebox{\plotpoint}}
\put(665.53,1205.00){\usebox{\plotpoint}}
\put(686.29,1205.00){\usebox{\plotpoint}}
\put(707.04,1205.00){\usebox{\plotpoint}}
\put(727.75,1206.00){\usebox{\plotpoint}}
\put(748.51,1206.00){\usebox{\plotpoint}}
\put(769.26,1206.00){\usebox{\plotpoint}}
\put(790.02,1206.00){\usebox{\plotpoint}}
\put(810.77,1206.00){\usebox{\plotpoint}}
\put(831.53,1206.00){\usebox{\plotpoint}}
\put(852.29,1206.00){\usebox{\plotpoint}}
\put(873.04,1206.00){\usebox{\plotpoint}}
\put(893.80,1206.00){\usebox{\plotpoint}}
\put(914.55,1206.00){\usebox{\plotpoint}}
\put(935.31,1206.00){\usebox{\plotpoint}}
\put(956.02,1206.80){\usebox{\plotpoint}}
\put(976.77,1207.00){\usebox{\plotpoint}}
\put(997.52,1207.00){\usebox{\plotpoint}}
\put(1018.28,1207.00){\usebox{\plotpoint}}
\put(1039.04,1207.00){\usebox{\plotpoint}}
\put(1059.79,1207.00){\usebox{\plotpoint}}
\put(1080.55,1207.00){\usebox{\plotpoint}}
\put(1101.30,1207.00){\usebox{\plotpoint}}
\put(1122.06,1207.00){\usebox{\plotpoint}}
\put(1142.81,1207.00){\usebox{\plotpoint}}
\put(1163.57,1207.00){\usebox{\plotpoint}}
\put(1184.31,1207.30){\usebox{\plotpoint}}
\put(1205.03,1208.00){\usebox{\plotpoint}}
\put(1225.79,1208.00){\usebox{\plotpoint}}
\put(1246.54,1208.00){\usebox{\plotpoint}}
\put(1267.30,1208.00){\usebox{\plotpoint}}
\put(1288.06,1208.00){\usebox{\plotpoint}}
\put(1308.81,1208.00){\usebox{\plotpoint}}
\put(1329.57,1208.00){\usebox{\plotpoint}}
\put(1350.32,1208.00){\usebox{\plotpoint}}
\put(1371.08,1208.00){\usebox{\plotpoint}}
\put(1391.83,1208.00){\usebox{\plotpoint}}
\put(1412.59,1208.00){\usebox{\plotpoint}}
\put(1425,1209){\usebox{\plotpoint}}
\sbox{\plotpoint}{\rule[-0.200pt]{0.400pt}{0.400pt}}%
\put(375,888){\usebox{\plotpoint}}
\put(375.00,888.00){\usebox{\plotpoint}}
\put(395.66,886.03){\usebox{\plotpoint}}
\put(416.37,885.00){\usebox{\plotpoint}}
\put(437.04,883.18){\usebox{\plotpoint}}
\put(457.75,882.20){\usebox{\plotpoint}}
\put(478.45,881.00){\usebox{\plotpoint}}
\put(499.13,879.29){\usebox{\plotpoint}}
\put(519.83,878.32){\usebox{\plotpoint}}
\put(540.53,877.00){\usebox{\plotpoint}}
\put(561.21,875.44){\usebox{\plotpoint}}
\put(581.92,874.51){\usebox{\plotpoint}}
\put(602.60,873.00){\usebox{\plotpoint}}
\put(623.30,871.61){\usebox{\plotpoint}}
\put(644.01,870.64){\usebox{\plotpoint}}
\put(664.68,869.00){\usebox{\plotpoint}}
\put(685.38,867.76){\usebox{\plotpoint}}
\put(706.09,866.79){\usebox{\plotpoint}}
\put(726.76,865.00){\usebox{\plotpoint}}
\put(747.46,863.87){\usebox{\plotpoint}}
\put(768.17,862.89){\usebox{\plotpoint}}
\put(788.84,861.01){\usebox{\plotpoint}}
\put(809.55,860.04){\usebox{\plotpoint}}
\put(830.26,859.00){\usebox{\plotpoint}}
\put(850.92,857.11){\usebox{\plotpoint}}
\put(871.63,856.14){\usebox{\plotpoint}}
\put(892.33,855.00){\usebox{\plotpoint}}
\put(913.00,853.27){\usebox{\plotpoint}}
\put(933.71,852.33){\usebox{\plotpoint}}
\put(954.41,851.00){\usebox{\plotpoint}}
\put(975.09,849.45){\usebox{\plotpoint}}
\put(995.80,848.47){\usebox{\plotpoint}}
\put(1016.49,847.00){\usebox{\plotpoint}}
\put(1037.18,845.58){\usebox{\plotpoint}}
\put(1057.88,844.61){\usebox{\plotpoint}}
\put(1078.56,843.00){\usebox{\plotpoint}}
\put(1099.26,841.70){\usebox{\plotpoint}}
\put(1119.97,840.73){\usebox{\plotpoint}}
\put(1140.64,839.00){\usebox{\plotpoint}}
\put(1161.35,837.87){\usebox{\plotpoint}}
\put(1182.06,836.90){\usebox{\plotpoint}}
\put(1202.72,835.00){\usebox{\plotpoint}}
\put(1223.42,833.96){\usebox{\plotpoint}}
\put(1244.14,833.00){\usebox{\plotpoint}}
\put(1264.80,831.11){\usebox{\plotpoint}}
\put(1285.51,830.15){\usebox{\plotpoint}}
\put(1306.21,829.00){\usebox{\plotpoint}}
\put(1326.89,827.28){\usebox{\plotpoint}}
\put(1347.60,826.31){\usebox{\plotpoint}}
\put(1368.29,825.00){\usebox{\plotpoint}}
\put(1388.97,823.40){\usebox{\plotpoint}}
\put(1409.68,822.43){\usebox{\plotpoint}}
\put(1425,821){\usebox{\plotpoint}}
\end{picture}
\end	{center}
\vskip 0.15in
\caption{Ratio of the tensions of the $k=2$ ($\bullet$) and $k=3$ 
($\circ$) strings to that of the fundamental ($k=1$) string 
in SU(6). Plotted against the square of the lattice spacing.
Continuum extrapolations are shown.}
\label{fig_kkn6}
\end 	{figure}
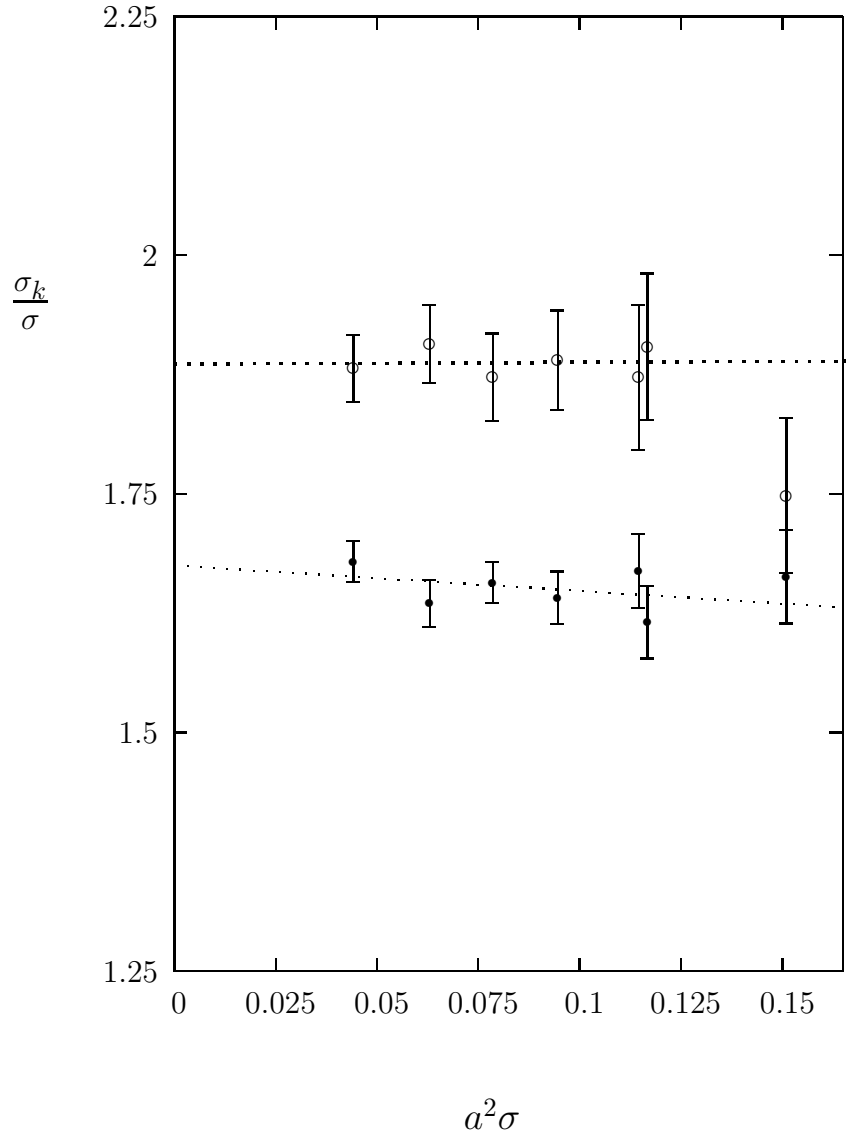

\begin	{figure}[p]
\begin	{center}
\leavevmode
\input	{plot_kkn8}
\end	{center}
\vskip 0.15in
\caption{Ratio of the tensions of the $k=2$ ($\bullet$), $k=3$ ($\circ$) 
and $k=4$ ($\times$) strings to that of the fundamental ($k=1$) string 
in SU(8). Plotted against the square of the lattice spacing.
Continuum extrapolations are shown.}
\label{fig_kkn8}
\end 	{figure}

\begin	{figure}[p]
\begin	{center}
\leavevmode
\setlength{\unitlength}{0.240900pt}
\ifx\plotpoint\undefined\newsavebox{\plotpoint}\fi
\sbox{\plotpoint}{\rule[-0.200pt]{0.400pt}{0.400pt}}%
\begin{picture}(1500,1800)(0,0)
\font\gnuplot=cmr10 at 12pt
\gnuplot
\sbox{\plotpoint}{\rule[-0.200pt]{0.400pt}{0.400pt}}%
\put(350.0,250.0){\rule[-0.200pt]{4.818pt}{0.400pt}}
\put(325,250){\makebox(0,0)[r]{\ \ {$0$}}}
\put(1405.0,250.0){\rule[-0.200pt]{4.818pt}{0.400pt}}
\put(350.0,500.0){\rule[-0.200pt]{4.818pt}{0.400pt}}
\put(325,500){\makebox(0,0)[r]{\ \ {$0.5$}}}
\put(1405.0,500.0){\rule[-0.200pt]{4.818pt}{0.400pt}}
\put(350.0,750.0){\rule[-0.200pt]{4.818pt}{0.400pt}}
\put(325,750){\makebox(0,0)[r]{\ \ {$1$}}}
\put(1405.0,750.0){\rule[-0.200pt]{4.818pt}{0.400pt}}
\put(350.0,1000.0){\rule[-0.200pt]{4.818pt}{0.400pt}}
\put(325,1000){\makebox(0,0)[r]{\ \ {$1.5$}}}
\put(1405.0,1000.0){\rule[-0.200pt]{4.818pt}{0.400pt}}
\put(350.0,1250.0){\rule[-0.200pt]{4.818pt}{0.400pt}}
\put(325,1250){\makebox(0,0)[r]{\ \ {$2$}}}
\put(1405.0,1250.0){\rule[-0.200pt]{4.818pt}{0.400pt}}
\put(350.0,1500.0){\rule[-0.200pt]{4.818pt}{0.400pt}}
\put(325,1500){\makebox(0,0)[r]{\ \ {$2.5$}}}
\put(1405.0,1500.0){\rule[-0.200pt]{4.818pt}{0.400pt}}
\put(350.0,1750.0){\rule[-0.200pt]{4.818pt}{0.400pt}}
\put(325,1750){\makebox(0,0)[r]{\ \ {$3$}}}
\put(1405.0,1750.0){\rule[-0.200pt]{4.818pt}{0.400pt}}
\put(350.0,250.0){\rule[-0.200pt]{0.400pt}{4.818pt}}
\put(350,200){\makebox(0,0){\ {$0$}}}
\put(350.0,1730.0){\rule[-0.200pt]{0.400pt}{4.818pt}}
\put(529.0,250.0){\rule[-0.200pt]{0.400pt}{4.818pt}}
\put(529,200){\makebox(0,0){\ {$0.05$}}}
\put(529.0,1730.0){\rule[-0.200pt]{0.400pt}{4.818pt}}
\put(708.0,250.0){\rule[-0.200pt]{0.400pt}{4.818pt}}
\put(708,200){\makebox(0,0){\ {$0.1$}}}
\put(708.0,1730.0){\rule[-0.200pt]{0.400pt}{4.818pt}}
\put(888.0,250.0){\rule[-0.200pt]{0.400pt}{4.818pt}}
\put(888,200){\makebox(0,0){\ {$0.15$}}}
\put(888.0,1730.0){\rule[-0.200pt]{0.400pt}{4.818pt}}
\put(1067.0,250.0){\rule[-0.200pt]{0.400pt}{4.818pt}}
\put(1067,200){\makebox(0,0){\ {$0.2$}}}
\put(1067.0,1730.0){\rule[-0.200pt]{0.400pt}{4.818pt}}
\put(1246.0,250.0){\rule[-0.200pt]{0.400pt}{4.818pt}}
\put(1246,200){\makebox(0,0){\ {$0.25$}}}
\put(1246.0,1730.0){\rule[-0.200pt]{0.400pt}{4.818pt}}
\put(350.0,250.0){\rule[-0.200pt]{258.967pt}{0.400pt}}
\put(1425.0,250.0){\rule[-0.200pt]{0.400pt}{361.350pt}}
\put(350.0,1750.0){\rule[-0.200pt]{258.967pt}{0.400pt}}
\put(150,1300){\makebox(0,0){\Large{$\frac{\sigma[C_2(0)]}{C_2(0)}$}}}
\put(862,25){\makebox(0,0){\large{$1/N^2$}}}
\put(350.0,250.0){\rule[-0.200pt]{0.400pt}{361.350pt}}
\put(1246.0,950.0){\rule[-0.200pt]{0.400pt}{3.854pt}}
\put(1236.0,950.0){\rule[-0.200pt]{4.818pt}{0.400pt}}
\put(1236.0,966.0){\rule[-0.200pt]{4.818pt}{0.400pt}}
\put(748.0,1206.0){\rule[-0.200pt]{0.400pt}{13.009pt}}
\put(738.0,1206.0){\rule[-0.200pt]{4.818pt}{0.400pt}}
\put(738.0,1260.0){\rule[-0.200pt]{4.818pt}{0.400pt}}
\put(574.0,1126.0){\rule[-0.200pt]{0.400pt}{8.913pt}}
\put(564.0,1126.0){\rule[-0.200pt]{4.818pt}{0.400pt}}
\put(564.0,1163.0){\rule[-0.200pt]{4.818pt}{0.400pt}}
\put(450.0,1031.0){\rule[-0.200pt]{0.400pt}{7.227pt}}
\put(440.0,1031.0){\rule[-0.200pt]{4.818pt}{0.400pt}}
\put(440.0,1061.0){\rule[-0.200pt]{4.818pt}{0.400pt}}
\put(406.0,990.0){\rule[-0.200pt]{0.400pt}{5.059pt}}
\put(396.0,990.0){\rule[-0.200pt]{4.818pt}{0.400pt}}
\put(1246,958){\circle*{12}}
\put(748,1233){\circle*{12}}
\put(574,1145){\circle*{12}}
\put(450,1046){\circle*{12}}
\put(406,1000){\circle*{12}}
\put(396.0,1011.0){\rule[-0.200pt]{4.818pt}{0.400pt}}
\put(350,966){\usebox{\plotpoint}}
\put(350.00,966.00){\usebox{\plotpoint}}
\put(367.04,977.84){\usebox{\plotpoint}}
\put(383.99,989.79){\usebox{\plotpoint}}
\put(400.46,1002.42){\usebox{\plotpoint}}
\put(417.24,1014.63){\usebox{\plotpoint}}
\put(434.37,1026.33){\usebox{\plotpoint}}
\put(451.27,1038.38){\usebox{\plotpoint}}
\put(467.74,1050.99){\usebox{\plotpoint}}
\put(484.48,1063.26){\usebox{\plotpoint}}
\put(501.27,1075.47){\usebox{\plotpoint}}
\put(518.51,1087.01){\usebox{\plotpoint}}
\put(535.29,1099.23){\usebox{\plotpoint}}
\put(551.72,1111.89){\usebox{\plotpoint}}
\put(568.51,1124.10){\usebox{\plotpoint}}
\put(585.61,1135.84){\usebox{\plotpoint}}
\put(602.54,1147.84){\usebox{\plotpoint}}
\put(619.04,1160.43){\usebox{\plotpoint}}
\put(635.75,1172.73){\usebox{\plotpoint}}
\put(652.95,1184.33){\usebox{\plotpoint}}
\put(669.78,1196.47){\usebox{\plotpoint}}
\put(686.56,1208.68){\usebox{\plotpoint}}
\put(702.99,1221.36){\usebox{\plotpoint}}
\put(719.81,1233.52){\usebox{\plotpoint}}
\put(737.02,1245.11){\usebox{\plotpoint}}
\put(753.81,1257.31){\usebox{\plotpoint}}
\put(770.33,1269.86){\usebox{\plotpoint}}
\put(787.02,1282.20){\usebox{\plotpoint}}
\put(804.19,1293.85){\usebox{\plotpoint}}
\put(821.05,1305.94){\usebox{\plotpoint}}
\put(837.83,1318.15){\usebox{\plotpoint}}
\put(854.26,1330.83){\usebox{\plotpoint}}
\put(871.50,1342.37){\usebox{\plotpoint}}
\put(888.29,1354.57){\usebox{\plotpoint}}
\put(905.07,1366.78){\usebox{\plotpoint}}
\put(921.86,1378.99){\usebox{\plotpoint}}
\put(938.39,1391.52){\usebox{\plotpoint}}
\put(955.53,1403.20){\usebox{\plotpoint}}
\put(972.32,1415.41){\usebox{\plotpoint}}
\put(989.10,1427.62){\usebox{\plotpoint}}
\put(1005.94,1439.76){\usebox{\plotpoint}}
\put(1022.80,1451.85){\usebox{\plotpoint}}
\put(1039.59,1464.06){\usebox{\plotpoint}}
\put(1056.37,1476.27){\usebox{\plotpoint}}
\put(1073.16,1488.48){\usebox{\plotpoint}}
\put(1090.07,1500.51){\usebox{\plotpoint}}
\put(1106.86,1512.71){\usebox{\plotpoint}}
\put(1123.64,1524.92){\usebox{\plotpoint}}
\put(1140.43,1537.13){\usebox{\plotpoint}}
\put(1157.26,1549.28){\usebox{\plotpoint}}
\put(1174.13,1561.37){\usebox{\plotpoint}}
\put(1190.91,1573.57){\usebox{\plotpoint}}
\put(1207.70,1585.78){\usebox{\plotpoint}}
\put(1224.72,1597.64){\usebox{\plotpoint}}
\put(1241.37,1610.00){\usebox{\plotpoint}}
\put(1258.16,1622.20){\usebox{\plotpoint}}
\put(1274.94,1634.41){\usebox{\plotpoint}}
\put(1291.73,1646.62){\usebox{\plotpoint}}
\put(1308.87,1658.29){\usebox{\plotpoint}}
\put(1325.40,1670.83){\usebox{\plotpoint}}
\put(1342.18,1683.04){\usebox{\plotpoint}}
\put(1358.97,1695.25){\usebox{\plotpoint}}
\put(1376.21,1706.79){\usebox{\plotpoint}}
\put(1392.64,1719.46){\usebox{\plotpoint}}
\put(1409.42,1731.67){\usebox{\plotpoint}}
\put(1425,1743){\usebox{\plotpoint}}
\end{picture}
\end	{center}
\vskip 0.15in
\caption{The fluctuation on the two point correlation function 
$C_2(t)$ defined in eqns(\ref{eqn_factcorr3} - \ref{eqn_factcorr8}). 
The $N$-dependence
on $10^4$ lattices at fixed lattice spacing, $a \simeq 1/5T_c$.} 
\label{fig_Nfluctcor}
\end 	{figure}
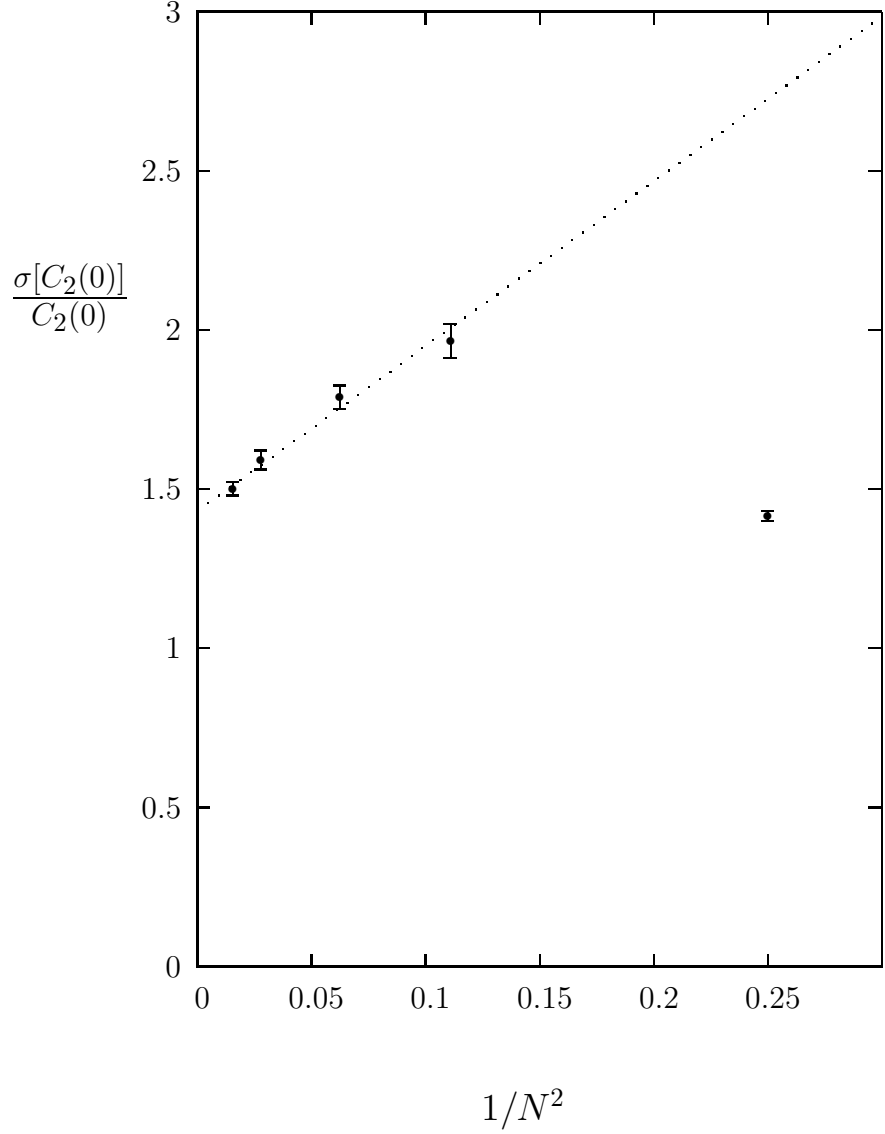

\end{document}